\documentclass[12pt,reqno]{amsart}
\usepackage{comment}
\usepackage{array}
\usepackage[mathscr]{euscript}
\usepackage[T2A]{fontenc}
\usepackage[utf8x]{inputenc}
\usepackage{breqn}
\usepackage{todonotes}
\usepackage{slashed}
\usepackage{youngtab}
\usepackage{amsmath}
\usepackage{amssymb}
\usepackage{amsxtra}
\usepackage{color}
\usepackage[margin=1.2in]{geometry}

\usepackage[pagebackref=true]{hyperref}
\hypersetup{
  linkbordercolor = blue,
   extension = notused, }

 \newcommand{\beq}{\begin{equation}}
                \newcommand{\bea}{\begin{eqnarray}}
                \newcommand{\eea}{\end{eqnarray}}
                 \newcommand{\eeq}{\end{equation}}

\newcommand{\pt} {\tt{p}}
\newcommand{\pyt} {\tt{p}_{\yt}}
\newcommand{\yt} {\tt{y}}
\newcommand {\bA}   {\mathbb A}

\newcommand {\BC}   {\mathbb C}
\newcommand{\sE} {\mathscr E}

\newcommand {\BR}   {\mathbb R}
\newcommand {\BP}   {\mathbb P}

\newcommand {\bR}   {\mathbf{R}}

\newcommand {\bV}   {\mathscr V}
\newcommand {\qe} {\mathfrak q}

\newcommand {\ii} {\mathrm{i}}
\newcommand {\half} {{\scriptstyle{\frac{1}{2}}}}
\newcommand {\qalf} {{\scriptstyle{\frac{1}{4}}}}

\newcommand {\ba}  {\mathfrak{a}}
\newcommand {\bh}  {\mathfrak{h}}

\newcommand {\bv} {\mathbf{v}}

\newcommand {\vt} {\vartheta}

\newcommand {\bnu} {\underline{\boldsymbol{\nu}}}

\newcommand {\bu}{ \mathbf{u}}
\newcommand {\bx}{  \mathbf{x}}

\newcommand {\bz}{  \mathbf{z}}

\newcommand {\BS}   {\mathbb S}

\newcommand {\BZ}   {\mathbb Z}

\newcommand {\CP}   {\mathbb C \mathbb P}

\newcommand {\zb} {{\bar z}}

\newcommand {\CalA} {\mathcal A}
\newcommand {\CalB} {\mathcal B}
\newcommand {\CalC} {\mathcal C}

\newcommand {\CalD} {\mathcal D}

\newcommand {\CalF} {\mathcal F}

\newcommand {\CalH} {\mathcal H}

\newcommand {\CalL} {\mathcal L}
\newcommand {\CalM} {\mathcal M}
\newcommand {\mfM} {\mathfrak{M}}
\newcommand {\CalN} {\mathcal N}
\newcommand {\CalO} {\mathcal O}
\newcommand {\CalP} {\mathcal P}
\newcommand {\CalQ} {\mathcal Q}
\newcommand {\CalR} {\mathcal R}
\newcommand {\CalS} {\mathcal S}

\newcommand {\CalU} {\mathcal U}
\newcommand {\CalV} {\mathcal V}
\newcommand {\CalX} {\mathcal X}
\newcommand {\CalY} {\mathcal Y}
\newcommand {\CalW} {\mathcal W}
\newcommand {\CalZ} {\mathcal Z}

\newcommand {\xt} {\mathrm{x}}
\newcommand {\gt} {\mathrm{g}}

\newcommand{\ve}{\varepsilon}
\newcommand{\ep}{\epsilon}

\renewcommand{\hat}{\widehat}

\begin{document}

\title[Blowups in BPS/CFT]{Blowups  in BPS/CFT correspondence, \\
and Painlev{\'e} VI}

\author{Nikita Nekrasov}

\address{Simons Center for Geometry and Physics,\\
C.~N.~Yang Institute for Theoretical Physics,\\
Stony Brook University, Stony Brook NY 11794-3636, USA\\
\footnote{on leave of absence from:
Center for Advanced Studies, Skoltech and IITP RAS, Moscow, Russia}
E-mail: nikitastring@gmail.com}

\begin{abstract}

We study four dimensional supersymmetric gauge theory in the presence of surface and point-like defects (blowups) and propose an identity relating partition functions at different values of $\Omega$-deformation parameters $({\ve}_{1}, {\ve}_{2})$. As a consequence, we obtain the formula conjectured in 2012 by O.~Gamayun, N.~Iorgov, and O.~Lysovyy, relating the tau-function  ${\tau}_{PVI}$ to $c=1$ conformal blocks of Liouville theory and propose its generalization for the case of Garnier-Schlesinger system. To this end we clarify the notion of the quasiclassical tau-function ${\tau}_{PVI}$ of Painlev{\'e} VI and its generalizations.
We also make some remarks about the sphere partition functions, the boundary operator product expansion in the ${\CalN}=(4,4)$ sigma models related to four dimensional ${\CalN}=2$ theories on toric manifolds, discuss crossed instantons on conifolds, elucidate some aspects of the BPZ/KZ correspondence, and applications to quantization. 
\end{abstract}

\maketitle

\section{Introduction}\label{aba:sec1}

Since the discovery of Kramers and Wannier \cite{KrWa} of the duality relating different temperature regimes of the Ising model the search for strong-weak coupling transformations mapping the regime of complicated dynamics in one quantum field theory to the regime of controlled computations of, possibly, another, has been ongoing. In the case of supersymmetric field theory the conjectured dualities e.g. \cite{Montonen:1977sn} can be tested with the help of localization techniques introduced in \cite{Nekrasov:2002qd}, see also \cite{Pestun:2007rz, Kapustin:2009kz}. One instance of such duality is the classical/quantum correspondence \cite{NW}, \cite{NRS}, 
relating spectrum of one quantum mechanical model to symplectic geometry of another, classical, model. For Hitchin systems \cite{Hitchin} this correspondence is connected to Langlands duality \cite{BD},\cite{Feigin:1994in}. Another example of the same duality, in Sine-Gordon theory, is found in \cite{LZSG}. Yet another interesting duality \cite{ising2} maps the correlation functions in one quantum system, e.g. spin-spin correlators in the abovementioned Ising model, to some classical dynamics, described, e.g. by the Painlev{\'e} III \cite{ising1} or Painlev{\'e} VI equations \cite{JMising}. 

In this paper, which is, in many ways a part VI of the series \cite{NekBPSCFT}, although the number VI in the title stands for something else\footnote{So that by taking the confluent limits from PVI to PV, PIV, PIII, etc. one will not recover the previous titles in \cite{NekBPSCFT}}, we explore a relatively recently found connection between a classical mechanical system and $c=1$ conformal blocks of a two dimensional conformal field theory. The conformal blocks behave in many respects as the wavefunctions of some quantum mechanical system. So we could loosely call this relation a classical/quantum correspondence.  Actually, we shall find it useful to deform both sides of the correspondence (in some sense, quantize the classical mechanical side, while $\beta$-deforming the conformal block side):
\beq
{\bf\Psi}({\bf a}, {\ve}_{1}, {\ve}_{2} , {\bf m} ; {\bf w}, {\qe}) = \sum_{{\mathfrak{n}}\in {\Lambda}}  {\bf\Psi}({\bf a} + {\ve}_{1} {\mathfrak{n}}, {\ve}_{1}, {\ve}_{2}-{\ve}_{1} , {\bf m} ; {\bf w}, {\qe})  Z ({\bf a} + {\ve}_{2} {\mathfrak{n}}, {\ve}_{1}-{\ve}_{2}, {\ve}_{2} , {\bf m} , {\qe}) \ .
\label{eq:blowsur}
\eeq
Here ${\bf\Psi}({\bf a}, {\ve}_{1}, {\ve}_{2} , {\bf m} ; {\bf w}, {\qe})$, 
$Z ({\bf a} , {\ve}_{1}, {\ve}_{2} , {\bf m} , {\qe})$ are the conformal blocks of the current algebra and ${\CalW}$-algebra, respectively. 
The natural habitat for \eqref{eq:blowsur} is the four dimensional ${\CalN}=2$ supersymmetric $\Omega$-deformed gauge theory, where it is the relation
between the (unnormalized) expectation value ${\bf\Psi}$ of a surface defect located at the surface $z_{2}=0$ with its own couplings ${\bf w}$, and
the supersymmetric partition function $Z$ of the theory on ${\BR}^{4}$, with the bulk coupling $\qe$:
\beq
{\qe} = e^{- \frac{8{\pi}^{2}}{g^{2}}} e^{\ii {\vt}}\ .
\eeq 
The relation \eqref{eq:blowsur} accompanies the well-known equivariant blowup formula
\beq
Z ({\bf a} , {\ve}_{1}, {\ve}_{2} , {\bf m} , {\qe}) = \sum_{{\mathfrak{n}}\in {\Lambda}}  Z ({\bf a} + {\ve}_{1} {\mathfrak{n}}, {\ve}_{1}, {\ve}_{2}-{\ve}_{1} , {\bf m} , {\qe})
 Z ({\bf a} + {\ve}_{2} {\mathfrak{n}}, {\ve}_{1}-{\ve}_{2}, {\ve}_{2} , {\bf m} , {\qe})
\label{eq:blowbulk}
\eeq
found in \cite{NY}. 

We shall exploit the consequences of the Eq.\eqref{eq:blowsur} in the context of the BPS/CFT correspondence, which we review in the next section. Our main focus will be on the limit ${\ve}_{1} \to 0$. The surface defect has the thermodynamic limit
\beq
{\bf\Psi}({\bf a}, {\ve}_{1}, {\ve}_{2} , {\bf m} ; {\bf w}, {\qe})  \sim {\exp}\ \frac{1}{{\ve}_{1}} {\bf S}({\bf a}, {\ve}_{2} , {\bf m} ; {\bf w}, {\qe}) 
\eeq
for some \emph{complexified free energy} ${\bf S}({\bf a}, {\ve}_{2} , {\bf m}; {\bf w}, {\qe})$, which is a twisted superpotential
of the effective ${\CalN}=(2,2)$ two dimensional theory. On the CFT side ${\bf\Psi}$ is identified with a certain conformal block obeying first order differential equation with respect to ${\qe}$, which becomes a Hamilton-Jacobi equation for ${\bf S}$ in the ${\ve}_{1} \to 0$ limit. We show, that the exponentiated ${\ve}_{2}$-derivative of the Hamilton-Jacobi potential ${\bf S}$, when expressed in appropriate variables, coincides with the Painlev{\'e} VI $\tau$-function (as well as formulate a conjecture about more general isomonodromy problems):
\beq
{\tau} ({\alpha}, {\beta} ;  {\vec\vt} ; {\qe} ) = {\exp} \, \frac{\partial {\bf S}}{\partial \ve_{2}} \, , 
\eeq
with the monodromy data ${\alpha}$, $\beta$, and ${\vec\vt}$ expressed in terms of ${\bf a}/{\ve}_{2}, {\bf w}$, and ${\bf m}/{\ve}_{2}$ (the identification of the monodromy data is achieved in the forthcoming companion paper \cite{SJNN}). For $SU(2)$ theory with $N_f =4$ hypermultiplets the relation \eqref{eq:blowsur}
becomes, in the limit ${\ve}_{1} \to 0$, the one found in 2012 by O.~Gamayun, N.~Iorgov and O.~Lysovyy (GIL for short).

The paper is organized as follows. First, we recall a few facts about the supersymmetric gauge theories, twists, $\Omega$-deformations, surface defects, and blowup formulas. Then we proceed with some remarks about non-stationary integrable systems, and their $\tau$-functions. We show, for a restricted class of such systems (which include $SL_2$ Schlesinger equations in genus $0$), how to relate the $\tau$-function to the Hamilton-Jacobi potential.
After that we return to the realm of supersymmetic gauge theories. We review the non-perturbative Dyson-Schwinger equation in ${\CalN}=2$ $SU(2)$ theory with $N_f=4$ fundamental hypermultiplets, and show that the surface defect partition function obeys the BPZ \cite{BPZ} equation, whose ${\ve}_{1} \to 0$ limit gives the Painlev{\'e} VI equation (PVI, for short) . We also study more sophisticated defects and show that their partition functions can be organized into a horizontal section ${\bf\Upsilon}$  of a meromorphic flat connection on a $4$-punctured sphere. Then we state the blowup formula in the presence of the surface defect and establish the relation between the PVI Hamilton-Jacobi potential and the partition functions of the bulk theory taken at the special $\Omega$-background locus ${\ve}_{1} + {\ve}_{2} = 0$. In this way we recover the GIL formula. 

In the appendix we recall a few facts about the geometry of the blown up ${\BC}^{2}$, and the arguments leading to the blowup formula.  
We also indicate how one extends the gauge origami constructions of \cite{NN2004,NekBPSCFT} to the simplest nontrivial toric Calabi-Yau fourfold ${\CalY}$: 
the product of resolved conifold ${\CalZ}$ and a complex line, ${\CalY} = {\CalZ} \times {\BC}$, in other words a total space of the 
${\CalO}(-1) \oplus {\CalO}(-1) \oplus {\CalO}$ bundle over ${\BP}^1$.  The key point is that the blowup $\widehat{{\BC}^{2}}$
is a hypersurface of the resolved conifold $\CalZ$. 

Many crucial points, in particular the relation between the potential ${\bf S}$, Coulomb moduli $\bf a$, and the monodromy of the meromorphic connections, as well as the detailed relation between ${\bf\Upsilon}$  and intersecting defects,  will be established in the companion paper \cite{SJNN}. 

{\bf Acknowledgements.} I thank M.~Bershtein and A.~Zamolodchikov for discussions in 2013 which helped to sharpen the formula \eqref{eq:blowsur}, and to S.~Jeong for numerous discussions and collaboration in \cite{SJNN}.  I am also grateful to C.~LeBrun, I.~Krichever,  H.~Nakajima
and A.~Okounkov  for their comments. My interest in blowup formulas  goes back to 1996-1997, thanks to the numerous discussions  with A.~Losev aimed at solving the constraints the electromagnetic duality of effective theory imposes on the contact terms between supersymmetric non-local observables, while the $qq$-characters, the main technical tool of our analysis, grew out of our work with V.~Pestun \cite{NP1}. 

When \cite{Gamayun:2012ma} came out, we immediately pointed out the connection of their formula to the ${\ve}_{2} = -{\ve}_{1}$, $SU(2)$, $N_{f}=4$ supersymmetric partition function and to the blowup equations\cite{NikPR}. Since then a few approaches to the formula \cite{Gamayun:2012ma} were put forward \cite{Bershtein:2014yia,Iorgov:2014vla,Bershtein:2018zcz}, as well as some generalizations \cite{Grassi:2016nnt}. Our approach is different and in some sense more direct as it produces the result in the GIL form. I  lectured about it at the IHES, Saint Petersbourg Steklov Mathematics Institute, Harvard University (CMS and Physics Department), Skoltech Center for Advanced Studies, Northeastern University, Department of Mathematics at the Higher School of Economics, at the Physics Department at Ludwig Maximilian University, at the Imperial College London, at the Department of Mathematics at Columbia University.  I am grateful to these institutions for their hospitality and to the audiences for useful discussions. 

The need to go to extra dimensions to make the formula look natural  may sound, to algebraically minded, like proving the main theorem of algebra using topology. 
But we always liked that proof \cite{Arnold}....

\vskip 1cm

\section{BPS/CFT correspondence}

{}In the early days 
of exact computations in supersymmetric gauge theories \cite{Vafa:1994tf,Nekrasov:2002qd,Losev:2003py,Nekrasov:2003rj,NekBPSCFT}, which were built on the works of H.~Nakajima \cite{Nakajima1,Nakajima2, Nakajima3, NakHilb}, it was observed, that
the correlation functions of selected observables, including the partition function, coincide with conformal blocks of some two dimensional conformal field theories, or, more generally, are given by the matrix elements of representations of some infinite dimensional algebras, such as Kac-Moody, Virasoro, or their $q$-deformations. 
In \cite{Nekrasov:2002qd} this phenomenon was attributed to the chiral nature of the tensor field propagating on the worldvolume of the fivebranes. The fivebranes ($M5$ branes in $M$-theory and $NS5$ branes in $IIA$ string) were used in \cite{Klemm:1996bj}, \cite{Witten5} to engineer, in string theory setup, the supersymmetric systems whose low energy is described by ${\CalN}=2$ supersymmetric gauge theories in four dimensions. This construction was
extended and generalized in \cite{Gaiotto:2009we}.  
This correspondence, named  \emph{the BPS/CFT correspondence} in \cite{NN2004}  has been supported by a large class of very detailed examples in
 \cite{NekBPSCFT,Alday:2009aq, Alday:2010vg}, and more recently in \cite{KT, Kimura:2015rgi, Haouzi:2019jzk, Haouzi:2020yxy}.

In this paper we shall be mostly dealing with the ${\CalN}=2$ supersymmetric theory with gauge group $SU(2)$ 
and $N_f =4$ matter hypermultiplets in fundamental representation. In \cite{Alday:2009aq} the partition function of this theory is identified with the $4$-point conformal block of Liouville theory. In \cite{Alday:2010vg,KT} the expectation value of a regular surface defect  in that theory is conjectured to be given by the
$\widehat{\mathfrak{sl}}_{2}$ conformal block. Earlier, in \cite{NN2004}, we presented less precise conjectures
about the expectation values of surface defects and 
Knizhnik-Zamolodchikov \cite{KZ} (KZ) equations. The surface defects or surface observables in Donaldson theory and its generalizations were studied in \cite{KM, Losev:1995cr,
Braverman:2004, NN2004, GukovWitten}. In \cite{NekBPSCFT}, using the construction of \cite{ChainSaw} and the theory of $qq$-characters, the expectation value of certain surface defects has been rigorously shown to obey the equations, which are equivalent to Belavin-Polyakov-Zamolodchikov  \cite{BPZ} and Knizhnik-Zamolodchikov \cite{KZ} equations
\cite{NT}.  

In this paper we utilize these results to deduce a nontrivial property of conformal blocks of two dimensional quantum field theories 
from a natural property of correlation functions of four dimensional supersymmetric gauge theories, adding one more piece to the puzzling complete set of constraints on a quantum field theory. 

\vskip 1cm

\section{Gauge theory and defects}

{}${\CalN}=2$ supersymmetric theory can be coupled to ${\CalN}=2$ supergravity. In the limit
$M_{Planck} \to \infty$ it becomes non-dynamical. Among the supergravity backgrounds 
the ones with global fermionic symmetries ${\CalQ}$  are important as the correlation functions
of ${\CalQ}$-invariant operators, in particular, the partition function, exactly coincide with those of a potentially much simpler theory whose fields are the ${\CalQ}$-invariant fields of the original theory. In other words, the superversion of Kaluza-Klein reduction is exact (recall that Kaluza-Klein reduction of gravity theory on the manifold $X$ with $G$-isometry is the approximation to the gravity theory on $B = X/G$ coupled to $G$-gauge theory and a sigma model valued in the space of $G$-invariant metrics on the generic fiber $F$ of the projection $X \to X/G$). The simplest, in which the background $SU(2)$ R-symmetry connection is taken to coincide with the $SU(2)_{R}$-part of the $Spin(4) = SU(2)_L \times SU(2)_R$ spin connection, is Donaldson-Witten twist. The resulting $(4|8)$-dimensional supermanifold on which the theory lives is split ${\CalS}_{M} = {\Pi}E_{M}$ where $E_{M}$ is the vector bundle over the ordinary four-dimensional manifold $M$, $E = T_{M} \oplus {\Lambda}^{2,+}T^*_{M}\oplus {\BR}$
is the sum of the tangent bundle, the bundle of self-dual two-forms, and a one dimensional trivial vector bundle. 

There are other backgrounds of ${\CalN}=2$ supergravity, admitting global fermionic symmetries, albeit for 
a restricted class of underlying four-manifolds $M$. These backgrounds, starting with the 
$\Omega$-background in \cite{Nekrasov:2002qd}, Pestun's background in 
\cite{Pestun:2007rz}, and their generalizations \cite{Closset:2014uda}, can be used to perform
exact computations in supersymmetric theories, learning about the effective low-energy action, gravitational topological susceptibility and other interesting aspects of the theory. 

In this paper we shall be dealing with the $\Omega$-backgrounds, which are associated with four-manifolds $M$ with isometries. Remarkably, the important parameters, such as the central charge $c$ or background charge $Q$, of the  two dimensional conformal field theories which are BPS/CFT duals of theories subject to such backgrounds, depend on the $\Omega$-background parameters $({\ve}_{1}, {\ve}_{2})$ \cite{Alday:2009aq}.

Moreover, the limit, where one of the parameters of the $\Omega$-background, e.g. ${\ve}_{1}$, goes to zero, 
corresponds to the semiclassical, $c \to \infty$, limit of the corresponding CFT. 
The relations \eqref{eq:blowsur}, \eqref{eq:blowbulk} then relate this limit to the limit ${\ve}_{1}+{\ve}_{2} \to 0$, 
which corresponds to the ``free fermion'' $c=1$ point.

\subsection{Gauge theory on the blowup}

Let us compare the gauge theory on the supermanifolds
${\CalS}_{M}$ and ${\CalS}_{M'}$ for which the underlying four-manifolds $M$ and $M'$
differ only in the neighborhood of a point $p \in M$. For example, let 
$M'$ be obtained by cutting a small ball $B^{4}_{\ve}(p)$ centered at $p$, with the boundary $S^3_{\ve} = {\partial}B^{4}_{\ve}(p)$, and contracting the orbits of Hopf $U(1)$ action on the latter. In other words, in the local model of $S^{3}_{\ve}$
\beq
|u_{1}|^2 + |u_{2}|^{2} = {\ve}^2
\eeq
with $(u_1, u_2) \in {\BC}^2 \approx {\BR}^{4} \approx T_{p}M$, we identify the point
$(u_1, u_2)$ with the point $(u_1 e^{\ii\vartheta}, u_2  e^{\ii\vartheta})$ for any $\vartheta$. 
The boundary $S^3_{\ve}$ disappears, leaving  a non-contractible copy of ${\BC\BP}^1$ called the exceptional curve (or, more generally, an exceptional set). 
The resulting space $M'$ is the blowup of $M$ at the point $p$. Different parameterizations of the neighborhood $B^4_{\ve}(p)$ of $p$ lead to different spaces, but for small $\ve$ they are all diffeomorphic. 

Now consider a quantum field theory on the manifolds $M$ and $M'$. Since $M \backslash B^{4}_{\ve}(p)$ and $M' \backslash \{ $ small, size $\sim \ve$ neighborhood of the exceptional set $\}$ are diffeomorphic, the topological quantum field theories on $M$ and $M'$ share the same space of states, defined on a small three-sphere. On $M$ this sphere is the boundary of $B^4_{\ve}(p)$ while on $M'$ it is a boundary of a tubular neighborhood of the exceptional curve, the total space of the bundle of small circles over ${\BC\BP}^{1}$.
The partition functions $Z_M$ and $Z_M'$ are, therefore, the matrix elements between the vector $\langle out |$ representing the path integral over  $M \backslash B^{4}_{\ve}(p)$
and the vectors $| B^4 \rangle$, and $| {\overline{\BC\BP}}^2 \backslash {\tilde B}^4 \rangle$, representing the path integrals over the fillings of the three-sphere inside $M$ and $M'$, respectively. 

Now let us use the state-operator correspondence, and represent these vectors by local operators. The state $| B^4 \rangle$ corresponds, naturally, to the unit operator $1$, while
the state $| {\overline{\BC\BP}}^2 \backslash {\tilde B}^4 \rangle$ is, in general, a nontrivial element of the topological ring. It was argued in \cite{Losev:1997tp} for asymptotically free theories, and shown in \cite{NY} for $SU(n)$ theory with up to $2n$ fundamental hypermultiplets, that the state $| {\overline{\BC\BP}}^2 \backslash {\tilde B}^4 \rangle$ also
corresponds to the unit operator in the topological ring:
\beq
Z_{M'} = Z_{M}
\label{eq:blow}
\eeq
Everything above can be generalized to the equivariant topological field theories, which are defined on four manifolds with some symmetry. The generalization, in which the topological nilpotent supercharge is deformed to the equivariant differential ${\CalQ}_{V}$, which squares to the infinitesimal isometry $V$, is called the $\Omega$-deformation, the corresponding supergravity background the $\Omega$-background. 

The computations of correlation functions of the ${\CalQ}_{V}$-closed observables in the $\Omega$-background on four-manifold $M^4$ are facilitated by the fact that the ${\CalQ}_{V}$-invariant field configurations tend to be concentrated near the $V$-fixed points on $M^4$, so that for noncompact $M^4$ (cf. \cite{NekLisbon}):
\beq
Z_{M^4} = \sum_{\rm fluxes} \prod_{\rm fixed\ points} Z_{\rm loc}  
\eeq
where by \emph{fluxes} we mean the magnetic fluxes (in the low-energy effective theory) passing through compact two-cycles on $M^4$. The noncompactness here is important, as the scalar ${\phi}$ in the vector multiplet has an expectation value fixed by the boundary conditions at infinity.

\subsection{Blowup formulas}

Since the work of Fintushel and Stern \cite{FS} on invariants of smooth $4$-manifolds, comparing the Donaldson invariants of $M$ and $M' \approx M \# {\overline{{\BC\BP}^2}}$ has been a fruitful exercise in Donaldson theory. In fact, it was in \cite{FS}  where the connection of Donaldson theory to a family of elliptic curves was found. The very same family of elliptic curves was later found in \cite{Seiberg:1994rs} to govern the exact low-energy effective action of the theory. In \cite{Losev:1997tp} and later in \cite{Lossev:1997bz} these observations were turned around into a way of testing Seiberg-Witten theory of a larger class of ${\CalN}=2$ gauge theories in four dimensions. Finally, in \cite{NY} the equivariant version of the blowup relations was proven, leading to the proof of the \cite{Nekrasov:2002qd} conjecture. We should point out that in the literature the term ``blowup formula'' \cite{FS}, \cite{Moore:1997pc}, \cite{GZ}, \cite{Losev:1997tp}, \cite{NY} usually refers to the representation of the expectation value of some observables
associated to the exceptional cycle on $M'$, as a local operator in the theory on $M$. This local operator
represents the point-like (codimension four) defect in supersymmetric gauge theory. In what follows we shall 
discuss other types of defects. 

\subsection{Surface defects in gauge theory}

Supersymmetric surface defects in ${\CalN}=2$ gauge theories 
are two dimensional sigma models with global $G$-symmetry coupled to four dimensional gauge theories with gauge group $G$.  The surface defect as an operator depends on some continuous parameters, such as the complexified K\"ahler moduli of the corresponding sigma model. It was expected from the early days of studies of dualities that these defects obey differential equations involving the parameters of the defect, and the bulk parameters, i.e. the parameters of the gauge theory in the ambient four dimensional space.  Moreover, these
equations are identical to the Ward identities of some two dimensional conformal theories. Some of these relations are rigorously established
\cite{NekBPSCFT}.

\subsubsection{Regular defects}

There are several practical (i.e. convenient for computations) realizations of surface defects in ${\CalN}=2$ gauge theories. 
The construction we use is inspired by \cite{ChainSaw}, \cite{Alday:2010vg} , it can be interpreted as an ${\BZ}_{n}$-orbifold (generalizing those studied in \cite{DouglasMoore}) of $U(n)$ gauge theory. 

The regular surface defect is defined by imposing an ${\BZ}_n$-orbifold projection on the space of fields of the $\Omega$-deformed ${\CalN}=2$ theory, where the group ${\BZ}_n$
acts both on the physical space by $\frac{2\pi}{n}$-rotation  in one of the two orthogonal two-planes ${{\BR}^{2} \subset \BR}^{2} \oplus {\BR}^{1,1} \approx {\BR}^{1,3}$ in physical spacetime
and in the color and flavor spaces. The path integral over the space of projected fields can be interpreted as the path integral in gauge theory defined on the quotient space
${\BR}^{1,3} /{\BZ}_{2} \approx {\BR}^{1,3}$ in the presence of the defect located at the surface $0 \times {\BR}^{1,1}$, where $0 \in {\BR}^{2}$ is the fixed point of the ${\BZ}_{n}$ action. We shall work in the Euclidean spacetime, with the complex coordinates $(z_1, z_2^{\frac 1n})$ on the covering space, and the coordinates $(z_1, z_2)$ on the quotient space, the defect being located at the $\{ (z_1, 0 ) \, | \, z_1 \in {\BC} \}$ plane.

\subsubsection{Parameters of the theory with defect}

{}The regular surface defect has both continuous and discrete
parameters. The continuous parameters are the fugacities of the fractional instanton charges. 
The latter are identified with the dimensions of the ${\BZ}_{n}$-isotypical components in the space $K$ of Dirac zeromodes on the covering space:
\beq
{\hat K} = {\tilde K}_{0} \otimes {\CalR}_{0} \oplus \ldots \oplus {\tilde K}_{n-1} \otimes {\CalR}_{n-1} 
\label{eq:kdec}
\eeq
where ${\CalR}_{\omega}$ is
 the nontrivial one dimensional irreducible representations of ${\BZ}_{n}$, in which the generator acts as the $\omega$'th root of unity:
 \beq
 T_{{\CalR}_{\omega}} [ {\Omega}_{n} ] = e^{\frac{2\pi\ii}{n} {\omega}}
 \eeq
 Similarly to \eqref{eq:kdec}, the color and flavor (Chan-Paton) spaces decompose as
\beq
\begin{aligned}
& {\hat N} = {\tilde N}_{0} \otimes  {\CalR}_{0} \oplus \ldots \oplus {\tilde N}_{n-1} \otimes {\CalR}_{n-1}\\
& {\hat M} = {\tilde M}_{0} \otimes  {\CalR}_{0} \oplus \ldots \oplus {\tilde M}_{n-1} \otimes {\CalR}_{n-1}\\
\end{aligned}
\label{eq:nmdec}
\eeq
 The contribution of the moduli space of instantons on the orbifold ${\BC}^{2}/{\BZ}_{n}$
 is weighed with the factor
 \beq
 \prod\limits_{{\omega}=0}^{n-1} \ {\qe}_{\omega}^{k_{\omega}}\, , \qquad k_{\omega} = {\rm dim}K_{\omega}
 \label{eq:fugak}
 \eeq 
 which can be further expressed as:
 \beq
 {\qe}^{k_{n-1}} \times \prod_{{\omega}=1}^{n-1} \ e^{2{\pi}{\ii} t_{\omega} (k_{\omega - 1} - k_{\omega})}  
 \eeq
 where ${\qe}$ is the bulk instanton fugacity, and $t_{\omega}$, ${\omega} =  1, \ldots , n-1$
 are the K{\"a}hler moduli (a sum of the theta angle and $\ii$ times a Fayet-Illiopoulos term)
 of the two dimensional theory on the surface of the surface defect. 
 There is also a perturbative prefactor which we define in the next section. 
 
On the covering space ${\BC}^{2}$ the $U(1)\times U(1)$-torus acts with the weights $(q_1, {\tilde q}_{2})$, where ${\tilde q}_{2}^{n} = q_{2}$, corresponding to the standard action on the quotient ${\BC} \times {\BC}/{\BZ}_{n} = Spec{\BC}[z_{1}, z_{2}]$.

In the main body of this paper we shall deal with the $n=2$ case (the Appendix and \cite{NekBPSCFT} present more general discussion). 
The tautological ADHM complex (see the appendix for more details and \cite{NekBPSCFT} for notations), in the ${\BZ}_{2}$-equivariant setting, splits as
\begin{multline}
{\hat S} = \left[ 
{\hat K} \otimes q_{1}{\tilde q}_{2} {\CalR}_{1} \longrightarrow {\hat K} \otimes (q_{1} {\CalR}_{0} \oplus  {\tilde q}_{2} {\CalR}_{1}) \oplus {\hat N} \longrightarrow {\hat K}  \right] = {\tilde S}_{0}[1] \otimes {\CalR}_{0} \oplus
{\tilde S}_{1}[1] \otimes {\CalR}_{1} \\
{\tilde S}_{0} = \left[  q_{1}{\tilde q}_{2} {\tilde K}_{0}  \to q_{1} {\tilde K}_{0} \oplus {\tilde q}_{2} {\tilde K}_{1} \oplus {\tilde N}_{0} 
\to   {\tilde K}_{1} \right] \\ 
{\tilde S}_{1} = \left[  q_{1}{\tilde q}_{2} {\tilde K}_{1}  \to q_{1} {\tilde K}_{1} \oplus {\tilde q}_{2} {\tilde K}_{0} \oplus {\tilde N}_{1} 
\to  {\tilde K}_{0} \right] \end{multline}
with ${\tilde N}_{0,1}$ being the ${\BZ}_{2}$-eigenspaces in the color space. 
We denote by  
$({\tilde a}_{0}, {\tilde a}_{1})$ the eigenvalues of the complex scalar in the vector multiplet (on the covering space) correspond to the ${\BZ}_2$ eigenspaces in the color space:
\beq
Ch({\tilde N}_{0})  =  e^{{\tilde a}_{0}}\, , \ Ch({\tilde N}_{1})  =  e^{{\tilde a}_{1}}\, , 
\eeq
while the masses are encoded by the characters of the flavor spaces:
\beq
{\rm Ch}({\tilde M}_{0}) = e^{{\tilde m}_{0}^{+}} +e^{{\tilde m}_{0}^{-}}\, , \ 
{\rm Ch}({\tilde M}_{1}) = e^{{\tilde m}_{1}^{+}} +e^{{\tilde m}_{1}^{-}} \, .
\eeq
To define the instanton contribution in the $\Omega$-deformed theory the global characters ${\rm Ch}({\tilde H}_{0})$, ${\rm Ch}({\tilde H}_{1})$ are useful:
\beq
{\hat H} = \frac{{\hat S}}{(1-q_{1}{\CalR}_{0})(1 - {\tilde q}_{2} {\CalR}_{1})} = {\tilde H}_{0} \otimes {\CalR}_{0} \oplus {\tilde H}_{1} \otimes {\CalR}_{1}
\eeq

\subsubsection{Instanton measure and bulk parameters}

The instanton measure, in the fixed point formula, has been computed in \cite{NekBPSCFT} (in $SU(2)$ ${\CalN}=2^{*}$ case it was done in \cite{Alday:2010vg, KT}). Using the plethystic exponent ${\sE}[ {\cdot} ]$, it can be written quite simply:
\beq
{\mu} \, = \, 
{\qe}_{0}^{-ch_0 ({\tilde H}_0)} \ {\qe}_{1}^{-ch_0 ({\tilde H}_1)}  \, {\sE} \left[ \, \left( \frac{- Ch{\hat S}\, Ch{\hat S}^{*} + Ch{\hat S}^{*}\, Ch{\hat M}}{(1-q_{1}^{-1}{\CalR}_{0})(1 - {\tilde q}_{2}^{-1} {\CalR}_{1})} \right)_{0} \, \right]
\eeq
where the expression in $\left( \dots \right)$ takes values in the representation ring of ${\BZ}_{2}$, which is isomorphic to ${\BC}[{\BZ}_{2}]$ (the trivial representation ${\CalR}_{0}$
is a unit, while ${\CalR}_{1}$ squares to ${\CalR}_{0}$), and $\left( \dots \right)_{0}$ stands for the ${\CalR}_{0}$ coefficient. Let us introduce the notation
\beq
{\qe}_{0} = - u \, , \ {\qe}_{1} = - {\qe} u^{-1} \,  , 
\label{eq:qw}
\eeq
for the fractional charge fugacities. The instanton with the fractional charges $(k_0, k_1)$ contributes
\beq
(-u)^{k_{0}-k_{1}} {\qe}^{k_{1}} \times {\rm one-loop \ contribution}
\label{eq:fuga}
\eeq
to the partition function ${\Psi}$.  
The measure factorizes 
\beq
{\mu} = {\mu}^{\rm bulk} {\mu}^{\rm surface}
\eeq
with
\beq
{\mu}^{\rm bulk} = {\qe}^{- \frac{{\tilde a}_{0}^2  + {\tilde a}_{1}^{2}-{\tilde a}_{0}{\ve}_{1} - {\tilde a}_{1} ({\ve}_{1} + {\ve}_{2})}{2{\ve}_{1}{\ve}_{2}} + k_{1}} \, {\sE} \left[ \, \frac{Ch(S)^{*}\left( Ch(M) - Ch(S) \right)}{(1-q_{1}^{-1})(1 - q_{2}^{-1})}  \, \right]\ , 
\label{eq:bulk}
\eeq
being the usual instanton measure of the $A_1$ theory ($U(n)$ theory with $2n$ fundamental hypermultiplets)
with
\beq
Ch(S) = Ch({\tilde S}_{0}) + Ch({\tilde S}_{1}) {\tilde q}_{2}^{-1} = Ch(N) - (1-q_{1})(1-q_{2})Ch({\tilde K}_{1}) {\tilde q}_{2}^{-1}\, , 
\eeq
\beq
Ch(M) = Ch({\tilde M}_{0}) +  {\tilde q}_{2}^{-1} Ch({\tilde M}_{1})\, , \ 
 Ch(N) = e^{{\tilde a}_{0}} + {\tilde q}_{2}^{-1} e^{{\tilde a}_{1}}
\eeq
and
\beq
{\mu}^{\rm surface} \ = \ u^{\frac{{\tilde a}_{0} - {\tilde a}_{1}}{2{\ve}_{1}} + k_{0}-k_{1}}\ 
{\sE} \left[ \, \frac{{\tilde q}_{2}}{1-q_{1}^{-1}}\, Ch({\tilde S}_{0} - {\tilde M}_{0})  Ch({\tilde S}_{1})^* \, \right]
\label{eq:surface}
\eeq
being the purely surface defect contribution. In \eqref{eq:bulk}, \eqref{eq:surface}
 we dropped a multiplicative constant independent of ${\tilde a}$'s and $k$'s.

\subsubsection{A plethora of regular defects}

By writing
\beq
Ch(N) = e^{a} + e^{-a}\, , \ Ch(M) = \sum_{f=1}^{4} e^{m_{f}}
\eeq
we identify: 
\begin{multline}
{\tilde a}_{0} = \pm a\, , \ {\tilde a}_{1} = \mp a + {\tilde\ve_2} \, , \\
{\tilde m}_{0}^{+} = m_{s(3)}\, , \ {\tilde m}_{0}^{-} = m_{s(4)} \, , \ {\tilde m}_{1}^{+} = m_{s(1)} + {\tilde\ve_2} \, , \ {\tilde m}_{1}^{-} = m_{s(2)} + {\tilde\ve_2} \ .
\label{eq:match}
\end{multline}
 for some permutations $\pm  \in S(2)$, $s \in S(4)$. Obviously, permuting the masses $3 \leftrightarrow 4$ or $1 \leftrightarrow 2$ does not change anything. The exchange $a \leftrightarrow -a$ is a residual gauge symmetry of the bulk theory. However, it is convenient to distinguish the surface defects corresponding to different $u \in S(2)$, as they are permuted by the monodromy in the fractional gauge couplings.

\section{Tau-function}

We take a pause now. From the world of the four dimensional physics we turn our attention to classical integrable systems, albeit the ones defined over complex numbers. This is partly motivated by the well-known, by now, connection between four dimensional ${\CalN}=2$ supersymmetric gauge theories and integrable systems
\cite{Gorsky:1995zq, Donagi:1995cf, Martinec:1995i, Martinec:1995ii, IM}. Actually, this connection is a ${\ve}_{1}, {\ve}_{2} \to 0$ limit of the BPS/CFT correspondence. In the present discussion we shall keep some of the $\Omega$-deformation parameters finite. 
Let us forget for a moment about gauge theories, conformal field theories, or supersymmetry. 

In this section we introduce a notion of $\tau$-function for the classical integrable system. We also formulate and prove the formula relating the $\tau$-function to the Hamilton-Jacobi potential of the same system, in the special cases. The Schlesinger isomonodromy deformation equations (in the case of $2 \times 2$-matrices) fall into that special class. In this case the $\tau$-function coincides with the one introduced by \cite{JM}. It plays an important r{\^o}le in the discussions of movable singularities of the solutions to isomonodromic deformation equations \cite{Malgrange, Bolibruch}. 

\subsection{Classical integrability in the non-stationary case}

Let $({\CalP}, {\varpi})$ be a complex symplectic manifold, ${\rm dim}{\CalP} = 2r$. Denote by $Symp({\CalP}) = Diff ({\CalP}, {\varpi})$ the group of holomorphic diffeomorphisms $\CalP$ preserving $\varpi$.   Let $(H_{i}({\qe}))_{i=1}^{r}$, ${\qe} \in {\CalU}$, $H_{i} : {\CalP}^{2r} \to {\BC}$ be a family of (generically) functionally-independent Poisson-commuting holomorphic functions, parameterized by a simply-connected $r$-dimensional complex manifold ${\CalU}$:
\beq
\{ H_{i} , H_{j} \}_{\varpi} = 0 \, .
\label{eq:poissh}
\eeq
Specifically, in some local Darboux coordinates $(y^{a}, p_{y^{a}})_{a=1}^{r}$,
\beq
{\varpi} = \sum_{a=1}^{r} dp_{y^{a}} \wedge dy^{a}
\eeq
on ${\CalP}$, the Eq. \eqref{eq:poissh} reads:
\beq
\sum_{a=1}^{r} \frac{{\partial}H_{j}(y,p; {\qe})}{{\partial}y^{a}} \frac{{\partial}H_{k}(y,p; {\qe})}{{\partial}p_{y^{a}}} 
= \sum_{a=1}^{r} \frac{{\partial}H_{k}(y,p; {\qe})}{{\partial}y^{a}} \frac{{\partial}H_{j}(y,p; {\qe})}{{\partial}p_{y^{a}}}
\label{eq:poissh2}
\eeq
A priori, \eqref{eq:poissh}, \eqref{eq:poissh2} admit some transformations of the Hamiltonians $H_{i}$, e.g. $H_{i} \to H_{i} + {\eta}_{i}^{jk} H_{j}H_{k} + \ldots$.

\subsection{The $\tau$-function}

Let us restrict this freedom by the additional nontrivial requirement: there exist a coordinate system $({\qe}_{i})_{i=1}^{r}$ on ${\CalU}$, such that 
\beq
\frac{{\partial} H_{j} (y, p; {\qe})}{{\partial}{\qe}_{i}} = \frac{{\partial} H_{i} (y, p; {\qe})}{{\partial}{\qe}_{j}}
\label{eq:comm}
\eeq 
Let us introduce the $1$-form ${\bh}$ on ${\CalU}$, valued in the Poisson algebra of functions on $\CalP$, viewed as Lie algebra: 
\beq
{\bh} = \sum_{i=1}^{r}  H_{i} d{\qe}_{i}
\eeq
Then \eqref{eq:poissh2}, \eqref{eq:comm} can be neatly summarized as:
\beq
d_{\CalU} {\bh} = 0 \, , \  [ {\bh} , {\bh} ] = 0 \, , 
\eeq
or, as the flatness of the $\lambda$-connection: ${\lambda} d_{\CalU} + {\bh}$. Let us denote by ${\bh}^{\vee}$
the corresponding $Vect({\CalP})$-valued $1$-form:
\beq
{\bh}^{\vee} = \sum_{i=1}^{r} \left( {\varpi}^{-1} d_{\CalP} H_{i} \right) d{\qe}_{i}
\label{eq:bhvee}
\eeq
One can, therefore, for any $\lambda$, look for $g ({\lambda}) \in Symp({\CalP})$, such that
\beq
{\bh}^{\vee} = {\lambda} g({\lambda})^{-1} d_{\CalU} g ({\lambda})
\label{eq:lamtau}
\eeq
For ${\lambda} \to \infty$ one has, $g({\lambda} ) = {\exp} \, {\lambda}^{-1} V + \ldots$, with the Hamiltonian vector field $V = {\varpi}^{-1} dS_{0}$, $S_{0}(y, p; {\qe})$ being the obvious potential for $\bh$, ${\bh} = d_{\CalU} S_{0}$, or, in coordinates:
\beq
H_{i}(y, p; {\qe}) = \frac{{\partial} S_{0} (y, p; {\qe})}{\partial {\qe}_{i}}
\label{eq:his0}
\eeq
In writing \eqref{eq:poissh2}, \eqref{eq:comm}, \eqref{eq:his0} we are explicitly showing the arguments of the functions involved, so as to make it clear which arguments are kept intact while taking partial derivatives. 
Now, the fun fact is that for finite ${\lambda} \neq 0, \infty$ the Eq. \eqref{eq:lamtau}
can be also solved. Let us set $\lambda = 1$ (otherwise rescale $H_{i}$'s). Fix a basepoint
${\qe}_{0} \in {\CalU}$, and define, for ${\qe} \in {\CalU}$ sufficiently close to ${\qe}_{0}$, 
a symplectomorphism $g_{{\qe}_{0}}^{\qe} \in Symp({\CalP})$, so that $ g_{{\qe}_{0}}^{{\qe}_{0}} = id$, as a 
solution $g_{{\qe}_{0}}^{\qe} (y_{(0)}, p_{y}^{(0)}) = (y, p_{y})$, $y = y \left( y_{(0)}, p_{y}^{(0)}; {\qe}\right)\, , \ p_{y} = p \left( y_{(0)}, p_{y}^{(0)}; {\qe}\right)$ of the system of
Hamilton equations:
\beq
\begin{aligned}
\frac{dy^{a}}{d{\qe}_{i}} = \frac{{\partial}H_{i}}{{\partial}p_{y^{a}}} \, , \ 
\frac{dp_{y^{a}}}{d{\qe}_{i}} = - \frac{{\partial}H_{i}}{{\partial}y^{a}} \\
\label{eq:hameqs}
\end{aligned}
\eeq
with the initial condition
\beq
y_{(0)} = y \left( y_{(0)}, p^{(0)}; {\qe}_{0}\right)\, , \ p_{y}^{(0)} = p \left( y_{(0)}, p_{y}^{(0)}; {\qe}_{0}\right)
 \eeq
Now consider the $(1,0)$-form on ${\CalU} \times {\CalP}$ (more precisely on a sufficiently small neighborhood of $\qe_0$)
\beq
{\tilde\bh} = \left( g_{{\qe}_{0}}^{\qe} \right)^{*} {\bh} = \sum_{i=1}^{r} {\tilde H}_{i} d{\qe}_{i}\, , \ 
{\tilde H}_{i} \left( y_{(0)}, p_{y}^{(0)}; {\qe} \right) = H_{i} \left( y \left( y_{(0)}, p^{(0)}; {\qe}
\right)\, , \, p \left( y_{(0)}, p_{y}^{(0)}; {\qe} \right) \right)
\eeq
 It is now a simple matter to verify that $d_{\CalU} {\tilde\bh} = 0$, therefore locally
 ${\tilde\bh} = d_{\CalU} {\rm log}{\tau}$. 

The family $g_{{\qe}_{0}}^{\qe}$ of symplectomorphisms can be  described, 
in $y$-polarization,  by 
the generating function ${\CalS} (y , y_{(0)}; {\qe}, {\qe}_{0})$, via:
\beq
\begin{aligned}
& p_{y^{a}} = \frac{{\partial} {\CalS}}{{\partial} y^{a}} \, ,  \  p_{y^{a}}^{(0)} = 
\frac{{\partial} {\CalS}}{{\partial} y^{a}_{(0)}}\, , \\ 
& \frac{{\partial} {\CalS}}{{\partial} {\qe}_{i}} = - 
H_{i} \left( y, \frac{{\partial}{\CalS}}{{\partial}y} ; {\qe} \right) \, , \
\frac{{\partial} {\CalS}}{{\partial} {\qe}_{0,i}} = 
H_{i} \left( y_{(0)}, \frac{{\partial}{\CalS}}{{\partial}y_{(0)}} ; {\qe}_{0} \right) 
\end{aligned}
\label{eq:yyqq}
\eeq
or, in the mixed e.g. $y-p$ polarization, by the generating function
${\Sigma} (y, p^{(0)}; {\qe}, {\qe}_{0})$, via:
\beq
\begin{aligned}
&
 p_{y^{a}} = \frac{{\partial} {\Sigma}}{{\partial} y^{a}} \, ,  \  y^{a}_{(0)} = -  
\frac{{\partial} {\Sigma}}{{\partial} p_{y^{a}_{(0)}}}\, , \\ 
& \frac{{\partial} {\Sigma}}{{\partial} {\qe}_{i}} = - 
H_{i} \left( y, \frac{{\partial}{\Sigma}}{{\partial}y} ; {\qe} \right)\, , \ 
\frac{{\partial} {\Sigma}}{{\partial} {\qe}_{0,i}} =  
H_{i} \left( - \frac{{\partial}{\Sigma}}{{\partial}p^{(0)}} , p^{(0)} ; {\qe}_{0} \right) \, , \\
\end{aligned}
\label{eq:ypqq}
\eeq
 with
 \beq
 {\Sigma} ( y, p^{(0)} ; {\qe}, {\qe}_{0}) \ = \ {\rm Crit}_{y_{(0)}} \left(   
 {\CalS}  ( y, y_{(0)} ; {\qe}, {\qe}_{0} )  - \sum_{a=1}^{r} p^{(0)}_{y^{a}} y_{(0)}^{a} \right)
 \eeq
What is the connection between ${\Sigma}$, ${\CalS}$, and ${\tau}$? 

 \subsection{$\tau$-function for non-relativistic Hamiltonians}
 
Let us further assume that the Hamiltonians $H_i$'s are quadratic functions
of the momenta $p_{y_{a}}$'s and auxiliary parameters ${\vt}_{a}$'s: 
\beq
H_{i} (y, p ; {\qe}; {\vt}) \ = \ \sum\limits_{a,b=1}^{r}\, g_{i}^{ab} (y; {\qe}) p_{y^{a}}p_{y^{b}} +  g_{i|b}^{a} (y; {\qe}) p_{y^{a}}{\vt}_{b} + g_{i| ab} (y; {\qe}) {\vt}_{a}{\vt}_{b}  \ . 
\label{eq:H2p}
\eeq
Let $S (y, {\alpha}; {\qe}; {\vt})$ be the generating function of 
the canonical transformation sending $(y, p)$ over ${\qe}\in {\CalU}$ to the constants
$({\alpha}, {\beta})$ of motions given by the Hamilton equations, 
\beq
\begin{aligned}
& \frac{dy^{a}}{d{\qe}_{i}} = \frac{{\partial}H_{i}}{{\partial} p_{y^{a}}} \, , \\
& \frac{dp_{y^{a}}}{d{\qe}_{i}} = - \frac{{\partial}H_{i}}{{\partial} y^{a}} \, , 
\end{aligned}
\eeq
aka the Hamilton-Jacobi potential:
\beq
\frac{{\partial} S(y, {\alpha}; {\qe}; {\vt})}{{\partial} {\qe}_{i}} = - 
H_{i} \left( y, \frac{{\partial} S(y, {\alpha}; {\qe}; {\vt})}{{\partial} y} ; {\qe} ; {\vt} \right) \label{eq:hjpot}
\eeq
with
\beq
p_{y^{a}} = \frac{{\partial} S(y, {\alpha}; {\qe}; {\vt})}{{\partial} y^{a}} \, , \quad 
{\beta}_{i} = \frac{{\partial} S(y, {\alpha}; {\qe}; {\vt})}{{\partial} \alpha_{i}} 
\eeq
For example, we can take ${\alpha} = y^{(0)}$, ${\beta} = p_{y}^{(0)}$. In our subsequent 
discussion $({\alpha}, {\beta})$ will parameterize the monodromy data in the
Riemann-Hilbert correspondence. 
We claim:
\begin{multline}
{\rm log}\, {\tau}({\alpha}, {\beta} ;  {\vec\vt} ; {\qe} ) \ = \ - \, \sum_{i=1}^{r} {\alpha}_{i} {\beta}_{i} \ + \\
 + \left(  S (y , {\alpha}; {\qe}; {\vt})  - \sum_{a=1}^{r}
{\vt}_{a} \frac{\partial}{\partial\vt_{a}} S (y, {\alpha}; {\qe}; {\vt} ) \right) \Biggr\vert_{y\, :\,   {\beta} = \frac{{\partial}S}{{\partial}{\alpha}}}  \ . 
\label{eq:rh1}
\end{multline}
{\bf Proof.} We should keep in mind that $y^{a}$'s, defined through ${\beta}_{i} = \frac{{\partial}S}{{\partial}{\alpha}_{i}}$, depends on $\qe$, $\alpha$ and $\vt$. This dependence is easy to compute, using \eqref{eq:hjpot}:
\begin{multline}
0 = \frac{{\partial}^{2}S}{{\partial}{\alpha_{j}} {\partial} y^{a}} \frac{dy^{a}}{d{\qe}_{i}} + \frac{{\partial}^{2}S}{{\partial}{\alpha_{j}} {\partial} {\qe}_{i}} = \\
= \frac{{\partial}^{2}S}{{\partial}{\alpha_{j}} {\partial} y^{a}} \left( \frac{dy^{a}}{d{\qe}_{i}} - \frac{{\partial} H_{i}}{{\partial}p_{y^{a}}} \right) \Longrightarrow \frac{dy^{a}}{d{\qe}_{i}} = \frac{{\partial} H_{i}}{{\partial}p_{y^{a}}} \Biggr\vert_{y\, :\,   p = {\partial}S/{\partial}y, \, {\beta} = {\partial}S/{\partial}{\alpha}}
\label{eq:dydq}
\end{multline}
as 
\beq
{\rm Det} \Biggl\Vert \frac{{\partial}^{2}S}{{\partial}{\alpha} {\partial} y} \Biggr\Vert \neq 0
\eeq
for $S$ to define a symplectomorphism in the $y - {\alpha}$ polarization. 

{}Now let us differentiate (with ${\alpha}, {\beta}, {\vt}$ fixed):
\begin{multline}
\frac{\partial}{{\partial}{\qe}_{i}} {\rm log}\,{\tau} ( {\alpha}, {\beta} ; {\vt}; {\qe} ) = - H_{i} (y, p; {\qe}; {\vt}) \Biggr\vert_{y\, :\,   p = \frac{{\partial}S}{{\partial}y}, \, {\beta} = \frac{{\partial}S}{{\partial}{\alpha}}}
+ \sum_{a=1}^{r} \frac{{\partial}S}{{\partial}y^{a}} \frac{dy^{a}}{d{\qe}_{i}} - \\
- \sum_{b=1}^{r} \left( 
{\vt}_{b} \frac{{\partial}^{2}S (y , {\alpha}; {\qe}; {\vt})}{{\partial}{\vt}_{b}{\partial}{\qe}_{i}}  \Biggr\vert_{y\, :\,  {\beta} = \frac{{\partial}S}{{\partial}{\alpha}}} + \sum_{a=1}^{r} \frac{dy^{a}}{d{\qe}_{i}} {\vt}_{b} \frac{{\partial}^{2}S (y , {\alpha}; {\qe}; {\vt})}{{\partial}{\vt}_{b}{\partial}y^{a}}   \Biggr\vert_{y\, :\,  {\beta} = \frac{{\partial}S}{{\partial}{\alpha}}}  \right) = \\
=\ \left( p \frac{\partial}{\partial p} + {\vt} \frac{\partial}{\partial\vt} - 1 \right) H_{i} (y, p ; {\qe} ; {\vt} ) \Biggr\vert_{y\, :\,   p = \frac{{\partial}S}{{\partial}y}, \, {\beta} = \frac{{\partial}S}{{\partial}{\alpha}}} = \\
\\
= \ H_{i} (y, p ; {\qe} ; {\vt} ) \vert_{y\, :\,   p = \frac{{\partial}S}{{\partial}y}, \, {\beta} = \frac{{\partial}S}{{\partial}{\alpha}}} 
\end{multline}
where we used \eqref{eq:H2p}, \eqref{eq:dydq}, and, most importantly:
\begin{multline}
{\vt} \frac{\partial}{\partial\vt} H_{i} ( y, \frac{\partial S (y , {\alpha}; {\qe}; {\vt})}{\partial y} ; {\qe} ; {\vt} ) = \\
\left( {\vt} \frac{\partial}{\partial\vt} H_{i} (y, p ; {\qe} ; {\vt} ) \right) \Biggr\vert_{p = {\partial S}/{\partial y}}
+ \sum_{a=1}^{r} {\vt} \frac{\partial H_{i}}{\partial p_{y_{a}}}\Biggr\vert_{p = {\partial S}/{\partial y}}
 \frac{{\partial}^{2} S(y , {\alpha}; {\qe}; {\vt})}{{\partial y^{a}}{\partial \vt}}
\end{multline}

\vskip 1cm

\section{Painlev{\'e} VI, Garnier, Gaudin, and Schlesinger systems}

\vfill

{}An important case of \eqref{eq:H2p} is given by the $SL(2)$ Schlesinger system, which we present in some detail. 

{}Let $p \geq 1$ denote a positive integer, and let  ${\bz} \subset {\BC\BP}^{1}$ denote a finite ordered set of cardinality $p+3$:
\beq
{\bz} = \{ z_{-1}, z_{0}, \ldots , z_{p+1} \}
\eeq
Sometimes we set $z_{p+1} = {\infty}$, $z_{0} = 1$, $z_{1} = {\qe}_{1}$, $z_{2}= {\qe}_{1}{\qe}_{2}, \ldots , z_{p} = {\qe}_{1}{\qe}_{2} \ldots {\qe}_{p}$, $z_{-1} = 0$, with ${\qe}_{1}, {\qe}_{2}, \ldots , {\qe}_{p} \in {\BC}^{\times}_{|{\cdot}| <1}$. 

There are two $SL(2, {\BC})$ symmetries in our story. One, which we shall denote by $H$, is the two-fold cover of the  symmetry group of the ${\BC\BP}^{1}$, parametrizing the points $z_{i}$ above.  Say, an element 
\beq
h = \left( \begin{matrix} a & b \\ c & d \end{matrix} \right) \in H
\label{eq:habcd}
\eeq
acts via 
\beq
 {\bz} \mapsto {\bz}^{h} = \left\{ \, \frac{az_{-1} + b}{cz_{-1} + d} \, , \, \frac{az_{0} + b}{cz_{0} + d}\, , \, \ldots \, , \,
 \frac{az_{p+1} + b}{cz_{p+1} + d} \, \right\} 
 \eeq
Let us denote by ${\CalC}_{p} = \left[ \left( {\BC\BP}^{1} \right)^{p+3} \backslash {\rm diagonals} \right]$ the space of the $\bz$'s. 
The quotient ${\CalC}_{p}/H = {\mfM}_{0,p+3}$ is the moduli space of smooth genus $0$
curves with $p+3$ punctures. 

The other $SL(2, {\BC})$ symmetry group, denoted by $G$ (or $G_1$ in later chapters), is the gauge group, meaning it does not act, in itself, on any physical observable. However, in some description of the system it is 
a symmetry acting on redundant variables. Its r{\^o}le is described momentarily. 

\subsection{The phase space ${\CalM}_{p}^{\rm alg}$}

To begin with, the phase space ${\CalM}_{p}^{\rm alg}$  is
the set  of $p+3$-tuples $(A_{\xi})_{\xi \in {\bz}}$, obeying the equations 
\beq
\sum_{{\xi} \in {\bz}} A_{\xi} = 0 \label{eq:mom}
\eeq
modulo the equivalence relation $(A_{\xi})_{\xi \in {\bz}} \sim ({\rm g} A_{\xi} {\rm g}^{-1})_{\xi \in {\bz}}$
for ${\rm g} \in G$. Each $A_{\xi} \in Lie (G)$ is a $2\times 2$ traceless, ${\rm Tr}A_{\xi} = 0$,  complex matrix, obeying:
\beq
{\half} {\rm Tr} A_{\xi}^{2} = {\Delta}_{\xi} = {\vt}_{\xi}^{2} \, . 
\label{eq:delxi}
\eeq

\subsubsection{Complexified two-sphere}

Before we proceed, we need to recall a few facts about the orbit of an individual matrix. 
 
Let ${\vt} \in {\BC}$. We denote by ${\CalO}_{\vt} \subset {\BC}^{3}$ the quadric surface 
\beq
x^{2} + y^{2} + z^{2} = {\vt}^{2}
\label{eq:csph}
\eeq
We identify ${\BC}^{3}$ with the space of $2\times 2$ traceless matrices of the form:
\beq
A (x,y,z)  = \left( \begin{matrix} z  & x + {\ii} y \\ x - {\ii} y 
& - z \end{matrix} \right)  \ , 
\label{eq:orb1}
\eeq
which makes ${\CalO}_{\vt}$ the space of $A(x,y,z)$, such that ${\rm Det}A(x,y,z) = - {\vt}^{2}$. 

It is a complex symplectic manifold with the holomorphic $(2,0)$ symplectic form:
\beq
{\Omega}_{\vt} = \frac{dx \wedge dy}{2 z}
\eeq
We have, for ${\vt} \neq 0$, two holomorphic maps: 
\beq
\begin{matrix}
& & & \quad {\CalO}_{\vt} & & & \\
& &  &    & & & \\
&  & \swarrow^{\kern -.3in p_{+}}  &  &  \searrow^{\kern .1in p_{-}}   & & \\
& &&&&& \\
&  & {\BC\BP}^{1}   & &    {\BC\BP}^{1} & &
\end{matrix}
\eeq 
which associate to $A(x,y,z)$ its eigenlines $L_{\pm} \subset {\BC}^{2}$ corresponding to the
eigenvalues $\pm \vt$, explicitly, 
\beq
p_{\pm}: (x, y, z) \mapsto ( z \pm {\vt} : x - {\ii}y ) = ( x + {\ii} y : - z \pm {\vt} ) \ . 
\label{eq:ppm}
\eeq
 In writing \eqref{eq:ppm} we associate to the vector
 $\left( \begin{matrix} {\psi}_{+} \\ {\psi}_{-} \end{matrix} \right)$ a point $({\psi}_{+} : {\psi}_{-}) \in {\BC\BP}^{1}$
 in homogeneous coordinates. This is well-defined as long as $({\psi}_{+}, {\psi}_{-}) \neq 0$, which is true 
 for either $( z \pm {\vt},  x - {\ii}y )$ or $ ( x + {\ii} y,  - z \pm {\vt} )$ in \eqref{eq:ppm}, or both. 
   
 By covering ${\BC\BP}^{1}$ with two coordinate charts $U_{+} \cup U_{-}$, with ${\gamma} \in {\BC}$ parameterizing $U_{+} = \{ \, ({\gamma}: 1) \, \}$ and ${\tilde\gamma} \in {\BC}$ parametrizing $U_{-} = \{ \, (1: {\tilde\gamma}) \, \}$, so that
 ${\gamma}{\tilde\gamma}=1$ on the intersection $U_{+} \cap U_{-} \approx {\BC}^{\times}$, we can produce two Darboux coordinate systems\footnote{We choose a non-standard normalization of Darboux coordinates, with an extra ${\ii}/2$-factor in the symplectic form, for later convenience} $\{ ({\gamma}_{\pm}, {\beta}) \, , \, ({\tilde\gamma}_{\pm}, {\tilde\beta}) \}$ on ${\CalO}_{\vt}$, where ${\gamma}_{\pm} = {\gamma} \circ p_{\pm}$, ${\tilde\gamma}_{\pm} = {\tilde\gamma} \circ p_{\pm}$:
 \beq
 ( z \pm {\vt} : x - {\ii}y ) = ( x + {\ii} y : - z \pm {\vt} ) = ( {\gamma}_{\mp} : 1 ) = ( 1 : {\tilde\gamma}_{\mp} ) \ , 
 \eeq
 so that
 \beq
 {\Omega}_{\vt} = -\frac{\ii}{2} d{\beta} \wedge d{\gamma}_{\pm} = -\frac{\ii}{2}  d{\tilde\beta} \wedge d{\tilde\gamma}_{\pm}\ , 
 \label{eq:omgvtbg}
 \eeq
 so that the global functions $x, y,z$ are expressed as: 
\beq
\left( \begin{matrix} z \\ 
\\
x + {\ii} y \\ 
\\
x - {\ii} y \end{matrix} \right) \, = \, \left( \begin{matrix}
\pm {\vt}-{\beta} {\gamma}_{\pm} \\
\\
{\gamma}_{\pm}^2 {\beta} \mp 2 {\vt} {\gamma}_{\pm} \\
\\
- {\beta} \end{matrix} \right) \, = \, \left( \begin{matrix}  {\tilde\gamma}_{\pm}{\tilde\beta} \mp {\vt} \\
\\
- {\tilde\beta} \\
\\
{\tilde \gamma}_{\pm}^{2} {\tilde\beta}  \mp 2{\vt}{\tilde\gamma}_{\pm} \end{matrix} \right) 
\label{eq:pxo}
\eeq
The presence of the shift ${\beta} \mapsto {\tilde\beta} = - {\beta} {\gamma}_{\pm}^{2} \mp 2{\vt}{\gamma}_{\pm}$ is what makes the bundle an affine line bundle, as opposed to the more familiar vector line bundle. We can solve \eqref{eq:pxo} to see how ``$+$'' coordinates translate to the ``$-$'' coordinates:
\beq
{\gamma}_{+} - 
{\gamma}_{-} = 2{\vt}/{\beta}\, , \ {\tilde\gamma}_{+} - {\tilde\gamma}_{-}  =  2{\vt}/{\tilde\beta} \ ,
\label{eq:pmtr}
\eeq
equivalently,  
the map $({\beta}, {\gamma}_{+}) \mapsto ({\beta}, {\gamma}_{-})$ is generated by the generating function 
\beq
F ({\gamma}_{+}, {\gamma}_{-}) = 2{\vt}\ {\rm log} ( {\gamma}_{+} - {\gamma}_{-} ) \ .
\label{eq:flipgf}
\eeq 
When $\vt = 0$ something special happens. The singular quadric $x^2+y^2+z^2 =0$ has a resolution 
${\tilde{\CalO}_{0}}$ of singularity at $x=y=z=0$, which is the space of pairs $\{ \, ( A(x,y,z), {\Upsilon} ) \, | \, 
{\rm Det}A(x,y,z) = 0, \, {\Upsilon} \approx {\BC} \subset {\rm ker}A(x,y,z) \, \}$. For $(x,y,z) \neq (0,0,0)$ the
kernel is one-dimensional and is uniquely determined by the matrix $A(x,y,z)$. For 
$(x,y,z) = (0,0,0)$ the choices of a one-dimensional kernel span a copy of ${\BC\BP}^{1}$. 
The projection ${\tilde{\CalO}_{0}} \to {\BC\BP}^{1}$ (forgetting $A(x,y,z)$) identifies
${\tilde{\CalO}_{0}}$ with the cotangent bundle $T^{*}{\BC\BP}^{1}$. 

The formulae \eqref{eq:pxo}
become, then, the standard coordinates on $T^{*}{\BC\BP}^{1}$ with ${\gamma}$ and ${\tilde\gamma}$ covering 
the base ${\BC\BP}^{1}$ in the standard way, with ${\gamma}{\tilde\gamma} = 1$ away from the North and South pole, ${\gamma} = 0$, and ${\tilde\gamma}=0$, respectively.

The group $G$ acts on ${\CalO}_{\vt}$ in the usual way:
\beq
{\rm g} = \left( \begin{matrix} a & b \\ c & d \end{matrix}\right) : \ A \mapsto {\rm g} A {\rm g}^{-1}\, , 
\eeq
where ${\rm g} \in G$, i.e. $ad - bc =1$,
which translates to the fractional-linear transformations of ${\gamma}_{\pm}$, and affine transformations of ${\beta}$:
\beq  
({\gamma}_{\pm}, {\beta} ) \mapsto \left( \, \frac{a {\gamma}_{\pm} +b}{c {\gamma}_{\pm} +d} \, , \, {\beta} ( c {\gamma}_{\pm}+d)^2 \mp  2{\vt}c
(c{\gamma}_{\pm}+d) \, \right)\, , \  
\label{eq:sl2act}
\eeq
 and ${\beta} ( c {\gamma}_{+}+d)^2 -  2{\vt}c
(c{\gamma}_{+}+d) = {\beta} ( c {\gamma}_{-}+d)^2 +  2{\vt}c
(c{\gamma}_{-}+d)$ thanks to \eqref{eq:pmtr}. 
Notice that the symplectic form written in the $\gamma_{\pm}$ coordinates has the manifestly $G$-invariant form:
\beq
{\Omega}_{\vt} = {\ii\vt} \, \frac{d{\gamma}_{+} \wedge d{\gamma}_{-}}{({\gamma}_{+} - {\gamma}_{-})^{2}}
\eeq
{}The symplectic form ${\Omega}_{\vt}$ is, for $\vt \neq 0$, cohomologically nontrivial. Indeed, there is a
compact two-cycle $S \subset {\CalO}_{\vt}$, given by: $(x,y,z) = {\vt} (n_{1}, n_{2}, n_{3})$, with $(n_{1}, n_{2}, n_{3}) \in {\BR}^{3}$, with
\beq
n_{1}^{2} + n_{2}^{2} + n_{3}^{2} = 1 \ .
\eeq
On this cycle ${\bar\gamma}_{+} = - {\tilde\gamma}_{-}$, ${\bar\gamma}_{-} = - {\tilde\gamma}_{+}$, ${\bar\gamma}_{-} {\gamma}_{+} = - 1$, ${\bar\gamma}_{+}{\gamma}_{-} = -1$.  
It is easy to calculate
\beq
\frac{1}{2\pi} \int_{S} {\Omega}_{\vt} = \pm {\vt}\ ,
\label{eq:peromg}
\eeq
the sign depending on the choice of the orientation of $S$. It is not canonical, as $S$ is not a holomorphic curve. The cycle $S$ is not, of course, $G$-invariant. However, it is preserved by the maximal compact subgroup of $G$, isomorphic to $SU(2)$.  

Now, let us discuss the geometry of the eigenbundles $L_{\pm}$ over ${\CalO}_{\vt}$, first, for $\vt \neq 0$. The line $L_{\pm}$
over $(x,y,z) \in {\CalO}_{\vt}$ is the $\pm \vt$ eigenline ${\Upsilon}^{\pm} \subset {\BC}^{2}$ of $A(x,y,z)$:
\beq
A(x,y,z) {\Upsilon}^{({\pm})} = \pm {\vt} {\Upsilon}^{({\pm})} \, , 
\label{eq:eiga}
\eeq
We can trivialize $L_{\pm}$ over $U_{-{\pm}}$ by:
\beq
{\Upsilon}^{({\pm})} \, = \, {\BC} \cdot \left( \begin{matrix} z \, {\pm} \, {\vt} \\ x - {\ii} y \end{matrix} \right)
\label{eq:eigb}
\eeq
and,  over $U_{+{\pm}}$ by:
\beq
{\Upsilon}^{({\pm})} \, = \, {\BC} \cdot \left( \begin{matrix} - x - {\ii}y \\  z \, {\mp} \, {\vt}  \end{matrix} \right) \ .
\label{eq:eigc}
\eeq
The scalar factors taking \eqref{eq:eigb} to \eqref{eq:eigc} 
\beq
\frac{z \,{\mp}\, {\vt}}{x-{\ii}y} = {\gamma}_{\pm}
\label{eq:lpm}
\eeq
are the transition functions defining the holomorphic line bundles $L_{\pm}$ over ${\CalO}_{\vt}$. They are well-defined on $\CalU$. Thus\footnote{This relation is responsible for the so-called Berry phase \cite{Berry}}, 
\beq
L_{\pm} = p_{\pm}^{*} \, {\CalO}(-1)
\eeq
{}The case $\vt = 0$ is special, in that $A(x,y,z)$ for $(x,y,z) \neq 0$ is not diagonalizable. It has a Jordan block form, with the eigenline ${\rm ker} A(x,y,z)$ spanned by 
\beq
{\Upsilon}  \propto \left( \begin{matrix} z  \\ x - {\ii} y \end{matrix} \right)
\label{eq:eigd0}
\eeq
away from the $x = {\ii}y$, $z=0$ line, and by
\beq
{\Upsilon}  \propto \left( \begin{matrix} -x-{\ii} y  \\ z \end{matrix} \right)
\label{eq:eigd1}
\eeq
away from the $x = - {\ii}y$, $z=0$ line. The bundle of eigenlines extends to the $(x,y,z) = 0$ locus
on the 
resolution $T^{*}{\BC\BP}^{1}$. Of course, the matrix $A(0,0,0) =0$ has the two dimensional space of
zero eigenstates, however a point $({\gamma}_{0} : {\gamma}_{1})$ of the zero section ${\BC\BP}^{1} \subset T^{*}{\BC\BP}^{1}$ singles out a one dimensional subspace ${\BC}  \left( \begin{matrix}
{\gamma} \\ 1 \end{matrix} \right) = {\BC} \left( \begin{matrix} 1 \\
{\tilde\gamma} \end{matrix} \right)
\subset {\BC}^{2}$ which is an eigenspace of the matrix 
\beq
{\beta} \left( 
\begin{matrix} - {\gamma}   &  {\gamma}^2  \\
-  1 &  {\gamma}    \end{matrix} \right) = {\tilde\beta}  \left( 
\begin{matrix} {\tilde\gamma}   &  -1   \\
{\tilde\gamma}^{2}  &  - {\tilde\gamma}    \end{matrix} \right)
\eeq
which, in the limit ${\beta} \to 0$ or ${\tilde\beta} \to 0$, becomes 
$A(0,0,0)$.

\subsubsection{${\CalM}_{p}^{\rm alg}$ as symplectic quotient}

We can now return back to the vicinity of the Eqs. \eqref{eq:mom} and write 
\beq
{\CalM}_{p}^{\rm alg} = \left(  \times_{{\xi} \in {\bz}} \ {\CalO}_{{\vt}_{\xi}} \right) // G
\label{eq:malg}
\eeq
where $A_{\xi} = A (x_{\xi}, y_{\xi}, z_{\xi}) \in {\CalO}_{\vt_{\xi}} \subset {\BC}^{3}$, 
is its moment map, and $//$ stands for taking the symplectic quotient,
which means imposing the moment map equation \eqref{eq:mom} and then taking the quotient (with subtleties having to do with the noncompactness of $G$ swept under a rug) with respect to the action of $G$. The product of the coadjoint orbits in \eqref{eq:malg} has the
symplectic form:
\beq
{\Omega} = \sum_{{\xi} \in {\bz}} {\Omega}_{{\vt}_{\xi}} \eeq
which descends to the symplectic form $\varpi$ on ${\CalM}_{p}^{\rm alg}$. 

{}Let us denote by ${\ba} \in {\CalM}_{r}$ the gauge equivalence class of the $p+3$-tuple ${\ba} = \left[ (A_{\xi})\sim ({\rm g}A_{\xi} {\rm g}^{-1}) \right]$. 

\subsubsection{Symplectic and Poisson structures}

Computing $\varpi$ is somewhat cumbersome, however it is easy to compute the Poisson brackets of functions of matrix elements of $A_{\xi}$'s:
\beq
\{ x_{\xi}, y_{\eta} \} = z _{\eta} \, {\delta}_{\xi, \eta}\, , \ \{ y_{\xi}, z_{\eta} \} = x _{\eta} \, {\delta}_{\xi, \eta}\, , \ 
\{ z_{\xi}, x_{\eta} \} = y _{\eta} \, {\delta}_{\xi, \eta}\, . 
\eeq
Let us define algebraic functions on ${\CalM}_{p}^{\rm alg}$:
\beq
h_{\xi\eta}({\ba}) = {\rm Tr}A_{\xi}A_{\eta} = h_{\eta\xi}({\ba}), , \qquad {\xi}, {\eta} \in {\bz} \, ,  {\ba} \in {\CalM}_{p}
\label{eq:hxe}
\eeq
out of which we build the functions on ${\CalM}_{p}^{\rm alg} \times {\CalC}_{p}$: 
\beq
H_{\xi}({\ba}, {\bz}) = \sum_{\eta \neq {\xi}} \frac{h_{\eta\xi}}{{\xi} - {\eta}}
\label{eq:sch}
\eeq
We shall view them, up to a small subtlety, as ``time''-dependent Hamiltonians on ${\CalM}_{p}^{\rm alg}$. 
It is easy to show that $H_{\xi}, H_{\eta}$ Poisson-commute, and obey ${\partial}H_{\xi}/{\partial}{\eta} = {\partial}H_{\eta}/{\partial}{\xi}$, as in \eqref{eq:comm}. 
Define $(1,0)$-form on ${\CalC}_{p}$ valued in functions on ${\CalM}_{p}^{\rm alg}$: 
\beq
{\mathfrak{T}}  = \sum_{\xi \in {\bz}} H_{\xi} d{\xi} = 
\frac 12 \sum_{{\xi} \neq {\eta} \in {\bz}} \, h_{\eta\xi}\, d{\rm log}({\xi} - {\eta}) \ , 
\label{eq:tausch}
\eeq
Thanks to $\sum\limits_{\xi \in \bz} A_{\xi} = 0$,  we have the relations, cf. \eqref{eq:delxi}:
\beq
\sum\limits_{\xi \in \bz} \left( \begin{matrix} H_{\xi} \\
 {\xi} H_{\xi} - {\Delta}_{\xi}  \\
{\xi}^2 H_{\xi} - 2 {\Delta}_{\xi} {\xi} \end{matrix} \right) = 0\, , 
\eeq 
which imply that 
\eqref{eq:tausch} is covariant with respect
to the
$H$-action, cf. \eqref{eq:habcd}:
\beq
h^{*}{\mathfrak{T}} = {\mathfrak{T}}+
2 \sum_{\xi \in \CalX} {\Delta}_{\xi} \, d{\rm log}({c \xi} + d )
\eeq
As in our general discussion, 
\beq
d_{\bz} {\mathfrak{T}} = 0\ , \ {\rm D} \, {\mathfrak{T}} = 0
\label{eq:intth}
\eeq
where $d_{\bz}$ stands for the exterior derivative on ${\CalC}_{p}$, 
and 
\beq
{\rm D} = d_{\bz} + \sum_{\xi \in \CalX} \, d{\xi} \wedge \{ H_{\xi}, \cdot \}_{\ba} 
\eeq 
where we put the subscript $\{ \cdot , \cdot \}_{\ba}$ to stress that the Poisson brackets are on ${\CalM}_{p}^{\rm alg}$. 

\subsection{Isomonodromy equations}

Let us now introduce the Schlesinger system \cite{Schles}. It is simply the collection of flows, generated by the Hamiltonians $H_{\xi}$: 
\beq
\frac{\partial A_{\xi}}{\partial \eta} = \frac{ [A_{\xi}, A_{\eta}]}{{\xi} - {\eta}} ( 1 - {\delta}_{\eta\xi}) - {\delta}_{\eta\xi} 
\sum_{\zeta \neq \xi} \frac{ [A_{\xi}, A_{\zeta}]}{{\xi} - {\zeta}}
\label{eq:schfl}
\eeq 
They remarkable property is that the monodromy data of
\beq
{\nabla} = dz \left( {\partial}_{z} + {\CalA}(z) \right)\, , \qquad {\CalA}(z) = \sum_{\xi \in \bz} \frac{A_{\xi}}{z-{\xi}}\, , 
\label{eq:conn1}
\eeq 
remains fixed along the flow. Indeed, \eqref{eq:schfl} imply, for every ${\xi} \in {\bz}$: 
 \beq
 \frac{\partial}{\partial \xi}  {\nabla}  = [ {\nabla} ,  \frac{A_{\xi}}{z-{\xi}} ] \ ,
 \eeq 
 In the case $r=1$ the equations \eqref{eq:schfl} read, more explicitly:
 \beq
\frac{DA_{0}}{D{\qe}} = \frac{[A_{0},A_{\qe}]}{\qe}\, , \qquad \frac{DA_{1}}{D{\qe}} = \frac{[A_{1}, A_{\qe}]}{{\qe}-1}\, , \qquad \frac{DA_{\qe}}{D{\qe}} = \frac{[A_{\qe},A_{0}]}{\qe} +\frac{[A_{1}, A_{\qe}]}{1-{\qe}}\, , 
\label{eq:isoflow}
\eeq
where $D/D{\qe} = d/d{\qe} + [B, {\cdot}]$, with $B$ a 
compensating gauge transformation (recall that the residues $A_{\xi}$ are defined 
up to an overall conjugation). To avoid dealing with $B$ we can consider the $\qe$-evolution of the gauge-invariant functions
$h_{\xi\eta}$, so that in the $r=1$ case:
\beq
\frac{d}{d{\qe}} h_{0{\qe}} = \frac{{\tilde h}_{01{\qe}}}{1-{\qe}}\, , \qquad \frac{d}{d{\qe}} h_{1{\qe}} = \frac{{\tilde h}_{01{\qe}}}{{\qe}}
\eeq
where
\beq
{\tilde h}_{01{\qe}} = {\rm Tr}\left(  [A_{0}, A_{1}]A_{\qe} \right)
\label{eq:cubinv}
\eeq
 Let ${\vec\alpha} = ({\alpha}_{1}, \ldots , {\alpha}_{r}), {\vec\beta}= ({\beta}_{1}, \ldots , {\beta}_{r})$
 denote the parameters of the gauge equivalence classes of the monodromy data\footnote{We hope there is no confusion between the ${\beta}_{i}$ coordinates on the moduli space of local systems and the Darboux coordinates on the orbits ${\CalO}_{\vt}$'s}, or, more specifically, Darboux  coordinates $({\vec\alpha}, {\vec\beta})$  on the 
 moduli space ${\CalM}^{\rm loc}_{p}$ of flat $G$-connections on the $p+3$-punctured sphere \cite{NRS}\footnote{The precise construction depends on additional combinatorial data}. 
 By sending the pair $\left( {\ba} \, , \, {\bz} \right) \in {\CalM}_{p}^{\rm alg}
\times {\CalC}_{p}$ to the monodromy data of $\nabla$ we get a map
\beq
m: \, {\CalM}_{p}^{\rm alg}
\times {\CalC}_{p} \longrightarrow {\CalM}^{\rm loc}_{p}
\eeq
 which commute with the flows \eqref{eq:schfl}, in other words, the orbits of the flows belong to the fibers of $m$. Now, \eqref{eq:intth} imply, that
locally on $m^{-1} ({\vec\alpha}, {\vec\beta})$ the $(1,0)$-form $\mathfrak{T}$ is exact:
\beq
\mathfrak{T} \ \Biggr\vert_{m^{-1} ({\vec\alpha}, {\vec\beta})} = d {\BS}
\label{eq:texact}
\eeq
where the potential ${\BS}$ can be viewed as a function on ${\CalC}_{r} \times 
{\CalM}^{\rm loc}_{r}$, ${\BS} = {\BS}({\bz}; {\vec\alpha}, {\vec\beta})$, obeying
\beq
{\rm D}\, {\BS}({\bz}; {\vec\alpha}, {\vec\beta}) = {\mathfrak{T}}
\eeq
Finally, note that the connection $\nabla$ is $H$-covariant, in that 
\beq
h^{*}{\nabla} ({\ba}, {\bz} ) = {\nabla} ({\ba}, {\bz}^{h})
\eeq 
 where we set $A_{h({\xi})} = A_{\xi}$ for all ${\xi} \in {\bz}$. Thus, the monodromy data is $H$-invariant (provided we transform the combinatorial data, such as the generating loops of ${\pi}_{1}(S^{2} \backslash {\bz}, pt)$). In this way the map $m$ descends to the quotient
 \beq
 {\tilde m} : {\CalM}_{p}^{\rm alg} \times {\tilde\mfM}_{0,p+3} \longrightarrow {\CalM}^{\rm loc}_{p}
 \eeq
where ${\tilde\mfM}_{0,p+3} \to {\mfM}_{0,p+3}$ is a finite cover, keeping track of the additional combinatorics involved in describing the monodromy data. 

 \subsubsection{Schlesinger $\tau$-function}

The $\tau$-function 
for the $p+3$-point Schlesinger problem is defined through the potential ${\BS}({\bz}; {\vec\alpha}, {\vec\beta})$:
\beq
{\tau} ({\qe}_{1}, \ldots, {\qe}_{r} ; {\vec\alpha}, {\vec\beta}) = e^{{\BS}({\bz}; {\vec\alpha}, {\vec\beta})}  \ \prod_{{\xi} \neq {\eta}} ( {\xi} - {\eta} )^{D_{\xi\eta}} \, ,   
\eeq
where the symmetric, $D_{\eta\xi} = D_{\xi\eta}$, matrix $\Vert D_{\eta\xi} \Vert$, linear in ${\vt}_{\xi}^2$, 
 obeys 
\beq
\sum\limits_{\eta \neq \xi} D_{\eta\xi} = 2 {\vt}_{\xi}^{2} \ . 
\eeq
The $\tau$-function is defined on the product
\beq
\widetilde{{\mfM}_{0,p+3}} \times {\CalM}^{\rm loc}_{p} \ 
\eeq
of the universal cover of the moduli space of complex structures of $S^{2} \backslash {\bz}$ and the moduli space of $G$-flat connections on  $S^{2} \backslash {\bz}$ with fixed conjugacy classes of monodromy around individual punctures. To stay away from complications \cite{Bolibruch}, we assume all conjugacy classes generic. 

\vskip 1cm

\section{Darboux coordinates}

Let us describe the system \eqref{eq:schfl} more explicitly, in some Darboux coordinates
on ${\CalM}_{p}^{\rm alg}$. Basically, there are two classes of coordinate systems on ${\CalM}_{p}^{\rm alg}$:
1) the $\bz$-dependent action-angle \cite{ArnoldCM} variables $(a_{i}, {\varphi}_{i})_{i=1}^{p}$, or separated variables \cite{sklyanin}, which we shall call the $w$-coordinates $(w_{i}, p_{w_{i}})_{i=1}^{p}$, 
and 2) what we call 
the background-independent, or $y$-coordinates $(y^{a}, p_{y^{a}})_{a=1}^{p}$. For the latter
there are several interesting possibilities, which we review below. See \cite{Babich} for some earlier work. 

Throughout this section we shall use the notation
\beq
{\vec\vt} = ( {\vt}_{0} , {\vt}_{\qe} , {\vt}_{1}, {\vt}_{\infty} ) \in {\BC}^{4}
\eeq

\subsection{Background independent coordinates}

\subsubsection{The $\yt$-coordinates}

Using the coordinates \eqref{eq:pxo} on ${\CalO}_{\vt}$ we can represent
the symplectic quotient ${\CalM}_{p}$ \eqref{eq:malg} as the set of $2|{\bz}| = 2(p+3)$-uples 
$({\gamma}_{{\pm}, \xi}, {\beta}_{\xi})_{{\xi} \in {\bz}}$, obeying 
\beq
\sum_{{\xi} \in {\bz}} {\beta}_{\xi} = 
\sum_{{\xi} \in {\bz}} {\gamma}_{{\pm}, \xi}{\beta}_{\xi} {\mp} {\vt}_{\xi} = \sum_{{\xi} \in {\bz}}
 {\gamma}_{{\pm}, \xi}^{2} {\beta}_{\xi} {\mp} 2{\vt}_{\xi}{\gamma}_{{\pm}, \xi} = 0
\label{eq:mommap}
\eeq
(and similarly in the  
$({\tilde\gamma}_{{\pm}, {\xi}}, {\tilde\beta}_{\xi})$-coordinates) 
modulo the diagonal $G$-action \eqref{eq:sl2act}:
\beq
({\gamma}_{{\pm}, \xi},{\beta}_{\xi})_{{\xi} \in {\bz}} \mapsto \left( \, \frac{a {\gamma}_{{\pm},\xi} +b}{c {\gamma}_{{\pm},\xi} +d} \, , \, {\beta}_{\xi} 
( c {\gamma}_{{\pm},\xi} +d)^2 {\mp}  2{\vt}_{\xi} c 
(c {\gamma}_{{\pm},\xi}+d) \, \right)_{{\xi} \in {\bz}} \,  . 
\label{eq:diagsl2}
\eeq
The idea behind the $\yt$-parametrization is the observation that the ${\gamma}_{{\pm},\xi}$-coordinates transform under the $G$-action independently of $\beta_{\xi}$'s. We thus may project an open subset ${\CalU}_{\ep}$, with ${\ep} = ({\ep}_{\xi})_{{\xi} \in {\bz}}$, ${\ep}_{\xi} = \pm$
of 
${\CalM}_{p}^{\rm alg}$ (the one where the ${\gamma}_{{\ep}_{\xi},\xi}$'s are well-defined, i.e. away from a codimension one locus, cf. the discussion below the Eq. \eqref{eq:pxo}, and also are different from each other) to ${\mfM}_{0,p+3}$, which, in turn, can be parameterized by 
\beq
{\yt}^{a}_{\ep} = 
\frac{({\gamma}_{a}-{\gamma}_{-1})({\gamma}_{0}-{\gamma}_{p+1})}{({\gamma}_{0}-{\gamma}_{-1})({\gamma}_{a}-{\gamma}_{p+1})}\ , \qquad a = 1, \ldots , p
\label{eq:ycor}
\eeq
where we use the short-hand notation ${\gamma}_{i} = {\gamma}_{{\ep}_{z_{i}}, z_{i}}$, $i=-1, \ldots, p+1$, 
and $p_{y^{a}}$, defined through the linear relations:
\beq
{\beta}_{i} = \sum_{i \neq j} \frac{{\Theta}_{ij}}{{\gamma}_{i} - {\gamma}_{j}} + \sum_{a=1}^{r}
\, p_{{\yt}^{a}} \, \frac{\partial {\yt}^{a}}{\partial {\gamma}_{i}} 
\label{eq:psor}
\eeq
where ${\Theta}_{ij} = {\Theta}_{ji}$ is any symmetric matrix, which we assume linear in 
$({\ep}_{z_{i}}{\vt}_{z_{i}})_{i=1}^{r+3}$, obeying
\beq
\sum_{j \neq i} {\Theta}_{ij} = 2{\ep}_{z_{i}}{\vt}_{z_{i}}\ .
\eeq
in order to 
solve \eqref{eq:mommap}. There are many choices for the ${\Theta}$-matrix, which makes the choice of $p_{{\yt}^{a}}$'s non-canonical:
\beq
p_{{\yt}^{a}} \longrightarrow p_{{\yt}^{a}} + {\partial}_{{\yt}^{a}} f({\yt})
\eeq
with
\beq
f({\yt}) = \sum\limits_{i \neq j} \, {\delta}{\Theta}_{ij} \, {\rm log}({\gamma}_{i} - {\gamma}_{j})
\eeq
for any symmetric matrix ${\delta}{\Theta}$ obeying $\sum\limits_{i \neq j} {\delta}{\Theta}_{ij} =0$ for any $j$.

{}For example, for $r=1$, ${\bz} = \{ 0, {\qe}, 1, {\infty} \}$, we may take, for ${\ep} = (+,+,+,+)$, 
\beq
{\yt} = {\yt}^{1}_{\ep} = \frac{{\gamma}_{+,\qe}- {\gamma}_{+,0}}{{\gamma}_{+,1}-{\gamma}_{+,0}} \frac{{\gamma}_{+,1} - {\gamma}_{+,\infty}}{{\gamma}_{+,\qe} - {\gamma}_{+,\infty}} \ , 
\eeq
and
\beq
\begin{aligned}
& {\beta}_{0} = \frac{{\vt}_{0}+{\vt}_{\qe} - {\vt}_{1} + {\vt}_{\infty}}{{\gamma}_{+,0}-{\gamma}_{+,\qe}} 
+ \frac{{\vt}_{0}-{\vt}_{\qe} + {\vt}_{1} - {\vt}_{\infty}}{{\gamma}_{+,0}-{\gamma}_{+,1}} 
+ \frac{\partial {\yt}}{\partial {\gamma}_{+,0}} p_{\yt} \\
& {\beta}_{\qe} = \frac{{\vt}_{0}+{\vt}_{\qe} - {\vt}_{1} + {\vt}_{\infty}}{{\gamma}_{+,\qe}-{\gamma}_{+,0}} + \frac{{\vt}_{0}-{\vt}_{\qe} - {\vt}_{1} + {\vt}_{\infty}}{{\gamma}_{+,1}-{\gamma}_{+,\qe}} + 
\frac{\partial \yt}{\partial {\gamma}_{+,\qe}} p_{\yt} \\
& {\beta}_{1} = \frac{{\vt}_{0}-{\vt}_{\qe} + {\vt}_{1} - {\vt}_{\infty}}{{\gamma}_{+,1}-{\gamma}_{+,0}} + \frac{{\vt}_{0}-{\vt}_{\qe} - {\vt}_{1} + {\vt}_{\infty}}{{\gamma}_{+,\qe}-{\gamma}_{+,1}} + 
\frac{2{\vt}_{\infty}}{{\gamma}_{+,1}-{\gamma}_{+,\infty}} + \frac{\partial \yt}{\partial {\gamma}_{+,1}} p_{\yt} \\ 
& {\beta}_{\infty} = \frac{2{\vt}_{\infty}}{{\gamma}_{+,\infty}-{\gamma}_{+,1}} + \frac{\partial \yt}{\partial {\gamma}_{+,\infty}} p_{\yt} \, , \\
\label{eq:pso}
\end{aligned}
\eeq 
The pair $({\yt}, p_{\yt})$ are the Darboux coordinates, i.e. ${\varpi} = dp_{\yt} \wedge d\yt$, on one patch of ${\CalM}_{1}^{\rm alg}$. 
Changing ${\gamma}_{+,{\xi}}$ to ${\gamma}_{-,{\xi}}$, accompanied by the change of the corresponding ${\vt}_{\xi}$ to $(-{\vt}_{\xi})$  leads to the canonical transformation $({\yt}, p_{\yt} ) \mapsto ({\yt}_{\xi}, p_{{\yt}_{\xi}})$ connecting to another patch on ${\CalM}_{1}^{\rm alg}$, as we show in some detail in the appendix. Also, 
choosing another ordering of $\bz$ maps $\yt$ to $1-{\yt}$, ${\yt}^{-1}$ etc.
and produces coordinates on other patches. The conjugate momentum $p_\yt$ changes accordingly, albeit non-canonically, as one can shift it by ${\partial}_{\yt} f$ for some $f({\yt})$. 

The functions $h_{0{\qe}}$ and $h_{1{\qe}}$, expressed in the $\yt$-coordinates, are
 simple polynomials:
\begin{multline}
h_{0{\qe}} = 
 {\yt}^{2} (1-{\yt}) p_{\yt}^{2} + {\yt} \left( {\yt} ({\vt}_{0} - {\vt}_{\qe} + {\vt}_{1} -{\vt}_{\infty}) + 2 ({\vt}_{\infty} - {\vt}_{1}) \right) p_{{\yt}} + \\
 + ({\vt}_{1} - {\vt}_{\infty})^{2} - {\vt}_{0}^{2} - {\vt}_{\qe}^{2} \, , \\
h_{1{\qe}} = 
{\yt} (1-{\yt})^{2} p_{{\yt}}^{2} - (1-{\yt}) \left( (1-{\yt}) ({\vt}_{0} - {\vt}_{\qe} + {\vt}_{1} - {\vt}_{\infty} ) - 2({\vt}_{0} + {\vt}_{\infty}) \right) p_{{\yt}} + \\
{\vt}_{0}^{2} - {\vt}_{\qe}^{2} - {\vt}_{1}^{2}- {\vt}_{\infty}^{2}
+ 2 {\vt}_{\infty} ( {\vt}_{1} - {\vt}_{\qe}) + 2{\vt}_{\infty} {\yt}^{-1} 
\left( {\vt}_{0} + {\vt}_{\qe} - {\vt}_{1} + {\vt}_{\infty} \right) \ . \\
\end{multline}
Now, the Hamiltonian $H_{\qe}  = \frac{h_{0{\qe}}}{\qe}  + \frac{h_{1{\qe}}}{{\qe}-1}$ contains both $p_{\yt}^{2}$ and
$p_{\yt}$ terms. In order to write down the equations of motion it is convenient to shift the $p_{\yt}$ variable
\beq
p_{\yt}\ = \ {\pt}_{\yt} \,  + \frac{{\hat\vt}_{\infty}-{\half}}{{\yt}} + \frac{{\hat\vt}_{1}}{{\yt}({\yt}-1)}+{\qe} \frac{{\hat\vt}_{\qe}-{\half}}{{\yt}({\yt}-{\qe})}\, , \label{eq:pythhat}
\eeq
where we defined
\beq
\begin{aligned}
 {\hat\vt}_{0} & \ = \ \half \left(  {\vt}_{0} + {\vt}_{\qe} - {\vt}_{1} - {\vt}_{\infty} \right) \, , \\
 {\hat\vt}_{1} &\ = \ \half \left(  {\vt}_{0} - {\vt}_{\qe} - {\vt}_{1} +  {\vt}_{\infty} \right) \, , \\
 {\hat\vt}_{\qe} -{\half} &\ = \ \half \left( {\vt}_{0} +{\vt}_{\qe} + {\vt}_{1} + {\vt}_{\infty} \right) \, , \\
 {\hat\vt}_{\infty}- {\half} &\  =\  \half \left( {\vt}_{0} -{\vt}_{\qe} + {\vt}_{1} - {\vt}_{\infty} \right) \, , \\
\end{aligned}
\label{eq:thehat}
\eeq
so as to eliminate the terms in $H_{\qe}$, linear in $\pyt$. In doing so, one should be careful as the shift depends on $\qe$, so the Hamiltonian will also change, according to:
\begin{multline}
p_{\yt} d{\yt} - H_{\qe} d{\qe} = {\pyt} dy - {\mathfrak{h}^{\yt}}_{PVI} ( {\yt}, {\pyt} ; {\qe}; {\vec{\hat\vt}}) d{\qe} +
d {\CalY}\ , \\
\qquad 
{\CalY} = \left( {\hat\vt}_{\infty} - {\half}\right) \, {\rm log}({\yt}) + \left( {\hat\vt}_{\qe} - {\half} \right)\, {\rm log} \left( 1 - {\qe}/{\yt} \right) + {\hat\vt}_{1}\, {\rm log} \left( 1- 1/{\yt} \right)  - \\
- \left( {\vt}_{0}^{2} + {\vt}_{\qe}^{2} 
\right) {\rm log}({\qe}) - {\half} \left( ({\vt}_{0}- {\vt}_{\infty})^{2}-({\vt}_{\qe} + {\vt}_{1})^{2} \right) {\rm log} ( 1- {\qe} ) \, , \label{eq:can1}
\end{multline}
and
\begin{multline}
{\mathfrak{h}^{\yt}}_{PVI}({\yt} , {\pyt} ; {\qe} ; {\vec{\hat\vt}}) \ = \ \frac{{\yt}({\yt}-1)({\yt}-{\qe})}{{\qe}({\qe}-1)} {\pyt}^{2} \ + \qquad\qquad\qquad\qquad\qquad\qquad\qquad\qquad \\ + 
\frac{{\hat\vt}_{0}^{2}}{{\yt}(1-{\qe})} + \frac{\frac 14 - {\hat\vt}_{\qe}^{2}}{{\yt}-{\qe}} + \frac{{\hat\vt}_{1}^{2}\, {\yt}}{{\qe}({\yt}-1)} + \frac{({\hat\vt}_{\infty} - {\half})^{2} {\yt}}{{\qe}(1-{\qe})}  \, . 
 \label{eq:ham1}
\end{multline}
The equation of motion, which follows from the Hamiltonian \eqref{eq:ham1}, is
\begin{multline}
\frac{d^{2} {\yt}}{d{\qe}^{2}} \ = \ \frac{1}{2} \left(\frac{d{\yt}}{d{\qe}} \right)^2
\left( \frac{1}{\yt} + \frac{1}{{\yt}-1} + \frac{1}{{\yt}-{\qe}} \right) -
\frac{d{\yt}}{d{\qe}} \left( \frac{1}{\qe} + \frac{1}{{\qe}-1} + \frac{1}{{\yt}-{\qe}} \right) + \\
+ \frac{2 {\yt}({\yt}-1)({\yt}-{\qe})}{{\qe}({\qe}-1)} \left( \frac{{\hat\vt}_{0}^{2}}{{\yt}^{2} (1-{\qe})}
+ \frac{\frac 14 - {\hat\vt}_{\qe}^{2}}{({\yt}-{\qe})^{2}} + \frac{{\hat\vt}_{1}^{2}}{{\qe}({\yt}-1)^2}
+ \frac{({\hat\vt}_{\infty} - \half )^{2}}{{\qe}({\qe}-1)} \right) 
\label{eq:p6i}
\end{multline}
This is the celebrated Painlev{\'e} VI equation \cite{Painleve}, or PVI, for short. However, this is not the only connection of the
isomonodromy deformation to PVI. 

In the $r>1$ case the Hamiltonian system we'd get in the $y$-coordinates
is known as the \emph{Garnier} system \cite{Garnier}. 

\subsubsection{Tau-function of Painlev{\'e} VI}

In \cite{JM} the  ${\tau}$-function ${\tau}_{PVI}({\qe} | {\alpha}, {\beta}; {\vec\vt} )$ was associated to PVI.  It is a holomorphic on the universal cover $\widetilde{{\CP}^{1} \backslash \{ 0, 1, \infty \}}_{\qe}$ function, defined by (cf. \cite{Gamayun:2013auu})
\beq
\frac{\partial}{{\partial}{\qe}} {\rm log} \left( {\qe}^{\frac{{\vt}_{0}^{2} + {\vt}_{\qe}^{2} - {\vt}_{1}^{2}-{\vt}_{\infty}^{2}}{2}} ( 1 - {\qe} )^{ \frac{-{\vt}_{0}^{2} + {\vt}_{\qe}^{2} + {\vt}_{1}^{2}-{\vt}_{\infty}^{2}}{2}} {\tau}_{PVI}({\qe} |  {\alpha}, {\beta} ; {\vec\vt}) \right) = H_{\qe}
\label{eq:tpvi}
\eeq
where the Schlesinger\footnote{Yes, Schlesinger, not Painlev{\'e}} Hamiltonian $H_{\qe}$ is to be evaluated on the trajectory characterized by the conserved
monodromy data $({\alpha}, {\beta})$ of the corresponding Schlesinger isomonodromy (= Painlev{\'e} Hamiltonian) flow.

\subsubsection{Polygon variables}

We can view $\ba$ as a complex version of the $r+3$-gon $P_{r+3}$ in the complex three dimensional Euclidean space
${\BC}^{3}= Lie(G)$, with the vertices $v_{1} = 0$, $v_{2} = A_{z_{1}}$, $v_{3} = A_{z_{1}}+A_{z_{2}}$, $\ldots$, $v_{r+3} = A_{z_{1}} + A_{z_{2}} + \ldots + A_{z_{r+2}}$, and the edges\footnote{An edge $e = \overline{v'v''}$ is a complex line $v' + {\BC}(v''-v') \subset {\BC}^{3}$.} $e_{i} = \overline{v_{i}v_{i+1}}$, $i = 1, \ldots , r+3$,\footnote{with $v_{r+4} = v_{1}$}. The ``length'' of an edge $e_i$ is the integral of the square root of the restriction of the holomorphic metric $ds^2 = \half {\rm Tr} dA^2$ on $e_i$, along any path connecting the points $v_i$ and $v_{i+1}$, and is clearly given by $\sqrt{{\half}{\rm Tr} (v_{i+1}-v_{i})^2} = \pm {\vt}_{z_{i}}$, as the complex length is defined up to a sign. Let us now partition $P_{r+3}$ into $r+1$ triangle by $r$ ``diagonals'' ${\delta}_{1}, \ldots , {\delta}_{r}$. One choice of partitioning is 
\beq
{\delta}_{1} = \overline{v_{1}v_{3}}, \, {\delta}_{2} = \overline{v_{1}v_{4}}, \, \ldots \, , \, {\delta}_{r} = \overline{v_{1}v_{r+2}}
\label{eq:diags1}
\eeq
The corresponding triangles are
\beq
{\Delta}_{1} = \widehat{v_{1}v_{2}v_{3}}, \, {\Delta}_{2} = \widehat{v_{1}v_{3}v_{4}}, \, \ldots
\, {\Delta}_{r+1} = \widehat{v_{1}v_{r+2}v_{r+3}}
\label{eq:triangles1}
\eeq
More generally, let $T$ be a tree with $r+3$ tails (vertices of valency $1$) $e_{1}, \ldots , e_{r+3}$, and all internal vertices of valency $3$. Euler formula 
$\#$ vertices $- \#$ edges $= 1$ implies that $T$ has $r+1$ internal vertex ${\Delta}_{1}, \ldots , {\Delta}_{r+1}$, and $r$ internal edges (i.e. edges whose both ends are internal) ${\delta}_{1}, \ldots , {\delta}_{r}$. 
We assign the tails of $T$ to the edges of the polygon, while the internal edges correspond to the diagonals. The internal vertices correspond to the triangles. The three sides of the triangle ${\Delta}_{j}$ are the edges of $P_{r+3}$ and diagonals corresponding to the three edges of $T$ attached to $\Delta_j$. 
The Darboux coordinates in the coordinate chart $U_{T}$ are $({\ell}_{i}, {\theta}_{i})_{i=1}^{r}$ where ${\ell}_{i}$
are the complex lengths of the diagonals ${\delta}_{i}$, and the angles ${\theta}_i$ between the two triangles
${\Delta}_{j'}$ and ${\Delta}_{j''}$ corresponding to the two ends of the edge $\delta_i$ of $T$. 

Let us describe this explicitly in the $r=1$ case (the general case reduces to this one). 
Again, we compute the algebraic functions $h_{\eta\xi}$, defined in \eqref{eq:hxe}, 
in terms of the length ${\ell}$ of the diagonal and the dihedral angle $\theta$. 
The functions  $h_{01}, h_{0{\qe}}, h_{1{\qe}}$ obey the linear relation
\beq
h_{01} + h_{0{\qe}} + h_{1{\qe}} = {\vt}_{\infty}^{2} - {\vt}_{0}^{2} - {\vt}_{\qe}^{2} - {\vt}_{1}^{2}
\eeq
and have nontrivial Poisson brackets in the symplectic form $\varpi$ (cf. \eqref{eq:cubinv}):
\beq
\{ h_{0{\qe}}, h_{1{\qe}} \} \  = \ {\ii} \, {\tilde h}_{01{\qe}} 
\label{eq:pb}
\eeq
where
\beq
\frac 12 {\tilde h}_{01{\qe}}^{2} =   {\vt}_{0}^{2} h_{1{\qe}}^{2} + {\vt}_{{\qe}}^{2} 
h_{01}^{2} + {\vt}_{1}^{2} h_{0{\qe}}^{2} - 
h_{01}h_{0{\qe}} h_{1{\qe}} - 4 {\vt}_{0}^{2}{\vt}_{\qe}^{2} {\vt}_{1}^{2}
\label{eq:rel}
\eeq
The length $\ell$ of the diagonal ${\delta}_{1}$, connecting $v_{1}=0$ to $v_{3} = A_{0} + A_{\qe}$, is given by:
\beq
{\ell}^{2} = \half {\rm Tr} ( A_{0} + A_{\qe} )^{2} = {\vt}_{0}^{2} + {\vt}_{\qe}^{2} + h_{0{\qe}}\ .
\label{eq:diag}
\eeq
It is not difficult to compute
\beq
\begin{aligned}
& {\half} {\rm Tr} [A_{0}, A_{\qe}]^{2} = ( {\ell}^{2} - ({\vt}_{0} - {\vt}_{\qe})^2 ) ( {\ell}^{2} - ({\vt}_{0} +{\vt}_{\qe})^{2}) \\
& {\half} {\rm Tr} [A_{1}, A_{\infty}]^{2} = ( {\ell}^{2} - ({\vt}_{1} - {\vt}_{\infty})^2 ) ( {\ell}^{2} - ({\vt}_{1} +{\vt}_{\infty})^{2}) \\
\label{eq:commnorm}
\end{aligned}
\eeq
The angle $\theta$ between the two triangles $\widehat{v_{1}v_{2}v_{3}}$ and 
$\widehat{v_{1}v_{3}v_{4}}$ which are formed by the diagonal ${\delta}_{1}$
can be computed via
\begin{multline}
{\rm cos}({\theta}) \, = \, \frac{- 2{\ell}^{2} h_{1{\qe}} -  ({\ell}^{2} + {\vt}_{\qe}^{2} - {\vt}_{0}^{2})({\ell}^{2} + {\vt}_{1}^{2} - {\vt}_{\infty}^{2})}{\sqrt{ \prod\limits_{\pm} \left( {\ell}^{2} - \left({\vt}_{0} \pm {\vt}_{\qe}\right)^{2} \right) \left( {\ell}^{2} - \left({\vt}_{1} \pm {\vt}_{\infty}\right)^{2} \right)}}   \,  \, , \\
 {\rm sin} ({\theta}) \, = \, \frac{{\ii} {\ell}\,  {\tilde h}_{01{\qe}}}{\sqrt{\prod\limits_{\pm} \left( {\ell}^{2} - \left({\vt}_{0} \pm {\vt}_{\qe}\right)^{2} \right) \left( {\ell}^{2} - \left({\vt}_{1} \pm {\vt}_{\infty}\right)^{2} \right)}}  \,  
\label{eq:angl}
\end{multline}
Using \eqref{eq:pb}, \eqref{eq:rel} it is easy to verify that  
$\{ {\ell} , {\theta} \} = 1$, as in \cite{klyachko, km}.  See also \cite{Babich}. 

\subsection{Time-dependent coordinates}

We now present two sets of coordinate systems on ${\CalM}_{r}^{\rm alg}$ which depend explicitly on the complex structure of the underlying genus zero curve with $r+3$ punctures. 

\subsubsection{Action-angle variables}

It was observed long time ago that the polygon coordinates are actually the extremal versions of the action-angle coordinates on the partially compactified ${\CalM}_{r}^{\rm alg}$, which is an algebraic integrable system, obtained by a degeneration \cite{Nekrasov:1995nq} of Hitchin system \cite{Hitchin}. Fix ${\bz}_{0} \in {\CalC}_{r}$. 
Define the analytic curve ${\Sigma} \subset T^{*}\left( {\BC\BP}^{1} \backslash {\bz}_{0} \right)$ by the equation
\beq
\begin{aligned}
& {\rm Det}\left( {\CalA}(z) - p \cdot 1 \right) = 0\, , \\
& p^{2} = \sum_{{\xi} \in {\bz}_{0}} \, \frac{{\Delta}_{\xi}}{(z - {\xi})^{2}} + \frac{H_{\xi}({\ba}, {\bz}_{0})}{z-{\xi}}  \\
\end{aligned}
\label{eq:speccurve}
\eeq
It is not difficult to see that $\Sigma$ has genus $r$, e.g. by tropical limit, or by counting the number of branch points of \eqref{eq:speccurve} (and adding $2(r+3)$ points $z = {\xi}$, $p (z-{\xi}) = \pm {\vt}_{\xi}$ to make $\Sigma$ compact). 
The eigenlines $L_{z,p}$ of ${\CalA}(z)$ corresponding to the eigenvalue $p$
found from \eqref{eq:speccurve} form a line bundle ${\CalL}$ (one needs to
work a little bit at the ramification point $z= z_{*}$
where $p =0$, where ${\CalA}(z_{*})$ generically has the Jordan block form)
over $\Sigma$. By appropriate non-canonical normalization one can make ${\CalL}$ to be a degree zero line bundle, $c_{1}({\CalL}) = 0$, thus defining a point $[{\CalL}] \in Jac({\Sigma})$ in the Jacobian. A holomorphic line bundle can be equivalently described by a ${\BC}^{\times}$-gauge equivalence class of a $(0,1)$-part ${\bar\partial} + {\bar a}$ of a ${\BC}^{\times}$-connection on a trivial bundle ${\BC} \times {\Sigma}$ over $\Sigma$:
\beq
{{\CalB}un}_{{\BC}^{\times}}^{0} ({\Sigma}) = Jac({\Sigma}) = H^{0,1}({\Sigma})/H^{1}({\Sigma}, {\BZ}) = \{ \, {\bar a} \sim {\bar a} + 
 {\bar\partial} {\phi} \, \}\ ,
 \label{eq:jac}
 \eeq
 with gauge transformations $g= e^{{\ii}{\phi}}: {\Sigma} \to {\BC}^{\times}$. The $1$-form ${\ii} d{\phi} = g^{-1} dg$ for such transformation is globally well-defined on $\Sigma$, but it is not, in general, exact. Its periods are in $2{\pi}{\ii}{\BZ}$. 

The action-angle coordinates are defined relative to the choice of symplectic
basis $A_i, B^j \in H_{1}({\Sigma}, {\BZ}) \approx {\BZ}^{2r}$, obeying 
$A_{i} \cap B^{j} = {\delta}_{i}^{j}$, $A_{i}\cap A_{j} = 0$, $B^{i} \cap B^{j} = 0$. 
Then the Darboux coordinates $(a_{i}, {\varphi}_{i})_{i=1}^{r}$ are defined via:
\beq
a_{i} = \frac{1}{2\pi} \oint_{A_{i}} pdz \, , 
\label{eq:aper}
\eeq 
 and $d{\varphi}_{i}$ are the holomorphic differentials on $Jac({\Sigma})$, defined e.g. by fixing a representative
 \beq
 {\bar a} \ = \ \sum_{i=1}^{r} \ {\varphi}_{i} \, {\varpi}_{i}^{*}\, , \ 
 \label{eq:a01}
 \eeq
where ${\varpi}_{i}$, $i = 1, \ldots , r$ are holomorphic differentials on $\Sigma$, which are normalized relative to the
 $A_i, B^j$ basis:
 \beq
 \frac{1}{2\pi} \oint_{A_{j}} {\varpi}_{i} = {\delta}_{i}^{j} 
\eeq
The coordinates ${\varphi}_{i}$ are defined up to the shifts by $2{\pi}{\BZ}$ (this is familiar from the classical Liouville integrability \cite{ArnoldCM}), and up to the shifts ${\varphi}_{i} \sim {\varphi}_{i} + 2{\pi}{\tau}_{ij} m^{j}$, with $m^j \in \BZ$ and 
\beq
 \frac{1}{2\pi} \oint_{B^{j}} {\varpi}_{i} = {\tau}_{ij}
 \eeq
the period matrix of $\Sigma$. This feature is the general property of algebraic integrable systems \cite{Hitchin}.   
In our case the algebraic integrable system is the genus zero version of the $G$-Hitchin system, often called the (classical limit of the ) Gaudin model. 

Now, the polygon variables are obtained by taking the degeneration limit $[{\bz}_{0}] \to [{\bz}_{T}] \in \overline{{\mfM}_{0,r+3}}$ in which the genus zero curve becomes a stable genus zero curve, corresponding to the
tree $T$ (it is built out of the three holed spheres $S_{i}$, corresponding to the internal vertices $i$ of $T$). In this limit 
the curve $\Sigma$ also degenerates, so that its Jacobian becomes an algebraic torus $\left( {\BC}^{\times} \right)^{r}$, and the Hamiltonian flows generated by $H_{\xi}({\ba}, {\bz}_{0})$ (recall, that ${\bz}_{0}$ here are fixed, not to be confused with the times) become the {\it bending} flows. The periods \eqref{eq:aper} then approach
${\ell}_{i}$, the complex lengths of the polygon diagonals. 

\subsubsection{Separated variables}

Define $w_{i}$, $i = 1, \ldots , r$
to be the zeroes of ${\CalA}(z)_{21}$ (using ${\CalA}(z)_{12}$ leads to equivalent theory), in the gauge where $A_{\infty} = {\rm diag}({\vt}_{\infty}, - {\vt}_{\infty})$ is diagonal, 
 i.e. 
\beq
\sum_{{\xi} \in {\bz}} \frac{{\beta}_{\xi}}{w_{i} - {\xi}} = 0 \, , \qquad i = 1, \ldots , r
\eeq
while setting $p_{w_{i}}$ to be equal to the specific eigenvalue of ${\CalA}(w_{i})$,
 namely the value of ${\CalA}(w_{i})_{22} = - {\CalA} (w_{i})_{11}$, 
 i.e.
\beq
p_{w_{i}} = \sum_{{\xi} \in {\bz}} \frac{-{\vt}_{\xi} + {\beta}_{\xi}{\gamma}_{\xi}}{w_{i} - {\xi}} \, . 
\label{eq:pcor}
\eeq
Comparing with the Eq. \eqref{eq:eigb} we understand  
the invariant definition of $(w_{i}, p_{w_{i}})_{i=1}^{r}$ is the locus of vanishing of
${\Upsilon}^{(+)}$\footnote{corresponding to $L_{+}$ in the discussion below
\eqref{eq:lpm}}, the section of the line-bundle $L_{(p,z)} = {\rm ker} ( {\CalA}(z) - p )$ over  the spectral curve $\Sigma$, \cite{sklyanin, Feigin:1994in, Gorsky:1999rb}, which is the same eigenbundle of the Higgs field ${\CalA}(z)$ as
in \eqref{eq:speccurve}, except that now one normalizes it to have degree $r$ (so that
its unique, up to a constant multiple, section has $r$ zeroes and no poles).

In the $r=1$ case 
\beq
\begin{aligned}
& H_{\qe}  = \frac{h_{0{\qe}}}{\qe} + \frac{h_{1{\qe}}}{{\qe}-1} =   \\
& \qquad\qquad = 
\frac{w(w-1)(w-{\qe})}{{\qe}({\qe}-1)} p_{w}^2 +  \frac{{\vt}_{0}^{2}}{w (1-{\qe})} +  \frac{
{\vt}_{\qe}^{2}}{{\qe}-w} +  \frac{{\vt}_{1}^{2}\, w}{{\qe}(w-1)} + \frac{{\vt}_{\infty}^{2} \, w}{{\qe}(1-{\qe})} - {\tilde u}_{0}({\qe}) \, , \\
\end{aligned}
\eeq
where
\beq
{\tilde u}_{0}({\qe}) = \frac{{\vt}_{\qe}^{2}+{\vt}_{0}^{2}}{\qe} + \frac{{\vt}_{0}^{2} - {\vt}_{\qe}^{2}-{\vt}_{1}^{2}+{\vt}_{\infty}^{2}}{1-{\qe}}
\eeq
Here comes the trap: 
the equations of motion, derived from Hamiltonian $H_{\qe}$ in the $(w, p_{w})$ coordinates, do not
coincide with the isomonodromic flow  \eqref{eq:isoflow}. Instead,
\eqref{eq:isoflow} is a Hamiltonian flow, 
generated, in the $(w,p_{w})$-frame, by the $\qe$-dependent Hamiltonian
\beq
H_{\qe} + {\delta}h (w, p_{w}; {\qe})
\label{eq:ham6}
\eeq
where
\beq
{\delta}h (w, p_{w}; {\qe}) 
=  \frac{-w(w-1)p_{w} + {\vt}_{\infty} w}{{\qe}({\qe}-1)}  \ , 
\label{eq:compens}
\eeq 
is compensating the  $\qe$-dependence of the coordinates $(w,p_{w})$, as we explain in the next subsection. Since \eqref{eq:ham6} also contains both $p_{w}^2$ and $p_w$ terms, we shift the momentum:
\beq
p_{w} = {\pt}_{w} + \frac{1}{2(w-{\qe})}
\eeq
and shift accordingly the Hamiltonian \eqref{eq:ham6}, arriving, finally, at the true Hamiltonian ${\mathfrak{h}}^{w}_{PVI}(w, {\pt}_{w}; {\qe}; {\vec{\vt}})$,
\beq
p_{w} dw  - \left( H_{\qe} + {\delta}h (w, p_{w}; {\qe}) \right) d{\qe} = {\pt}_{w} dw - {\mathfrak{h}}^{w}_{PVI}(w, {\pt}_{w}; {\qe})d{\qe} + d{\CalW}
\eeq  
of the \eqref{eq:isoflow} flow\footnote{cf. \cite{Teschner}, \cite{Litvinov:2013sxa} 
 ${\delta}_{1} = {\qalf} - {\vt}_{0}^2$, ${\delta}_2 = {\qalf} - {\vt}_{\qe}^{2}$, ${\delta}_{3} = {\qalf} - {\vt}_{1}^{2}$, 
 ${\delta}_{5} = {\qalf} - \left({\vt}_{\infty} - {\half}\right)^2$}:
\begin{multline}
{\mathfrak{h}}^{w}_{PVI}(w, {\pt}; {\qe}; {\vec{\vt}}) \ = \\
= \ \frac{w(w-1)(w-{\qe})}{{\qe}({\qe}-1)} {\pt}^2 +\frac{{\vt}_{0}^{2}}{w(1-{\qe})} +
\frac{{\qalf} - {\vt}_{\qe}^2}{w-{\qe}} +  \frac{w\, {\vt}_{1}^{2}}{{\qe}(w-1)} +\frac{w\, \left({\vt}_{\infty} - {\half}\right)^2}{{\qe}(1-{\qe})}
\label{eq:pviham}
\end{multline}
 with
\beq
{\CalW} = - {\half} {\rm log}(w - {\qe}) + {\rm function\ of\ } {\qe} \, . 
\label{eq:can2}
\eeq
 A glance at \eqref{eq:pviham} and \eqref{eq:ham1} shows that the isomonodromy flow \eqref{eq:isoflow} is, again, a solution of the PVI equation, but now with the exponents $\left( {\vt}_{0}, {\vt}_{\qe}, 
 {\vt}_{1}, {\vt}_{\infty} \right)$ instead of 
 $\left( {\hat\vt}_{0}, {\hat\vt}_{\qe}, 
 {\hat\vt}_{1}, {\hat\vt}_{\infty} \right)$ given by \eqref{eq:thehat}.
 Note that the Hamiltonian \eqref{eq:pviham} is invariant under the flips ${\vec\vt} \mapsto  ( \pm {\vt}_{0}, \pm {\vt}_{\qe}, \pm {\vt}_{1}, {\vt}_{\infty} )$, which is to be expected, since the eigenvalues of the residues $A_{0}, A_{\qe}, A_{1}$ are defined up to a permutation. However, the flip of the last exponent
 ${\vt}_{\infty}$ does not leave \eqref{eq:pviham} invariant, instead the symmetry is ${\vt}_{\infty} \mapsto 1 - {\vt}_{\infty}$. This is a reflection of background-dependence of the $(w,{\pt}_{w})$-coordinate system. 
 
\subsection{Canonical transformations}

Let us discuss the background-dependence and the canonical transformations
$(y, {\hat p}_{y}) \to (w, {\tilde p}_{w})$ in more detail. 
So we set $r=1$ throughout this
section.

{}We shall use the involutions 
\beq
z \mapsto z^{\circ} \ = \ {\qe}\ \frac{1-z}{{\qe}-z}\, , \ z \mapsto z^{\ii} \ = \ \frac{\qe}{z} \, , \ z \mapsto z^{\ii\ii} = \frac{z-{\qe}}{z-1}
\label{eq:zcirc}
\eeq 
 of ${\BC\BP}^{1}$, $\left( z^{\circ} \right)^{\circ} = z$, 
 which induce the three nontrivial involutions of 
${\bz}$: 
\beq
\left( 0 \leftrightarrow 1 , {\qe} \leftrightarrow {\infty} \right) , \left( 0 \leftrightarrow {\infty} , {\qe} \leftrightarrow 1\right), \ {\rm and} 
\left( 0 \leftrightarrow {\qe} , 1 \leftrightarrow {\infty} \right) \ , 
\eeq
respectively. 
Each has two fixed points $z_{\pm}^{\circ} = {\qe} \pm \sqrt{{\qe}(1-{\qe})}$, $z_{\pm}^{\ii} = \pm \sqrt{\qe}$, $z_{\pm}^{\ii\ii} = 1 \pm \sqrt{1-{\qe}}$, respectively. 
The transformations $z \mapsto z^{\circ}, z^{\ii}, z^{\ii\ii}$
together with the identity map represent the action on ${\BC\BP}^{1} \backslash {\bz}$
of the celebrated Klein $4$-group $D_2$, 
isomorphic to ${\BZ}_{2} \times {\BZ}_{2}$. 

\subsubsection{Background dependence of $w$-coordinates}

Let us write
\beq
- {\CalA}(z)_{21} = \frac{{\beta}_{0}}{z} + \frac{{\beta}_{\qe}}{z-{\qe}} + \frac{{\beta}_{1}}{z-1} = 
{\kappa} \frac{z-w}{z(z-{\qe})(z-1)}\ , 
\label{eq:ca21}
\eeq
in the basis, where $A_{\infty}$ is diagonal, 
 from which we deduce: $w^{\circ} =  - \left( A_{1} \right)_{21} / \left( A_{\qe} \right)_{21}$, or
\beq
w =  \left( - {\beta}_{1}/{\beta}_{\qe} \right)^{\circ} \ . 
\eeq
In other words, the $w^{\circ}$ variable is $\qe$-independent, while $w$ is. 
Now substituting
 into \eqref{eq:pcor}, we get 
 \beq 
 p_{w} = \frac{(w^{\circ} -{\qe})^{2}}{{\qe} (1-{\qe}) w^{\circ}} 
 \left( \frac{(A_{0})_{11}}{w^{\circ}-1} +
 (A_{\qe})_{11}   +   \frac{{\qe} {\vt}_{\infty}}{{\qe}-w^{\circ}} \,  \, \right) \ . 
 \eeq
Accordingly, the functions $h_{\xi\eta}$, when expressed through $(w,p_{w})$, 
acquire an explicit $\qe$-dependence. 

Note that as ${\qe} \to 0, 1, \infty$, with fixed $A_{\xi}$'s, and $w^{\circ}$,  
the coordinate $w$ approaches $0, 1, w^{(\infty)} = 1 -w^{\circ}$, respectively. 
The coordinate $p_{w}$ is singular when ${\qe} \to 0,1$, while as ${\qe} \to \infty$ it approaches
\beq
p_{w}^{(\infty)} = - \frac{1}{w^{\circ}} \left( \frac{(A_{0})_{11}}{w^{\circ}-1} +
 (A_{\qe})_{11}   +    {\vt}_{\infty}  \,  \, \right) \ .
\eeq
Thus, we can solve for $(w,p)$ in terms of $(w^{({\infty})}, p_{w}^{(\infty)}, {\qe})$:
\beq
w =  \frac{{\qe} \, w^{(\infty)}}{w^{(\infty)} + {\qe} -1}\, , \qquad p_{w} = \frac{p_{w}^{(\infty)}  (w^{(\infty)}+{\qe}-1)^{2}  - 
{\vt}_{\infty} \left( w^{(\infty)}+{\qe}-1 \right)}{{\qe}({\qe}-1)} 
\label{eq:pwflow}
\eeq
The map $(w^{(\infty)}, p_{w}^{(\infty)}) \mapsto (w, p_{w})$ preserves the symplectic form $dp \wedge dw$:
\beq
dp_{w} \wedge dw =  dp_{w}^{(\infty)} \wedge dw^{(\infty)}
\eeq
and is generated, ${\delta}h(w, p_{w}; {\qe})  = - \frac{{\partial} L}{{\partial} {\qe}}$, by our friend \eqref{eq:compens}, 
 which naturally 
obeys ${\delta}h(w, p_{w}; {\qe}) 
= {\delta}h (w^{(\infty)}, p_{w}^{(\infty)}; {\qe})$.
Here 
\beq
L (w, p_{w}^{({\infty})}; {\qe}) = p_{w}^{({\infty})}  w \frac{1-{\qe}}{w-{\qe}} + {\vt}_{\infty} \, {\rm log} \left( 1 - \frac{w}{\qe} \right) 
\eeq
 is the generating function
of the map $(w^{({\infty})}, p_{w}^{({\infty})}) \mapsto (w, p_{w})$, is
\beq
p_{w} = \frac{{\partial}L}{{\partial}w}\, , \ w^{({\infty})} =  \frac{{\partial}L}{{\partial}p_{w}^{({\infty})}}
\eeq

\subsubsection{From $\yt$ to $w$ -- Okamoto transformations}

It is instructive to find the explicit map between $({\yt}, p_{\yt})$ and $(w, p_{w})$ coordinates.
Let us use the notation
\beq
{\vt} =  {\vt}_{0} + {\vt}_{\qe} + {\vt}_{1} + {\vt}_{\infty} \ .
\label{eq:sumth}
\eeq
We diagonalize $A_{\infty}$, i.e. find ${\rm g}_{\infty}$, such that
${\gt}_{\infty}^{-1} A_{\infty} {\gt}_{\infty} = {\rm diag} ( {\vt}_{\infty}, - {\vt}_{\infty} )$, 
meaning we set ${\gamma}_{\infty} = 0$, ${\beta}_{\infty} = 0$, which implies, cf. \eqref{eq:ycor}, \eqref{eq:pso}, 
that
\beq
{\yt} = \frac{1 - {\gamma}_{0}/{\gamma}_{\qe}}{1 - {\gamma}_{0}/{\gamma}_{1}}\, , \qquad
\frac{{\gamma}_{1}}{{\gamma}_{\qe}} =  1 + \frac{2{\vt}_{\infty}}{y p_{y}} \, , \qquad  
\frac{{\gamma}_{1}}{{\gamma}_{0}} = 1  - \frac{2{\vt}_{\infty}}{y(y-1)p_{y}} 
\label{eq:ygg}
\eeq
Next, setting the zero of the $21$-matrix element of ${\gt}_{\infty}^{-1}{\CalA}(z) {\gt}_{\infty}$ to be equal to $w$,
\beq
\left[ {\gt}_{\infty}^{-1}\left( \frac{A_{0}}{w}+\frac{A_{\qe}}{w-{\qe}} + \frac{A_{1}}{w-1} \right) {\gt}_{\infty} \right]_{21} = 0 \, , 
\eeq 
we get: 
\beq
{\beta}_{0} = - {\kappa} \frac{w}{\qe} \, , \ {\beta}_{\qe} = {\kappa} \frac{w-{\qe}}{{\qe}(1-{\qe})} \, , \ {\beta}_{1} = {\kappa} \frac{w-1}{{\qe}-1}
\eeq
for some $\kappa$. Finally, identifying $p_{w}$ with the eigenvalue 
 $\left[ {\gt}_{\infty}^{-1}\left( \frac{A_{0}}{w}+\frac{A_{\qe}}{w-{\qe}} + \frac{A_{1}}{w-1} \right) {\gt}_{\infty} \right]_{11}$ of ${\CalA}(w)$, and imposing the rest of the moment map equations: ${\beta}_{0}{\gamma}_{0} + {\beta}_{\qe} {\gamma}_{\qe} + {\beta}_{1} {\gamma}_{1} = {\vt}$, ${\beta}_{0}{\gamma}_{0}^2 + {\beta}_{\qe} {\gamma}_{\qe}^2 + {\beta}_{1} {\gamma}_{1}^2 = 2 {\theta}_{0} {\gamma}_{0} + 2{\theta}_{\qe} {\gamma}_{\qe} + 2{\theta}_{1} {\gamma}_{1}$ 
leads to the relations
\beq
\begin{aligned}
& p_{w} \ =   \frac{{\vt}_{0}}{w} + \frac{{\vt}_{\qe}}{w-{\qe}} + \frac{{\vt}_{1}}{w-1} - \frac{\vt}{w - x} \, ,  \quad
 p_{\yt} = \frac{2{\vt}_{\infty}}{{\yt}- x^{\circ}} - \frac{2{\vt}_{\infty}}{\yt} \, , \\
 & \\
&  \frac{\vt}{x-w}  \ = \ p_{x} \ = \  \frac{2{\vt}_{0}-{\vt}}{x} + \frac{2{\vt}_{\qe} - {\vt}}{x - {\qe}} + \frac{2{\vt}_{1} -{\vt}}{x-1} -
\frac{2{\vt}_{\infty}}{x - {\yt}^{\circ}} 
 \end{aligned}
\label{eq:wmap1}
\eeq
which can be interpreted as a composition of the
symplectomorphism: $({\yt},p_{\yt}) \mapsto (x, p_{x})$, which is generated by
\begin{multline}
S_{0}(x,{\yt}) = 2{\vt}_{\infty} \, {\rm log} \left( x {\yt} - {\qe} ( x + {\yt} - 1) \right) - 
2{\vt}_{\infty} \, {\rm log} \left( {\yt} \right) + \\
+ \left( 2{\vt}_{0} - {\vt} \right)\, {\rm log}(x) +
\left( 2{\vt}_{\qe} - {\vt} \right) \, {\rm log} ( x - {\qe} ) + \left(  2{\vt}_{1} - {\vt} \right) \, {\rm log}(x-1) 
\label{eq:yx}
\end{multline}
so that
\beq
p_{x}dx - p_{\yt} d{\yt} - h_{x}^{\yt} d{\qe} = - dS_{0} \, ,
\eeq
and the symplectomorphism $(x, p_{x}) \mapsto (w, p_{w})$, generated by
\beq
S_{1}(x, w) = {\vt}_{0}\, {\rm log}(w) + {\vt}_{\qe} \, {\rm log}(w- {\qe}) + {\vt}_{1}\, {\rm log}(w-1) - {\vt}\, {\rm log}(w-x)\, , 
\label{eq:xw}
\eeq
so that
\beq
p_{w} dw - p_{x}dx  - h_{w}^{x} d{\qe} = dS_{1}
\eeq
respectively. 
This is one of the so-called Okamoto transformations \cite{Okamoto}, usually discussed in the context of Painlev{\'e} equations \cite{IKSY, SL}. See also \cite{Giribet:2009hm} for some gauge theory applications. 

\subsubsection{Vierergruppe: the other coordinate changes}

Let us perform the involutions $z \mapsto z^{\circ}$, $z \mapsto z^{\ii}$, $z\mapsto z^{\ii\ii}$ and see what happens with the meromorphic $1$-form ${\CalA}(z)$, e.g.
\begin{multline}
{\CalA}^{\circ} = 
A_{0} d {\rm log}(z^{\circ}) + A_{\qe} d {\rm log} ( z^{\circ} - {\qe} ) + A_{1} d {\rm log}(z^{\circ} - 1) 
\ = \\
= \ A_{1} d {\rm log}(z)  + A_{\infty} d {\rm log} ( z - {\qe})   + A_{0} d {\rm log} (z-1)
\end{multline}
Thus, it would appear that we can permute the $\vt$-parameters $({\vt}_{0}, {\vt}_{\qe}, {\vt}_{1}, {\vt}_{\infty}) \mapsto ( {\vt}_{1}, {\vt}_{\infty}, {\vt}_{0} , {\vt}_{\qe})$ simply by a change of coordinates on a base curve, assuming, ${\CalA}(z)$ is a connection on a trivial vector bundle ${\CalO} \oplus {\CalO}$. 
However, the canonical transformation 
\beq
(w, \ {\pt}_{w}) \mapsto \left( w^{\circ}, \ {\pt}_{w^{\circ}} = \frac{(w-{\qe})^2}{{\qe}(1-{\qe})} \left( {\pt}_{w} + {\partial}_{w} f \right) \right)
\label{eq:circ1}
\eeq
where we tune $f =  - \half {\rm log} (w- {\qe}) + $ a  function  of $\qe$, to eliminate linear in ${\pt}_{w^{\circ}}$ terms in the transformed Hamiltonian:
\beq
{\pt}_{w}dw - {\mathfrak{h}}^{w}_{PVI} ( w, {\pt}_{w} ; {\qe}; {\vec{\vt}}^{\circ}) d{\qe} + d f =
{\pt}_{w^{\circ}}dw^{\circ} - {\mathfrak{h}}^{w^{\circ}}_{PVI} ( w^{\circ}, {\pt}_{w^{\circ}} ; {\qe}) d{\qe} \ ,
\label{eq:circh6} 
\eeq
gives us
 \begin{multline}
{\mathfrak{h}}^{w^{\circ}}_{PVI} ( w, {\pt} ; {\qe}; {\vec{\vt}}^{\circ}) \ = \\
= \ \frac{w (w-1)(w-{\qe})}{{\qe}({\qe}-1)} {\pt}^2 +\frac{{\vt}_{1}^{2}}{w(1-{\qe})} -
\frac{{\vt}_{\infty} ( {\vt}_{\infty} -1 )}{w-{\qe}} +  \frac{w\, {\vt}_{0}^{2}}{{\qe}(w-1)} +\frac{w\, {\vt}_{\qe}^2}{{\qe}(1-{\qe})} \, , 
 \end{multline}
 which encodes a slightly different transformation of $\vt$-parameters:
 \beq
 \left( {\vt}_{0}, {\vt}_{\qe} , {\vt}_{1} , {\vt}_{\infty} \right) \mapsto  \left( {\vt}_{0}, {\vt}_{\qe} , {\vt}_{1} , {\vt}_{\infty} \right)^{\circ} = \left( {\vt}_{1}, {\vt}_{\infty} - {\half}, {\vt}_{0} , {\vt}_{\qe} + {\half} \right)
 \label{eq:circth}
 \eeq 
 The associated transformations of the PVI variables $w \mapsto w^{\ii} = {\qe}/w$ and
$w \mapsto w^{\ii\ii} = \frac{w-{\qe}}{w-1}$ induce transformations
\beq
( {\vt}_{0}, {\vt}_{\qe}, {\vt}_{1} , {\vt}_{\infty})^{\ii} = \left( {\vt}_{\infty} - {\half} , {\vt}_{1} , {\vt}_{\qe}, {\vt}_{0} + {\half} \right) 
\eeq
and
\beq
( {\vt}_{0}, {\vt}_{\qe}, {\vt}_{1} , {\vt}_{\infty})^{\ii\ii} = \left( {\vt}_{1}, {\vt}_{\infty} - {\half}, {\vt}_{0} , {\vt}_{\qe} + {\half} \right)^{\ii} = \left(  {\vt}_{\qe}, {\vt}_{0} , {\vt}_{\infty} - {\half} , {\vt}_{1} + {\half}  \right) 
\eeq
The transformations ${\vec\vt} \mapsto {\vec\vt}, {\vec\vt}^{\circ}, {\vec\vt}^{\ii}, {\vec\vt}^{\ii\ii}$
together with the identity form the celebrated Klein $4$-group $D_2$, 
isomorphic to ${\BZ}_{2} \times {\BZ}_{2}$.

\subsection{Monodromy data: coordinates on ${\CalM}^{\rm loc}_{r}$}

The monodromy of $\nabla$ around the punctures ${\bz}$, defined through a choice
of a marked point $pt \notin {\bz}$ and a basis in the fundamental group
${\pi}_{1} ( {\BC\BP}^{1} \backslash {\bz}, pt)$,  are the group elements $g_{\xi} \in G$, obeying
\beq
\prod^{\kern -.1cm \to}\limits_{{\xi} \in {\bz}} g_{\xi} = 1
\eeq
where the order in the product is dictated by the choice of the basis loops in ${\pi}_{1} ( {\BC\BP}^{1} \backslash {\bz}, pt)$. 
The conjugacy class of each $g_{\xi}$ is fixed:
\beq
 c_{\xi} := \frac 12 {\rm Tr}g_{\xi} = {\rm cos} ( 2{\pi}{\vt}_{\xi} )\ , 
\eeq
 as near $z \sim {\xi}$ only the pole term in ${\CalA}(z)$ matters. The fact that choices of 
the base point etc. are arbitrary forces one to declare the collection $(g_{\xi})_{{\xi} \in {\bz}}$
 and the conjugate collection  $({\rm g} g_{\xi}{\rm g}^{-1})_{{\xi} \in {\bz}}$, for any 
${\rm g} \in G$, equivalent. This motivates the following definition:

The space of monodromy data is the moduli space 
\beq
{\CalM}^{\rm loc}_{r} = {\rm Hom} \left(  {\pi}_{1} ( {\BC\BP}^{1} \backslash {\bz}, pt), G \right)^{c}/G
\label{eq:flatconn}
\eeq
of $G$-representations of the fundamental group ${\pi}_{1} ( {\BC\BP}^{1} \backslash {\bz}, pt)$, restricted by the values of $c_{\xi}$, modulo the conjugation.
 
{}It is again a symplectic manifold, with the symplectic form descending from the topologically
defined Atiyah-Bott form
\beq
\frac{1}{4\pi} \int {\rm Tr} {\delta}A \wedge {\delta} A
\eeq
This symplectic manifold has a system of Darboux coordinates $({\alpha}_{i}, {\beta}_{i})_{i=1}^{r}$
on ${\CalM}^{\rm loc}_{r}$, introduced in \cite{NRS}.

For $r=1$, for the specific ordering of ${\bz} = \{ 0, {\qe}, 1, {\infty} \}$ the monodromy matrices
 $g_{0}, g_{\qe}, g_{1}, g_{\infty}$, obey
\beq
g_{\infty} g_{1} g_{\qe} g_{0} = 1
\,  , \qquad  \, \label{eq:mondat}
\eeq 
The Darboux atlas of ${\CalM}^{\rm loc}_{1}$  has the coordinate charts labeled by
the splittings ${\bz} = {\bz}_{0} \amalg {\bz}_{1}$, $|{\bz}_{0,1}| = 2$, plus some
additional discrete data, such as ordering of ${\bz}_{0,1}$.  
In the case ${\bz}_{0} = \{ 0, {\qe} \}, {\bz}_{1} = \{ 1, {\infty} \}$ the coordinates
are $({\alpha},{\beta})$, where
\beq
c_{0\qe} := {\half} {\rm Tr} g_{0}g_{\qe} = {\rm cos} (  2\pi {\alpha} ), 
\label{eq:mon0t}
\eeq
while $\beta$, the remaining monodromy parameter, can be found from 
\beq
c_{1\qe} = {\half} {\rm Tr} g_{1}g_{\qe}  = {\rm cos} ( 2{\pi}{\tilde a})
\label{eq:mon1t}
\eeq
via:
\begin{multline}
  e^{\pm {\beta}} \, \left( {\rm cos} 2{\pi}({\alpha} \pm {\vt}_{1}) - {\rm cos} 2{\pi} {\vt}_{\infty} \right) \left( {\rm cos} 2{\pi}({\alpha} \pm {\vt}_{\qe}) - {\rm cos} 2{\pi} {\vt}_{0} \right) = \\
= c_{\qe}c_{1} + c_{0} c_{\infty} \pm {\ii}c_{01} {\rm sin} ( 2\pi {\alpha} )- 
\left( c_{0}c_{1} + c_{\qe}c_{\infty} \mp {\ii} c_{1\qe} {\rm sin} 2\pi {\alpha} \right) e^{\pm 2\pi\ii {\alpha}} \, , \\
c_{0\qe}^2 + c_{1\qe}^2 + c_{01}^2 + 2 c_{0\qe}c_{1\qe}c_{01} + c_{0}^2 + c_{\qe}^2 + c_{1}^2 + c_{\infty}^2 + 4 c_{0}c_{\qe}c_{1}c_{\infty} = \\
1+ 2 c_{01} (c_{0}c_{1}+c_{\qe} c_{\infty} ) + 2 c_{1\qe} (c_{1}c_{\qe}+c_{0}c_{\infty}) + 2 c_{0\qe} (c_{0}c_{\qe}+c_{1}c_{\infty}) 
\end{multline}
which is equivalent to the definition of the $\beta$-coordinate in \cite{NRS} (in our notations):
\begin{multline}
{\rm cosh}({\beta}) \ = \ \frac{- c_{1\qe}\, {\rm sin} (2{\pi}{\alpha})^2 + c_{\qe}c_{1}+c_{0}c_{\infty} - c_{0\qe} (c_{0}c_{1} + c_{\qe}c_{\infty})}{4\prod\limits_{\pm, \pm} \sqrt{{\rm sin} \, 2{\pi}( {\alpha} \pm {\vt}_{0} \pm {\vt}_{\qe} )  {\rm sin} \, 2{\pi}( {\alpha} \pm {\vt}_{1} \pm {\vt}_{\infty} )}}  \ , \\
{\rm sinh}({\beta}) \ = \ \, \frac{{\ii} \, {\rm sin}(2\pi {\alpha}) \left(   c_{01} + c_{0\qe}c_{1\qe} - c_{0}c_{1} - c_{\qe}c_{\infty} \right) }{
4 \,  \prod\limits_{\pm, \pm} \sqrt{{\rm sin} \, 2{\pi}( {\alpha} \pm {\vt}_{0} \pm {\vt}_{\qe} )  {\rm sin} \, 2{\pi}( {\alpha} \pm {\vt}_{1} \pm {\vt}_{\infty} )}} \, . \end{multline}
The $\alpha$-coordinate, up to a reflection $\alpha \mapsto - \alpha$ and integer shifts ${\alpha} \sim {\alpha} + {\BZ}$
does not depend on the ordering of ${\bz}_{0,1}$, while ${\beta}$ has a simple ordering
dependence (see \cite{NRS} for more detail).

\section{The nonperturbative Dyson-Schwinger equations}

In this section we return to the study of surface defects in supersymmetric $U(n)$ gauge theory. 
We relate them to Painlev{\'e} VI for $n=2$. The higher $n$ case  will be studied elsewhere.  

\subsection{Fractional $qq$-characters}

For the $U(n)$ theory with $N_f = 2n$ flavors, 
introduce the fractional $Y$-observables:
\beq
Y_{\omega}(x) = ( x - {\tilde a}_{\omega}) \frac{Q_{\omega}(x)}{Q_{{\omega}-1}(x-{\tilde\ve}_{2})}\, , \qquad {\omega} \sim {\omega} + n \, ,\ {\omega} \in {\BZ}/n{\BZ} 
\label{eq:yobs}
\eeq
where
\beq
Q_{\omega}(x) = \prod_{k \in {\kappa}_{\omega}} \left( 1 - \frac{{\ve}_{1}}{x- k} \right) \, , \qquad Ch({\tilde K}_{\omega})  = \sum_{k \in {\kappa}_{\omega}} e^{k}
\label{eq:qobs}
\eeq
For large $x$, 
\beq
Q_{\omega}(x) = 1 - \frac{{\ve}_{1}k_{\omega}}{x} + \ldots 
\eeq

\subsection{Dyson-Schwinger equation for the surface defect}

\subsubsection{Mass and Coulomb manipulations}

{}Introduce the notations
\beq
\begin{aligned}
& {\tilde m}_{\omega} = {\tilde m}_{\omega}^{+} + {\tilde m}_{\omega}^{-}\, , \\
& {\mu}_{\omega}^{\pm} = \frac{{\tilde m}_{\omega}^{\pm} - {\tilde a}_{+}}{2{\tilde\ve}_{2}}  \, , \\ & {\mu}_{\omega} =
{\mu}_{\omega}^{+}+{\mu}_{\omega}^{-} \, , \ {\mu}^{\pm} = {\mu}^{\pm}_{0} + {\mu}^{\pm}_{1}\ .    
\label{eq:massom}
\end{aligned}
\eeq 
for the masses of the hypermultiplets, which transform under the ${\CalR}_{\omega}$ representation of the orbifold group, 
where
\beq
{\tilde a}_{+} = \frac{1}{n} \sum_{\omega} {\tilde a}_{\omega}
\eeq
For $n=2$ we also use:
\beq
{\tilde a}_{-} = \frac{{\tilde a}_{0} - {\tilde a}_{1}}{2}\, .
\eeq
With \eqref{eq:match} in place, we have ${\tilde a}_{+} = {\half} {\tilde\ve_2}$ and:
\beq
\ {\mu}_{0}^{+} = \frac{m_{s(3)}}{{\ve}_{2}} - \frac 14\, , \ {\mu}_{0}^{-} = \frac{m_{s(4)}}{{\ve}_{2}} - \frac 14\, , \
{\mu}_{1}^{+} = \frac{m_{s(1)}}{{\ve}_{2}} + \frac 14\, , \ 
{\mu}_{1}^{-} = \frac{m_{s(2)}}{{\ve}_{2}} + \frac 14\ . 
\label{eq:mubulk}
\eeq

\subsubsection{DS equation in the orbifold variables}

Recall the expression for the $qq$-characters of the ${\CalN}=2$ theory in the presence of the surface defect:
\beq
{\bz}_{\omega}(x) = Y_{{\omega}+1}( x + {\ve}_{1} + {\tilde\ve}_{2} ) + {\qe}_{\omega} \, P_{\omega}(x) \, Y_{\omega}(x)^{-1}
\, , \qquad   {\omega} \sim {\omega} + n, \, {\omega} \in {\BZ}/n{\BZ}
\, ,
\label{eq:qqchar}
\eeq
with
\beq
P_{\omega}(x) = (x - {\tilde m}_{\omega}^{+}) ( x - {\tilde m}_{\omega}^{-}) \ .
\eeq

The main property of the operators \eqref{eq:qqchar} is the cancellation of poles in $x$ of their expectation values, cf. \eqref{eq:qw}:
\beq
\langle {\bz}_{\omega}(x) \rangle = (1 + {\qe}_{\omega})  \left( x + \langle {\delta}_{\omega} \rangle \right)  +{\ve}_{1} + {\tilde\ve}_{2}- ( {\tilde a}_{\omega} + {\tilde a}_{\omega +1} )   + {\qe}_{\omega} {\tilde m}_{\omega}\, , \qquad {\omega} = 0 ,1 \, 
\label{eq:qqcharvev}
\eeq
where 
\beq
{\delta}_{\omega} = {\tilde a}_{\omega} + {\ve}_{1} (k_{\omega} -k_{{\omega}+1})  \, . \quad \eeq
{}For $n=2$, cf. \eqref{eq:fuga}, \eqref{eq:bulk}, \eqref{eq:surface}:
\beq
\langle {\delta}_{0,1} \rangle = \left(  {\tilde a}_{+} \pm {\ve}_{1} u {\partial}_{u} \right) {\Psi} \, , \ \langle {\delta}_{0,1}^2 \rangle = \left(  {\tilde a}_{+} \pm {\ve}_{1} u {\partial}_{u} \right)^2 {\Psi}\ , 
\eeq
where we used the notation \eqref{eq:qw} for the fractional couplings. 

{}The vanishing of the $x^{-1}$-term in the large $x$ expansion of the
 equations \eqref{eq:qqcharvev} implies the relation between the correlators:
\begin{multline}
0 = (1  -{\qe}) \left(  {\half} \langle {\delta}_{0}^{2} \rangle +  {\half} \langle {\delta}_{1}^{2} \rangle - {\tilde a}_{+}^{2} -  {\tilde a}_{-}^{2} + {\ve}_{1}{\tilde\ve}_{2} \langle k_{0} + k_{1} \rangle   \right)  + \\
+ ({\qe} -u ) \left( {\tilde m}_{0}^{+}{\tilde m}_{0}^{-} - {\tilde m}_{0}\langle {\delta}_{0} \rangle + \langle {\delta}_{0}^{2} \rangle \right) +{\qe} ( 1 - u^{-1} )  \left( {\tilde m}_{1}^{+}{\tilde m}_{1}^{-} - {\tilde m}_{1} \langle {\delta}_{1} \rangle + \langle {\delta}_{1}^{2} \rangle \right)\ . \label{eq:npds}
\end{multline}

Now, with the help of the relation
\beq
{\ve}_{1}{\tilde\ve}_{2} \left(   u{\partial}_{u} + 2 {\qe} {\partial}_{\qe} \right) {\Psi} \, = \,  \langle {\tilde a}_{+} \left(  {\ve}_{1} + {\tilde\ve}_2 - {\tilde a}_{+} \right) -  {\tilde a}_{-}^{2} + {\ve}_{1}{\tilde\ve}_{2} (k_{0}+k_{1}) \rangle \ , 
\eeq
we can rewrite \eqref{eq:npds} as the differential equation on ${\Psi}$:
\begin{multline}
(1-{\qe}) \left( {\ve}_{1}{\ve}_{2} \left( {\half} u{\partial}_{u} + {\qe}{\partial}_{\qe} \right) + {\tilde a}_{+}({\tilde a}_{+} - {\ve}_{1} - {\tilde\ve}_{2}) \right)  {\bf\Psi} + \\
\frac{(1-u)(u-{\qe})}{u} \left( {\ve}_{1} u {\partial}_{u} \right)^2 {\bf\Psi} + {\ve}_{1}{\ve}_{2} \left( {\mu}_{0} u (u - {\qe}) + {\mu}_{1} {\qe} (u-1) \right)   {\partial}_{u}  {\bf\Psi} + \\
+ \frac{{\ve}_{2}^{2}}{u} \left( (u-1) {\qe}{\mu}_{1}^{+}{\mu}_{1}^{-}  - u(u - {\qe}) {\mu}_{0}^{+}{\mu}_{0}^{-} \right)  {\bf\Psi} = 0
\label{eq:bpz}\end{multline}
This is the $n=2$ case of the more general fact, established in the \emph{BPS/CFT V} paper in \cite{NekBPSCFT}, that  the
regular surface defect ${\Psi}$ of the $U(n)$ theory with $2n$ fundamental hypermultiplets obeys the Knizhnik-Zamolodchikov/Belavin-Polyakov-Zamolodchikov \cite{KZ},\cite{BPZ} equation.

\subsubsection{Enters Painlev{\'e} VI}

{}In the limit ${\ve}_{1} \to 0$, the surface defect partition function has the asymptotics:
\beq
{\bf\Psi} \sim e^{\frac{{\ve}_2}{{\ve}_{1}} S \left( u ; {\qe} ;  \frac{\bf m}{{\ve}_{2}} | \frac{a}{{\ve}_{2}} \, \right)} \, u^{-\frac{{\ve}_{2} {\mu}_{1}}{2{\ve}_{1}}} (u-1)^{\frac{{\ve}_{2}({\mu}_{0}+{\half})}{2{\ve}_{1}}} (u-{\qe})^{\frac{{\ve}_{2} ( {\mu}_{1}-{\half})}{2{\ve}_{1}}} 
  \, , 
  \label{eq:psias}
\eeq
where $\sim$ means that we dropped a $u$-independent factor (it leads to a $\qe$-dependent shift in $S$),  
so that the Eq. \eqref{eq:bpz} becomes the Hamilton-Jacobi equation:
\beq
{\partial}_{\qe} S( u ; {\qe} ; {\vt} | {\alpha} )  = - H \left( u, {\partial}_{u} S( u ; {\qe} ; {\vt} | {\alpha} ); {\qe} ; {\vt} \right)
\label{eq:hj1}
\eeq
with the Hamiltonian 
\begin{multline}
H (u, p ; {\qe}; {\vt} )  = 
\frac{u(u-1)(u-{\qe})}{{\qe}({\qe}-1)} p^{2} + {\bf U} (u, {\qe}; {\vt})\, ,  \\
{\bf U} (u, {\qe}; {\vt}) =  \frac{u\, \left( {\vt}_{\infty} - {\half} \right)^{2}}{{\qe}(1-{\qe})}  + \frac{{\vt}_{0}^{2} }{u(1-{\qe})} 
+ \frac{u \, {\vt}_{1}^{2}}{{\qe}(u-1)} - 
 \frac{{\vt}_{\qe}^{2} - \frac{1}{4}}{u-{\qe}} \label{eq:hamp6}
\end{multline}
with
\begin{multline}
\pm 2 {\vt}_{0} 
= {\mu}_{1}^{+} - {\mu}_{1}^{-} \, , \ 
\pm 2 {\vt}_{\qe} = {\mu}_{1} ^{+} + {\mu}_{1}^{-} + {\half}  \, , \\ 
\pm 2 {\vt}_{1} = {\mu}_{0}^{+} + {\mu}_{0}^{-} + {\half} \, , \ 
\pm \left( 2 {\vt}_{\infty}  - 1 \right) = {\mu}_{0}^{+} - {\mu}_{0}^{-}  \, , \ 
\label{eq:th1}
\end{multline}
We recognize in \eqref{eq:hamp6} our friend Painlev{\'e} VI Hamiltonian \eqref{eq:pviham}. 
The signs $\pm$ in \eqref{eq:th1} reflect the ambiguity of the relation
between the parameters of the Hamiltonian \eqref{eq:ham6}
and the 
${\vt}$-parameters, which stems from the symmetries of \eqref{eq:pviham} we discussed at the end of the section {\bf 6.2}.  The Hamiltonian \eqref{eq:ham6} is invariant under the permutations ${\tilde m}^{+}_{\omega} \leftrightarrow {\tilde m}^{-}_{\omega}$, which do not change the surface defect.

\subsubsection{DS equation in the parameters of the bulk theory}

Using \eqref{eq:mubulk}, we get: 
\begin{multline}
{\vt}_{0} = \pm {\theta}_{0}\, , {\vt}_{\qe} = \pm \left( {\theta}_{\qe} +  {\half} \right)\, , \ {\vt}_{1} =  \pm {\theta}_{1}\, ,  \ {\vt}_{\infty} = {\half} \pm {\theta}_{\infty}\, ,    \qquad 
{\rm where} \\
{\vec\vt}\, = \, \frac{1}{2{\ve}_{2}} 
 \left(   m_{s(1)}-m_{s(2)}    \, , \   m_{s(1)} + m_{s(2)}   \, , \   m_{s(3)} + m_{s(4)}   \, , \ 
  m_{s(3)}-m_{s(4)} \right)  \  .
 \label{eq:th11}
 \end{multline}

\subsubsection{Playing with the fugacity}
 
Armed with our knowledge of  \eqref{eq:circth} we can transform the DS equation for ${\bf\Psi}$
to the form of Hamilton-Jacobi potential 
\beq
{\partial}_{\qe} S^{\circ}( w ; {\qe};  a/{\ve}_{2}, {\bf m}/{\ve}_{2} ) = - {\mathfrak h}^{w}_{PVI} \left( w , {\partial}_{w} S^{\circ} ; {\qe} , {\vec\vt}^{DS} \right) 
\label{eq:ham6m}
\eeq
with 
\beq
w = u^{\circ} = {\qe}_{1} \frac{1+{\qe}_{0}}{1+{\qe}_{1}}
\eeq
and the $\vt$-parameters
\begin{multline}
{\vt}_{0}^{DS} =  \frac{m_{s(3)} + m_{s(4)}}{2{\ve}_{2}} = \, , \ {\vt}_{\qe}^{DS} = \frac{m_{s(3)} - m_{s(4)}}{2{\ve}_{2}} \, , \\ 
{\vt}_{1}^{DS} = \frac{m_{s(1)} - m_{s(2)}}{2{\ve}_{2}} \, , \ {\vt}_{\infty}^{DS} = \frac{m_{s(1)} + m_{s(2)}}{2{\ve}_{2}}
\label{eq:thfrmasses}
\end{multline}
which are linear in the masses of fundamental multiplets. This property is crucial in our argument below. 

 {}The Hamilton-Jacobi potentials $S^{\circ}( w ; {\qe};  a/{\ve}_{2}, {\bf m}/{\ve}_{2} )$ and $S( u ; {\qe};  a/{\ve}_{2}, {\bf m}/{\ve}_{2} )$ are related by, cf. \eqref{eq:circ1}:
\beq
S^{\circ}( w ; {\qe};  a/{\ve}_{2}, {\bf m}/{\ve}_{2} ) = S( u ; {\qe};  a/{\ve}_{2}, {\bf m}/{\ve}_{2} )  - {\half} {\rm log}( u - {\qe} ) + {\rm function\  of} \ {\qe}
\eeq
We interpret $e^{S^{\circ}/ {\ve}_{1}}$ as the partition function of the surface defect in $SU(2)$ gauge theory, which differs from ${\bf\Psi}$ by a $U(1)$-factor. 
 
 \subsubsection{Initial conditions}

In order to solve for the surface defect partition function we need to recover the $a$-dependence of $S$. One can do this by fixing a small $\qe$ asymptotics\footnote{Technically this is similar to the analysis in \cite{Litvinov:2013sxa}}. First of all, it is easy to compute, for ${\qe} \to 0$ with $u$ fixed, i.e. ${\qe}_{0}$ is finite, while ${\qe}_{1} \to 0$, that
\begin{multline}
{\bf\Psi} \sim u^{\frac{{\tilde a}_{-}}{2{\ve}_{1}}} \sum_{d=0}^{\infty} u^{d} {\sE} \left[ \left( \frac{e^{{\tilde a}_{1}}+ {\tilde q}_{2} e^{{\tilde a}_{0}} ( 1 -q_{1}^d)}{1-q_{1}} \right)^{*} {\tilde q}_{2} \left( e^{{\tilde a}_{0}}  q_{1}^{d} - e^{{\tilde m}_{0}^{+}} - e^{{\tilde m}_{0}^{-}} \right)\right] = \\
u^{\frac{{\tilde a}_{-}}{2{\ve}_{1}}} \, {\sE} \left[ \left( \frac{e^{{\tilde a}_{1}}}{1-q_{1}} \right)^{*} {\tilde q}_{2} \left( e^{{\tilde a}_{0}}  - e^{{\tilde m}_{0}^{+}} - e^{{\tilde m}_{0}^{-}} \right)\right] \times \\
\sum_{d=0}^{\infty} u^{d} 
{\sE} \left[ \left( \frac{{\tilde q}_{2} e^{{\tilde a}_{0}} ( 1 -q_{1}^d)}{1-q_{1}} \right)^{*} {\tilde q}_{2} \left( e^{{\tilde a}_{0}}  q_{1}^{d} - e^{{\tilde m}_{0}^{+}} - e^{{\tilde m}_{0}^{-}} \right)\right]
{\sE} \left[ \frac{e^{{\tilde a}_{0} - {\tilde a}_{1} + {\tilde\ve}_{2}}  (q_{1}^{d} -1 )}{1-q_{1}^{-1}} \right]
\end{multline}
since only partitions ${\lambda}^{(0)}= 1^{d}$, ${\lambda}^{(1)} = {\emptyset}$ can contribute, for which
\beq
Ch({\tilde K}_{0}) = e^{{\tilde a}_{0}} \frac{1-q_{1}^{d}}{1-q_{1}}\, , \ Ch({\tilde K}_{1}) = 0\, ,\
Ch({\tilde S}_{0}) = e^{{\tilde a}_{0}}  q_{1}^{d}\, , \ Ch({\tilde S}_{1}) = e^{{\tilde a}_{1}} + {\tilde q}_{2} e^{{\tilde a}_{0}} ( 1 -q_{1}^d)
\eeq
Thus, the surface defect partition function, for ${\qe}_{1} = 0$, is essentially a hypergeometric function
\beq
{\bf\Psi} = u^{\frac{a - {\half}{\tilde\ve}_{2}}{{\ve}_{1}}} \, \frac{
{\Gamma}\left( \frac{a - {\tilde m}^{+}_{0}}{{\ve}_{1}}\right) {\Gamma} \left(  
\frac{a - {\tilde m}^{-}_{0}}{{\ve}_{1}}\right)}{{\Gamma} \left( 1 + \frac{2a}{{\ve}_{1}} \right)}\, _{2}F_{1} \left( \frac{a - {\tilde m}^{+}_{0}}{{\ve}_{1}} , 
\frac{a - {\tilde m}^{-}_{0}}{{\ve}_{1}}; 1 + \frac{2a}{{\ve}_{1}} ; u \right)
\eeq
  See \cite{SJNN} for more detail.

\subsection{Two dimensional connection from the four dimensional gauge theory}

{}As we discussed in the chapter $\bf 5$, the Painlev{\'e} VI equation describes the isomonodromic deformation of
a meromorphic connection on a $4$-punctured sphere, i.e. Schlesinger evolution. However, we saw that the Schlesinger-Painlev{\'e} correspondence is not one-to-one. We have the $\yt$-coordinates, the $w$-coordinates, the $w^{\circ}$-coordinates, the ${\yt}^{\circ}$-coordinates etc.

It  would be useful to identify an observable in the four dimensional gauge theory which can be identified with the meromorphic connection whose isomonodromic deformation is described by \eqref{eq:ham6m} or
\eqref{eq:hj1}. Indeed, the $U(n)$ $N_f=2n$ theory has the observables ${\bf\Upsilon}(z)$, which are the horizontal sections of the meromorphic connection similar to \eqref{eq:conn1} (for $r=1$, $n=2$ in our example). Here is the construction.

\subsubsection{Crossed surface defect}

Recall that the $Y(x)$-observable represent the regularized characteristic polynomial ${\rm det}(x - {\Phi})$ of the complex scalar in the vector multiplet, so it is a local observable. In the $\Omega$-deformed theory the supersymmetric non-local observables can be related to the local ones and vice versa. It is sometimes useful to work with the observable $Q_{\Sigma}(x)$ associated with the surface $\Sigma$ in the four dimensional space. 
One way to think about $Q_{\Sigma}(x)$ is that it is a partition function of a certain two dimensional theory interacting with the four dimensional one, i.e. a surface defect. For example, it can be engineered using the folded instanton construction \cite{NekBPSCFT}. In the theory with matter fields we have a choice of boundary conditions in the matter sector.   In the theory on ${\BR}^4$ with $\Omega$-deformation the $Q$-observable associated with the surface ${\Sigma} = {\BR}^{2}$ given by the equation $z_1 = 0$ solves the operator equation 
\beq
Y(x) = \prod_{f} (x - m_{f}^{+}) \, \frac{Q(x)}{Q(x- {\ve}_{2})} \, , 
\eeq
where $f$ labels the fundamental hypermultiplets, which have propagating modes along the $z_1 = 0$ surface, $m_{f}^{+}$ being the corresponding masses. 
The analogous observable associated with the $z_2=0$ surface obeys a similar equation with ${\ve}_{2}$ replaced by ${\ve}_{1}$.

{}Now let us add the surface defect, of the regular orbifold type, that we were studying in the previous sections.
The type of the boundary conditions which are imposed on the matter fields is encoded
in the way  
the $Q(x)$ observable, brought near the surface defect, fractionalizes to ${\tilde Q}_{\omega}(x)$, ${\omega} = 0, 1$,  cf. \eqref{eq:yobs}. There are four types in total. The
Vieregruppe $D_2$ acts on the set of regular surface defects, 
by exchanging ${\tilde m}_{0}^{+}$ with ${\tilde m}_{0}^{-}$
and/or ${\tilde m}_{1}^{+}$ with ${\tilde m}_{1}^{-}$.  

In what follows we use the $({\tilde m}_{0}^{+}, {\tilde m}_{1}^{+})$ choice (other choices will be important
in the computation of the monodromy data in \cite{SJNN}). The ${\tilde Q}_{\omega}(x)$-observables are related to $Y_{\omega}(x)$ in this case by:
\beq
Y_{\omega}(x) = (x - {\tilde m}_{\omega}^{+}) \frac{{\tilde Q}_{\omega}(x)}{{\tilde Q}_{{\omega} - 1} (x - {\tilde\ve}_{2})}
\label{eq:qfromy}
\eeq
The observables ${\tilde Q}_{\omega}(x)$ can be expressed as the ratio of the Euler class of the bundle of zero modes
of chiral fermions propagating along $z_1=0$ surface in the instanton background, and that of the matter fields:
\beq
{\tilde Q}_{\omega}(x) = {\sE} \left[ - e^{x} \left( \frac{{\hat S}^{*} - ({\hat M}^{+})^{*}}{1-{\tilde q}_{2}^{-1} {\CalR}_{-1}} \right)_{\omega} \, \right]
\eeq
By comparing \eqref{eq:qfromy} to the Eq. \eqref{eq:yobs} we can write:
\beq
{\tilde Q}_{\omega}(x) = Q_{\omega}(x) {\gamma}_{\omega}(x)
\eeq
where the functions ${\gamma}_{\omega}(x)$ solve the functional equation:
\beq
\frac{{\gamma}_{\omega}(x)}{{\gamma}_{\omega -1}(x - {\tilde\ve}_{2})} = \frac{x- {\tilde a}_{\omega}}{x-{\tilde m}^{+}_{\omega}} \ .
\label{eq:gom1}
\eeq
The Eq. \eqref{eq:gom1} can be easily solved in $\Gamma$-functions, since \eqref{eq:gom1} implies:
\beq
\frac{{\gamma}_{\omega}(x)}{{\gamma}_{\omega}(x - {\ve}_{2})} = \prod_{j=0}^{n-1} \frac{x- {\tilde a}_{\omega -j}- j{\tilde\ve}_{2}}{x-{\tilde m}^{+}_{\omega-j}- j{\tilde\ve}_{2}}
\label{eq:gom2}
\eeq
In the limit ${\ve}_{1} \to 0$ the fractional $qq$-character becomes the fractional $q$-character obeying the functional equation, for $n=2$:
\beq
\begin{aligned}
& Y_{0}(x+{\tilde\ve}_{2}) + {\qe}_{1}\, \frac{P_{1}(x)}{Y_{1}(x)} = (1 + {\qe}_{1}) (x - {\tilde a}_{+} - {\rho}) + 
{\ve}_{2} ({\half} - {\qe}_{1}{\mu}_{1}) \, , \\ 
& Y_{1}(x+{\tilde\ve}_{2}) + {\qe}_{0}\, \frac{P_{0}(x)}{Y_{0}(x)} = (1 + {\qe}_{0}) (x - {\tilde a}_{+} + {\rho}) + 
{\ve}_{2} ({\half} - {\qe}_{0}{\mu}_{0}) \, , \\
\end{aligned}
\label{eq:qchar}
 \eeq
 with (cf. \eqref{eq:surface},\eqref{eq:psias})
 \beq
 {\rho} = {\tilde a}_{-} + \langle {\ve}_{1} (k_{0}-k_{1})  \rangle \, =\,  u {\partial}_{u} S - \frac{{\mu}_{1}}{2} +  u  \frac{{\mu}_{0} + {\half}}{2 (u-1)} + 
 u \frac{{\mu}_{1} - {\half}}{2 (u - {\qe})}\ , 
 \label{eq:rhomap}
 \eeq 
 the last equality holding
 in the ${\ve}_{1} \to 0$ limit. 
 The Eqs. \eqref{eq:qchar} are equivalent to the system of linear functional-difference equations for ${\tilde Q}$'s. Denoting: 
 \beq
 {\hat Q}_{\omega} ({\xt}) = {\tilde Q}_{\omega} ( {\tilde a}_{+} + {\xt} {\ve}_{2} )\ , 
 \eeq
 and  recalling \eqref{eq:massom}
  we rewrite \eqref{eq:qchar} as:
 \beq
\begin{aligned}
& ({\xt} + \half - {\mu}^{+}_{0}) {\hat Q}_{0}({\xt} + \half ) + {\qe}_{1} ( {\xt} - {\mu}_{1}^{-} ) {\hat Q}_{0}({\xt} - \half) \ = \\
& \qquad\qquad\qquad\qquad \left( ( 1+ {\qe}_{1} ) ( {\xt} - {\rho} ) + \half - {\qe}_{1}{\mu}_{1} \right) {\hat Q}_{1}({\xt}) 
 \, , \\ 
&  ({\xt} + \half - {\mu}^{+}_{1}) {\hat Q}_{1}({\xt} + \half ) + {\qe}_{0} ( {\xt} - {\mu}_{0}^{-} ) {\hat Q}_{1}({\xt} - \half) \ = \\
& \qquad\qquad\qquad\qquad \left( ( 1+ {\qe}_{0} ) ( {\xt} + {\rho} ) + \half - {\qe}_{0}{\mu}_{0} \right) {\hat Q}_{0}({\xt}) 
 \, .\\ 
\end{aligned}
\label{eq:qchar2}
 \eeq 
 Let us now Fourier transform the ${\hat Q}_{\omega}$-functions:
 \beq
 {\bf\Upsilon}(z) = {\psi}^{\rm ab} (z) \, \sum_{\xt \in L} \, \left( \begin{matrix} {\hat Q}_{0}({\xt}) \, \\   \\ \,
 {\hat Q}_{1}({\xt}+{\half}) \, \\ \end{matrix} \right) \ (z/{\qe})^{\xt} \, , \ 
  \label{eq:ftr}
 \eeq
 with
 \beq
  {\psi}^{\rm ab} (z) = z^{{\qalf} - \frac{\mu^+}{2}}   
  (z-1)^{-\frac{{\mu}^{-}}{2}} (z - {\qe})^{\frac{1+{\mu}^{+}}{2}}
  \eeq
  and
  $L = L + {\BZ}$, a ${\BZ}$-invariant lattice $L \subset {\BC}$, such that the sum in \eqref{eq:ftr} converges. 
  There are several choices for $L$, dictated by the asymptotics of the sets of zeroes or poles of ${\tilde Q}_{\omega}$'s, respectively. The zeroes are the Chern roots of the bundle of zero modes, while the poles are determined by the masses. The different choices of $L$ leads to the solutions of \eqref{eq:horzs} which are convergent in various domains
of the $4$-punctured sphere, such as $|z| < |{\qe}|$, $|{\qe} | < |z| < 1$, and $1 < |z|$, respectively. 

\vskip 1cm
{}In terms of 
 ${\bf\Upsilon}(z)$ the Eq. \eqref{eq:qchar2} is the system of first order differential equations:
 \beq
\left( \frac{\partial}{\partial  z} + {\CalA}(z) \right) {\bf\Upsilon} \, = \, 0 \, , \qquad {\CalA}(z) = \frac{A_{0}}{z} + \frac{A_{\qe}}{z - {\qe}} + \frac{A_{1}}{z - 1} \, \ 
\label{eq:horzs}
\eeq
with the residues $A_{\xi}$ in the form \eqref{eq:pxo}, with the parameters summarized in the Table \eqref{eq:bgar}. 
\begin{table}
{
\caption{The residues $A_{0}, A_{\qe}, A_{1}, A_{\infty}$}
\label{eq:bgar}
}
{
\begin{tabular}{lllllll}
\hline
 & & & & &  \\[-6pt]
\qquad ${\xi}$ \quad & \quad & \qquad\qquad $2{\vt}_{\xi}$ \quad & \qquad ${\gamma}_{+,\xi}$ \quad & \qquad ${\gamma}_{-,\xi}$ \quad & \qquad\qquad ${\beta}_{\xi}$   \\
& & & &  \\[-6pt]
\hline
 & & & &  \\[-6pt]
\qquad $0$    & \quad & ${\mu}_{1}^{+} - {\mu}_{0}^{+} - {\half}$   & \qquad $0$ &  & ${\half} 
+ {\mu}_{0}^{+} + u {\mu}_{0}^{-} +  {\rho} ( 1- u) $ &   \\[3pt]
\qquad $\qe$ & \quad & $-  {\mu}^{+} - 1$ & \qquad $1$ & \ &  \qquad $- {\beta}_{1} - {\beta}_{0}$  &   \\[3pt]
\qquad $1$   \quad & & \ \ ${\mu}^{-}$ & \qquad $\frac{1}{u^{\circ}}$  & \ & $- \frac{{\qe}(1-u)}{u ( 1- {\qe})} ( {\mu}_{1}^{-} +  u {\mu}_{0}^{-}  + {\rho}(1-u))$ &     \\[3pt]
\qquad $\infty$      & \quad & ${\mu}_{0}^{-} - {\mu}_{1}^{-} - {\half}$ &\qquad  ${\infty}$ & $1 - 
\frac{ {\mu}^{+}_{1} + {\rho}  + \frac{u}{\qe} ( {\mu}^{-}_{0} - {\rho} )}{2{\vt}_{\infty}}$ &\qquad\qquad  $0$ & \\[3pt]
\hline
\end{tabular}
}
\end{table}
Notice that ${\gamma}_{+, {\infty}} = {\infty}$, since ${\gamma}_{-,{\infty}}$ is finite while
${\beta}_{\infty} = 0$. Therefore the ${\yt}$-variable, computed using ${\gamma}_{+,{\xi}}$ is equal to $u^{\circ}$
\beq
{\yt} = \frac{{\gamma}_{+,{\qe}} - {\gamma}_{+,0}}{{\gamma}_{+,1}-{\gamma}_{+,0}} \cdot \frac{{\gamma}_{+,1} - {\gamma}_{+,{\infty}}}{{\gamma}_{+,{\qe}}-{\gamma}_{+,{\infty}}} = u^{\circ} = w
\eeq 
Now, computing  $p_{\yt}$ from \eqref{eq:pso}, ${\hat{\vec{\vt}}}$ from  \eqref{eq:thehat}, and
${\pt}_{\yt}$  from \eqref{eq:pythhat} we get:
\beq
{\pt}_{\yt} = \frac{(u - {\qe})^{2}}{{\qe}({\qe}-1)} \left( {\partial}_{u}S  + \frac{1}{2 ({\qe} - u)} \right)
 = \frac{{\partial} S^{\circ}}{{\partial} w} 
\eeq
and
\beq
\begin{aligned}
& {\hat\vt}_{0} = -\frac{{\half} + {\mu}_{0}}{2} = - {\vt}_{0}^{DS}\, , \\ 
& {\hat\vt}_{\qe} = \frac{{\mu}_{0}^{-} - {\mu}_{0}^{+}}{2} = - {\vt}_{\qe}^{DS}\, , \\ 
& {\hat\vt}_{1} = \frac{{\mu}_{1}^{+} - {\mu}_{1}^{-}}{2} = {\vt}_{1}^{DS}\, , \\ 
& {\hat\vt}_{\infty} = 
\frac{{\mu}_{1}+\frac 32}{2} = 1 - {\vt}_{\infty}^{DS} \\
\end{aligned}
\eeq
We have thus figured out the meaning of the parameters of the Painlev{\'e} VI
equation which is the Dyson-Schwinger equation obeyed by the $SU(2)$ surface defect partition function. The coordinate $w = {\qe}_{1} \frac{1+{\qe}_{0}}{1+{\qe}_{1}}$ is identified with the ${\yt}^{\circ}$-coordinate, so that the isomonodromy problem for the connection \eqref{eq:horzs} is the Painlev{\'e} VI equation governed by the Hamiltonian of the form \eqref{eq:H2p} with $\vt$'s strictly linear in the masses of the fundamental hypermultiplets.

\section{The GIL conjecture}

The only formulas worth working on these days can be named with one or two three-letter abbreviation(s). So here is the GIL formula\footnote{Also known as the Kiev $\tau$-function formula}, proposed in \cite{Gamayun:2012ma} in
2012. It  expresses the $\tau$-function \eqref{eq:tpvi} of the PVI equation 
through the $c=1$ conformal blocks of Virasoro algebra:
\beq
{\tau}({\qe}; {\alpha}, {\beta}; {\vec\vt}) = {\qe}^{-{\vt}_0^2-{\vt}_{\qe}^2} (1-{\qe})^{2{\vt}_{\qe}{\vt}_{1}}  \sum_{n \in {\BZ}} \ e^{n {\beta}} \, {\CalC}_{\hbar} (a +n{\hbar}, {\vec m}) \, {\CalC}_{\hbar} (-a-n{\hbar}, {\vec m})\,  Z^{\rm inst}_{\hbar}(a +n {\hbar} , {\vec m}; {\qe}) \ , 
\label{eq:gil}
\eeq
where  $a = {\hbar}{\alpha}$
\beq
{\CalC}_{\hbar}(a,{\vec m}) = \frac{1}{G_{\hbar}({\hbar}+2a)}\prod_{f=1}^{4} G_{\hbar}({\hbar}+m_{f}+a)\, , 
\label{eq:pert1}
\eeq
\beq 
Z^{\rm inst}_{\hbar} (a, {\vec m}; {\qe}) =  {\qe}^{{\alpha}^2} \sum_{{\lambda}, {\mu} \in {\CalP}} \, {\qe}^{|{\lambda}|+|{\mu}|} \,  \frac{\prod\limits_{f=1}^{4} \, P_{\lambda} ( m_{f}+ a) P_{\mu} (m_{f}-a)}{\left( H_{\lambda} \cdot H_{\mu} \cdot {\mathfrak{H}(2a)}_{\lambda}^{\mu}  \cdot {\mathfrak{H}(-2a)}_{\mu}^{\lambda} \right)^2} \, ,  \\ \,
\label{eq:gil1}
\eeq
with ${\CalP}$ denoting the set of all partitions ${\lambda} = ( {\lambda}_{1} \geq {\lambda}_{2} \geq \ldots \geq {\lambda}_{{\ell}({\lambda}} \geq 0)$ of an integer $|{\lambda}| = {\lambda}_{1} + {\lambda}_{2} + \ldots $, or Young diagrams, with $(i,j) \in {\lambda}$
meaning $1 \leq j \leq {\lambda}_{i}$, and  
\begin{multline}
H_{\lambda} = \prod_{(i,j) \in \lambda} {\hbar} \left(  {\lambda}_{j}^{t} - i  + {\lambda}_{i} - j +1 \right)\ , \
P_{\lambda}(x) =  \prod_{(i,j) \in \lambda} \left( x +{\hbar}( i - j ) \right) \, , \\
{\mathfrak{H}(x)}_{\lambda}^{\mu} = 
\prod_{(i,j)\in {\lambda}} \left( x+ {\hbar} ( {\lambda}_{j}^{t} - i +{\mu}_{i} - j +1)  \right)\, , 
\label{eq:hprh}
\end{multline}
and
\beq
\frac{m_{1}}{\hbar} = {\vt}_{0}+{\vt}_{\qe}\, ,\ \frac{m_{2}}{\hbar} = {\vt}_{0} - {\vt}_{\qe} \, , \ \frac{m_{3}}{\hbar} = {\vt}_{1} + {\vt}_{\infty}\, , \ \frac{m_{4}}{\hbar} = {\vt}_{1} - {\vt}_{\infty}
\label{eq:masses}
\eeq 

\subsection{Symmetries, manifest, and less so}

The Ref. 
\cite{Alday:2009aq} observed that by writing
\begin{multline}
{\CalZ}_{U(2)} (a, - a, {\ve}_{1}, {\ve}_{2}, m_{1}, m_{2}, m_{3}, m_{4} ; {\qe}) = \\ ( 1- {\qe})^{\frac{(m_{1}+m_{2})(2{\ve}_{1}+2{\ve}_{2} - m_{3} - m_{4} )}{2{\ve}_{1}{\ve}_{2}}} Z_{SU(2)} (a , {\ve}_{1}, {\ve}_{2}, m_{1}, m_{2}, m_{3}, m_{4}; {\qe}) 
\label{eq:u1dec}
\end{multline}
we obtain the partition function of the $SU(2)$ gauge theory which is invariant under the $a \to -a$, and
under the transformations $m_{1} \leftrightarrow m_{2}$, or $m_{3} \leftrightarrow m_{4}$ (these are manifest in the $U(2)$ theory as well), and $m_{1} \leftrightarrow {\ve}_{1} + {\ve}_{2} - m_{2}$ or $m_{3} \leftrightarrow {\ve}_{1} + {\ve}_{2} - m_{4}$. The identification with the Liouville conformal blocks also suggests
a symmetry ${\qe} \to {\tilde\qe} = -{\qe}/(1- {\qe})$ ( it corresponds to the global transformation
$z \mapsto (z-{\qe})/(1-{\qe})$, sending $0,{\qe},1,{\infty}$ to ${\tilde\qe}, 0,  1, {\infty}$), accompanied by the exchange ${\Delta}_{0} \leftrightarrow {\Delta}_{\qe}$ $\Leftrightarrow$ ${\ve}_{1} + {\ve}_{2}  - m_{2} \leftrightarrow  m_{2}$ (while keeping $m_{1}, m_{3}, m_{4}$ intact). 

In the case ${\ve}_{1} = - {\ve}_{2} = {\hbar}$ the decoupling factor in \eqref{eq:u1dec} becomes:
\beq
(1-{\qe})^{2{\vt}_{0}{\vt}_{1}}
\eeq 
{}Note that the right hand side of the Eq. \eqref{eq:gil} is, up to a simple prefactor ${\qe}^{\#} (1-{\qe})^{\#}$ invariant under the permutations of $m_{1}, m_{2}, m_{3}, m_{4}$, 
as well as the overall change of sign ${\vec m} \to - {\vec m}$ (${\lambda} \leftrightarrow {\mu}^{t}$),  whereas the Painlev\'e VI equation is invariant under the transformations $\left( {\vt}_{0}\, , \, {\vt}_{\qe}\, , \, {\vt}_{1}\, , \, {\vt}_{\infty} \right) \to \left( \pm {\vt}_{0}\, , \ \pm {\vt}_{\qe}\, , \ \pm {\vt}_{1}\, , \ 1- {\vt}_{\infty} \right)$. 
The permutation $m_{2} \leftrightarrow m_{3}$ corresponds to the Okamoto transformation,  similar to \eqref{eq:thehat}:
\beq
\left( {\vt}_{0}\, , \, {\vt}_{\qe}\, , \, {\vt}_{1}\, , \, {\vt}_{\infty} \right) \mapsto \left( {\tilde\vt}_{0}\, , \, {\tilde\vt}_{\qe}\, , \, {\tilde\vt}_{1}\, , \, {\tilde\vt}_{\infty} \right) = \left( {\vt}_{0} - \frac 12 {\vt}\, , \, {\vt}_{\qe}- \frac 12 {\vt}\, , \, {\vt}_{1}- \frac 12 {\vt}\, , \, {\vt}_{\infty}- \frac 12 {\vt} \right)\, , 
\label{eq:newexp}
\eeq
with
\beq 
{\vt} =  {\vt}_{0} + {\vt}_{\qe} + {\vt}_{1} + {\vt}_{\infty} \, , 
\eeq
while sending the canonical coordinates $(w, {\pt})$ of \eqref{eq:pviham} to $({\tilde w}, {\tilde\pt})$, related by, cf. \eqref{eq:wmap1}:
\beq
\frac{\vt}{{\tilde w} - w} \,  = \,  {\pt} - \frac{{\vt}_{0}}{w} - \frac{{\vt}_{1}}{w-1} - \frac{{\vt}_{\qe}+{\half}}{w-{\qe}}\, . 
\eeq 
The map is a canonical transformation, with the generating function ${\sigma}(w, {\tilde w}; {\qe})$ :
\begin{multline}
p = \frac{\partial {\sigma}}{\partial w}\, , \ {\tilde p} = - \frac{\partial {\sigma}}{\partial {\tilde w}}\, , \  
{\tilde H} - H = \frac{\partial {\sigma}}{\partial \qe} \, , \\
e^{\sigma} =  (1-{\qe})^{\frac{{\vt}( {\vt}_{0} - {\vt}_{\qe} - {\vt}_{1} + {\vt}_{\infty} -1)}{2}} \left( \frac{w}{\tilde w} \right)^{{\vt}_{0}}  \left( \frac{1-w}{1-{\tilde w}} \right)^{{\vt}_{1}} \left( \frac{w - {\qe}}{{\tilde w} - {\qe}} \right)^{{\vt}_{\qe}+{\half}}  \left( \frac{{\tilde w} ( {\tilde w}-1 ) ({\tilde w}- {\qe})}{(w-{\tilde w})^{2}}  \right)^{\frac{\vt}{2}} \end{multline}

\subsection{The Mystery of GIL}

The mystery of the GIL formula is that the right-hand side of  the Eq. \eqref{eq:gil}
is a simple sum\footnote{Similar to, but not identical to the magnetic partition function of \cite{Nekrasov:2003rj}} of the partition functions of $\Omega$-deformed $A_1$-type \cite{NP1} ${\CalN}=2$ gauge theory $Z \left( {\vec a}, {\ve}_{1} , {\ve}_{2}; {\vec m} ; {\qe} \right)$ on ${\bR}^{4}$, 
with the following specifications: the gauge group is $U(2)$, the Coulomb parameters are fixed at the $SU(2)$ point, i.e. ${\vec a} = (a, -a)$, there are precisely $N_f = 4$ fundamental hypermultiplets, with the masses given by 
\beq
(m_{1}, m_{2}, m_{3}, m_{4} )  =   \left( {\vt}_{\qe}+{\vt}_0 , {\vt}_{\qe} - {\vt}_0, {\vt}_1 + {\vt}_{\infty}, {\vt}_1 - {\vt}_{\infty} \right) \  ,  \ {\rm up\ to\ a} \ S(4) \ {\rm permutation}, 
\eeq
and the $\qe$-parameter of the PVI equation is related to the complexified gauge coupling in the simplest possible way:
\beq
{\qe} = {\exp} \, 2\pi\ii \tau_{\rm uv} \, , \qquad \tau_{\rm uv} = \frac{\theta}{2\pi} + \frac{4\pi\ii}{g^2}
\eeq
Finally, the $\Omega$-deformation parameters correspond to the $c=1$ value of the Virasoro central charge in the dictionary of \cite{Alday:2009aq}: 
\beq
{\ve}_{1} = 1 = - {\ve}_{2}
\eeq
On the other hand, it is in the limit ${\ve}_{1} \to 0$, ${\ve}_{2} \to {\hbar}$ that
the surface defect partition function ${\bf\Psi}_{SU(2)}$ of the same $A_1$ theory, with $U(1)$-factor stripped, has the exponential asymptotics
\beq
{\bf\Psi}_{SU(2)}(a, {\ve}_{1}, {\ve}_{2} , {\vec m} ; w, {\qe}) \sim {\exp}\, \frac{S^{\circ}(a,  {\ve}_{2} , {\vec m} ; w, {\qe})}{{\ve}_{1}}\ , 
\eeq
where $S^{\circ}(a,  {\ve}_{2} , {\vec m} ; w, {\qe})$
obeys the Painlev{\'e} VI equation in the Hamilton-Jacobi form \eqref{eq:ham6m}. 
\vskip 1cm
 \centerline{$\bullet\sim\bullet\sim\bullet\sim\bullet\sim\bullet\sim\bullet$}
 \vskip 1cm
\centerline{So, what is going on?}
\vskip 1cm
 \centerline{$\bullet\sim\bullet\sim\bullet\sim\bullet\sim\bullet\sim\bullet$}
 \vskip 1cm

\subsection{Four dimensional perspective}

The secret of the GIL relation gets uncovered in the four dimensional picture. To relate the
$({\ve}_{1}, {\ve}_{2}) \to (0, {\hbar})$ limit to the $({\ve}_{1}, {\ve}_{2}) \to ({\hbar}, -{\hbar})$
limit, we use the blowup technique applied to the four dimensional gauge theory. 

Namely, let us study the partition function of the surface defect in gauge theory on ${\BR}^4 = {\BC}^{2}$
and its blowup ${\widehat{\BC}^{2}}$, in the $\Omega$-backgrounds which are identical at infinity. 

The partition functions are expected to coincide,
\beq
{\bf\Psi}_{{\BC}^{2}} = {\bf\Psi}_{\widehat{\BC}^{2}}\ ,
\eeq  as we reviewed in the introduction. 

The localization computation, however, gives a non-trivial identity
\beq
{\bf\Psi}_{\widehat{\BC}^{2}} = {\bf\Psi}_{{\BC}^{2}} \star Z_{{\BC}^{2}} \ .
\eeq
Here $\star$ stands for the convolution involving the sum over the fluxes through the single exceptional two-cycle on $M'={\widehat{\BC}^{2}}$ of the products of local contributions to the partition function from the two fixed points $p_0$ and $p_\infty$ of the torus $U(1)\times U(1)$ acting on ${\widehat{\BC}^{2}}$ (see appendix for detail). 
The analysis identical to \cite{NY, NekLisbon} gives the formula \eqref{eq:blowsur}, which we now specialize to the case of $SU(2)$ theory with $N_f = 4$ hypermultiplets (and divided both sides by ${\bf\Psi}_{{\BC}^{2}}$:
\beq
1 = 
\sum_{{\mathfrak{n}}\in {\BZ}}  \frac{{\bf\Psi}( a+ {\ve}_{1} {\mathfrak{n}}, {\ve}_{1}, {\ve}_{2}-{\ve}_{1} , {\vec m} ; w, {\qe})}{{\bf\Psi}(a, {\ve}_{1}, {\ve}_{2} , {\vec m} ; w, {\qe}) }  Z (a + {\ve}_{2} {\mathfrak{n}}, {\ve}_{1}-{\ve}_{2}, {\ve}_{2} , {\vec m} ;  {\qe}) \ .
\label{eq:blowsur2}
\eeq
Recall that the difference of the effective Coulomb parameters in the contributions of the two
fixed points in \eqref{eq:blowsur2} is due to the contribution of the gauge curvature to the localization equation on the scalar in the vector multiplet:
\beq
D_{A} {\sigma} + \iota_{V} F_{A} = 0
\eeq
which in the limit of abelian theory becomes: 
\beq
d{\sigma} + \iota_{V} F = 0\ .
\label{eq:sigmavf}
\eeq
The $U(n)$ gauge theory on $M' = {\widehat{{\BC}^{2}}}$ is dominated, at low energy, by the gauge field configurations, where $F = {\rm diag} \left( \mathfrak{n}_{1}, \ldots , \mathfrak{n}_{n} \right) F_{\bA}$, with $F_{\bA}$ the curvature of a $U(1)$-instanton on $M'$ (see the Appendix), with small instanton modifications near the zeroes of $V$. The reason is that the fixed points of the combined  $U(1)\times U(1)$ and $U(1)^n$ action (the latter being the group of constant $U(n)$ transformations, whose equivariant parameter is the asymptotic
value $a$ of the scalar $\sigma$ in the vector multiplet) are the point-like instantons sitting at the fixed points of $U(1)\times U(1)$ action on $M'$ (we assume the gauge bundle is endowed with some lift of the $U(1) \times U(1)$ action), grafted onto the $U(1)\times U(1)$-invariant abelian instantons. This rough picture is made more precise using the algebraic-geometric description, where the holomorphic rank $n$ vector bundle (the structure  defined by the instanton gauge field thanks to the $F^{0,2}=0$ equation) is extended to the rank $n$ torsion free sheaf, which in turns  splits, thanks to the $U(1)^{n}$-invariance, as a sum of $n$ ideal sheaves twisted by holomorphic line bundles $L_{i}$ (see \cite{NekLisbon} for more details). On $M$ such bundles are classified by their first Chern classes $\mathfrak{n}_{i} = \int_{S^2} c_{1}(L_{i})$. The Eq. \eqref{eq:sigmavf} can be now solved by observing that (see the Appendix
$A$ for detail)
\beq
\iota_{V} F_{\bA} = {\ve}_{1} d{\Sigma}_{1} + {\ve}_{2} d{\Sigma}_{2}
\eeq
with the functions $({\Sigma}_{1}, {\Sigma}_{2})$ taking values $(1,0)$ and $(0,1)$ at the fixed points $p_{0}$ and $p_{\infty}$, respectively. At infinity both $\Sigma_{1,2}$ approach zero. 

In this way the scalar ${\sigma}$ equals 
\beq
{\sigma} = a + \left( {\ve}_{1} {\Sigma}_{1} + {\ve}_{2} {\Sigma}_{2} \right) {\rm diag} \left( \mathfrak{n}_{1}, \ldots , \mathfrak{n}_{n} \right)
\eeq
with $a$ being the asymptotic value at $r \to \infty$.

Now we are in business -- let us take the ${\ve}_{1} \to 0$ limit of both sides of \eqref{eq:blowsur2}:
\begin{multline}
1 = 
\sum_{{\mathfrak{n}}\in {\BZ}} 
e^{\frac{S^{\circ}(a+ {\ve}_{1} {\mathfrak{n}},  {\ve}_{2}-{\ve}_{1} , {\vec m} ; w, {\qe})-S^{\circ}(a,  {\ve}_{2} , {\vec m} ; w, {\qe})}{{\ve}_{1}} + \ldots}
 \times   \left( Z (a + {\ve}_{2} {\mathfrak{n}}, -{\ve}_{2}, {\ve}_{2} , {\vec m} ;  {\qe}) + {\ldots } \right) = \\
e^{-\frac{{\partial} S^{\circ}(a, {\ve}_{2}, {\vec m}; w, {\qe})}{{\partial} {\ve}_{2}}} \sum_{{\mathfrak{n}}\in {\BZ}}  e^{\mathfrak{n} {\beta}}
\cdot  Z (a + {\ve}_{2} {\mathfrak{n}}, -{\ve}_{2}, {\ve}_{2} , {\vec m} ;  {\qe}) + {\ldots }    \ ,
\label{eq:blowsur3}
\end{multline}
where
\beq
{\beta} = \frac{{\partial}S^{\circ}}{{\partial}a}\ . 
\eeq
Now, let us use the homogeneity of $Z$ and ${\bf\Psi}$ under the simultaneous rescaling of
$a$, ${\ve}_{1}$, ${\ve}_{2}$, $m_{1}, m_{2}, m_{3}, m_{4}$ to bring \eqref{eq:blowsur3}
to the form:
\beq
e^{\frac{1}{{\ve}_{2}} \left( 1 - a \frac{\partial}{{\partial}a} - \sum\limits_{f=1}^{4} m_{f}\frac{\partial}{{\partial}m_{f}} \right) S^{\circ}} = \sum_{n\in {\BZ}}\ e^{n{\beta}} \
Z (a + {\hbar} n, {\hbar}, - {\hbar} , {\vec m} ;  {\qe})
\label{eq:gil2}
\eeq
Now, recall that $S^{\circ}$ solves the Hamilton-Jacobi equation whose Hamiltonian
is quadratic in momenta and masses, and use \eqref{eq:rh1} to obtain \eqref{eq:gil}\footnote{Also, recall that $Z (a) = {\CalC}(a) {\CalC}(-a) Z^{\rm inst}(a)$}. 

\section{Conclusions and open questions}

In this paper we have been discussing the applications of the four dimensional side of the BPS/CFT correspondence to the ``CFT'' side. Specifically, we applied the blowup equations, expressing the behavior of the correlation functions of four dimensional supersymmetric (twisted, $\Omega$-deformed) gauge theories under the simplest surgeries of the underlying four-manifold, to produce quite unusual identities among the conformal blocks of two dimensional conformal field theories. To fully appreciate these identities the conformal blocks must be analytically continued in their parameters, such as the central charges and spins. However, certain limits make perfect sense within the realm of two dimensional unitary theories. 

We mostly concentrated on the example of the four-point conformal block of the $SL(2)$-current
algebra and the five point conformal block of Virasoro algebra involving a $(1,2)$ degenerate field. The classical, by now, relation \cite{FZ}, which in the large $k$, large $c$ limit becomes the relation between the Painlev{\'e} VI equation and the isomonodromy problem, does not require much sophisticated reasoning. However, the recently discovered \cite{Gamayun:2012ma} unexpected connection between the same isonomondromy problem, Painlev{\'e} VI and the four-point $c=1$ Virasoro conformal blocks becomes explained using the simple blowup formula applied to the $SU(2)$ $N_{f}=4$ gauge theory, in the presence of a surface defect. 

Below we list a few other potential applications of these ideas.

\subsection{Quantization and extremal correlators}

The recent work \cite{Grassi:2019txd} 
  on extremal correlators in ${\CalN}=2$ superconformal field theories makes 
  an extensive use of the $S^4$ partition function of the theory, which was originally computed
by V.~Pestun in \cite{Pestun:2007rz} in terms of the $Z$-function \cite{Nekrasov:2002qd}
of the $\Omega$-deformed theory with ${\ve}_{1} = {\ve}_{2} = \frac{1}{R}$, $R$ being the sphere
radius. Althought the analysis of \cite{Grassi:2019txd} mostly relies on the perturbative, one-loop, part of the $Z$-function, it is conceivable a full non-perturbative treatment hides an interesting story. 
As we reviewed in \cite{NekBPSCFT}, the instanton partition function $Z^{\rm inst}({\bf a}, {\ve}_{1}, {\ve}_{2}, {\bf m}; {\qe})$ for ${\ve}_{1}/{\ve}_{2}$ a positive rational number is not given by the sum
over $N$-tuples of Young diagrams. The reason is that the fixed points of the rotational symmetry on the moduli space of instantons are not isolated, albeit the fixed points sets are compact. The blowup formula gives a way to partly circumvent this difficulty. Let us use the homogeneity of the partition function of the asymptotically conformal theory, and the symmetry ${\ve}_{1} \leftrightarrow {\ve}_{2}$ to write it as:
\beq
Z( {\bf a}, {\ve}_{1}, {\ve}_{2}, {\bf m}; {\qe} ) = z ({\bf a}/{\ve}_{1}, {\ve}_{2}/{\ve}_{1}, {\bf m}/{\ve}_{1}; {\qe}) = z ({\bf a}/{\ve}_{2}, {\ve}_{1}/{\ve}_{2}, {\bf m}/{\ve}_{2}; {\qe}) 
\eeq
The small ${\ve}_{2}$ expansion of $Z$ defines the effective twisted superpotential and the effective dilaton coupling:
\beq
Z( {\bf a}, {\ve}_{1}, {\ve}_{2}, {\bf m}; {\qe}) = {\exp}\, \frac{{\ve}_{1}}{{\ve}_{2}} w_{0} ( {\bf a}/{\ve}_{1}, {\bf m}/{\ve}_{1} ; {\qe}) + w_{1} ( {\bf a}/{\ve}_{1}, {\bf m}/{\ve}_{1} ; {\qe}) + \ldots
\eeq
The function ${\ve}_{1} w_{0}$ is the effective twisted superpotential of an effective two dimensional ${\CalN}=2$ theory living on the fixed plane of the remaining ${\ve}_{1}$-rotation. The \emph{genus one} term $w_{1}$
is the dilaton coupling, computed for a class of two dimensional theories
in \cite{Nekrasov:2014xaa}, and presumably computable along the lines of the recent discussion in \cite{Manschot:2019pog}. 

Then the chiral block of $S^4$ partition function, with the radius of the sphere
$R$, and some prescription for the masses $\bf m$ of the matter fields (see
\cite{Okuda:2010ke} for the relevant discussion),  
\begin{multline}
z ({\bf a}R,  1, {\bf m} R; {\qe} ) \ = \\
\begin{matrix}
{\rm Lim}\\
\scriptstyle{{\ve}_{1}, {\ve}_{2} \to \frac{1}{R}} \end{matrix}\qquad \sum_{{\mathfrak{n}} \in {\Lambda}} 
e^{\frac{{\ve}_{1}}{{\ve}_{2}-{\ve}_{1}} w_{0} \left( \frac{\bf a}{{\ve}_{1}} + \mathfrak{n}, \frak{\bf m}{{\ve}_{1}} \right)  + w_{1} \left( \frac{\bf a}{{\ve}_{1}} + \mathfrak{n}, \frak{\bf m}{{\ve}_{1}} \right)  + \frac{{\ve}_{2}}{{\ve}_{1}-{\ve}_{2}} w_{0} \left( \frac{\bf a}{{\ve}_{2}} + \mathfrak{n}, \frak{\bf m}{{\ve}_{2}} \right)  + w_{1} \left( \frac{\bf a}{{\ve}_{2}} + \mathfrak{n}, \frak{\bf m}{{\ve}_{2}} \right) + \ldots} = \\
\sum_{{\mathfrak{n}} \in {\Lambda}} \, e^{ - w_{0}({\alpha}, {\mu}; {\qe}) + 2 w_{1} ( {\alpha}, {\mu}; {\qe}) + R {\bf m} \cdot {\partial}_{\mu} w_{0}  ({\alpha}, {\mu}; {\qe}) +
R {\bf a}  \cdot {\partial}_{\alpha} w_{0} ({\alpha}, {\mu}; {\qe})} \Biggr\vert_{{\alpha} = {\bf a}R + \mathfrak{n}\, , \ {\mu} = {\bf m}R}
 \ .
\end{multline}
In this way the problems of computing the extremal correlators, the study of $c=25$ Liouville theory etc.  are connected to the quasiclassical conformal blocks, geometry of the variety of opers, and quantization of algebraic integrable systems.

\subsection{Blowups and boundary operator product expansion}

Recall that the supersymmetric ${\CalN}=2$ theories in four dimensions have an intimate connection with hyper-K{\"a}hler geometry. More specifically, the moduli space ${\CalM}$ of vacua  contains branches (the Higgs branches) which are hyper-K{\"a}hler manifolds, perhaps with singularities, which connect to other branches (the Coulomb branches), which are special K{\"a}hler manifolds. Once the theory is compactified on a circle, these other branches get enhanced to hyper-K{\"a}hler manifolds \cite{Seiberg:1996nz} ${\CalM}_{H}$. 

The four dimensional ${\CalN}=2$ theory on a manifold $M^{4}$ with $T^{2} = U(1) \times U(1)$-isometry, subject to the $\Omega$-deformation associated with this isometry, can be viewed, at low energy, as a two dimensional sigma model with the worldsheet ${\Sigma} = M^{4}/T^{2}$ and ${\CalM}_{H}$ as a target space. The fixed points or irregular orbits of $T^2$ give rise to the boundaries and corners of $\Sigma$. The study of boundary conditions in the sigma model, or branes, descending from the torus fixed loci, and surface defects added at these loci, is a rich a fruitful subject \cite{NW}. 

In this note we would like to point out that the blow-up of a fixed point of $T^2$ action modifies $\Sigma$ in a simple way, by cutting a little piece of the corner (and creating two corners instead, see Fig. in the appendix A.1.2). The values of the $\Omega$-deformation parameters correspond to the type of the brane on ${\CalM}_{H}$ a piece of the boundary of $\Sigma$ supports. The corners are, typically, boundary-changing vertex operators (such as the c.c.-brane/brane of opers in the case of ${\CalM}_{H}$ being the Hitchin moduli space). The transformation $({\ve}_{1}, {\ve}_{2}) \mapsto ({\ve}_{1} - {\ve}_{2}, {\ve}_{2})$ which we see at one of the corners after the blowup can be viewed as turning on a unit flux of a $U(1)$ Chan-Paton gauge field on a c.c.-brane. It would be nice to work out this correspondence in full detail. 

The blowup formulae \eqref{eq:blowsur}, \eqref{eq:blowbulk} then teach us something about
the operator product expansion of these special (geometric) vertex operators.

\subsection{Parts I, II, III, IV, and V}

An asymptotically conformal field theory can be made asymptotically free by sending some masses to infinity while keeping some relevant energy scales finite. For example, by sending $m_{4} \to \infty$ while keeping ${\Lambda}_{3}^{2} = m_{4}^{2} e^{2\pi \ii \tau}$ finite we get the $SU(2)$ gauge theory with $N_{f}=3$ massive hypermultiplets in the fundamental representation. Similarly, we can furthermore send $m_{3}$, $m_{2}$, and, finally, $m_{1} \to \infty$,  
leaving us with the pure $SU(2)$ ${\CalN}=2$ theory, the original model studied in \cite{Seiberg:1994rs}. 

The corresponding limits can be applied to the partition function $Z(a, {\vec m} , {\ve}_{1}, {\ve}_{2}, {\tau})$, as was done in \cite{Nekrasov:2002qd, Nekrasov:2003rj}. The blowup formulas hold for the asymptotically free theories (in fact, they can be proven with less pain by ghost number anomaly considerations, \cite{Losev:1997tp}). One expects therefore the formulas similar to the GIL formula for the Painlev{\'e} V, IV, etc. $\tau$-functions, as indeed proposed in \cite{Gamayun:2013auu}. The stumbling blocks in these identifications are the good Darboux 
coordinates on the moduli spaces of flat connections with irregular singularities (which is what the asymptotically free limits amount to). Fortunately a nice theory seems to be emerging \cite{CMR}. 

\subsection{Quasinormal modes}

Another application of the blowup equations might be the study of the black hole quasinormal modes. It was observed in \cite{Nekrasov:2018pqq}, that the Schr{\"o}dinger equation obtained by separating variables in the wave equation describing particles in the black hole background, has an interesting WKB geometry, related to Seiberg-Witten integrable systems. In recent papers 
\cite{Aminov:2020yma} this analogy was shown to hold on a much more detailed level, with the relevant ${\CalN}=2$ four dimensional theory being an $SU(2)$ gauge theory with $N_f = 2,3$ fundamental hypermultiplets. The quantization problem, which emerges in the ${\ve}_{2} \to 0$, ${\ve}_{1} = {\hbar}$ NS limit \cite{Nekrasov:2009rc} of that theory is related in \cite{Aminov:2020yma} to the ${\ve}_{1} = - {\ve}_{2}$ limit of the same theory. This is surely another manifestation of the blowup relations we explored in this paper, adapted to the asymptotically free theory case. It is remarkable that \cite{Aminov:2020yma}  also connects this story to the resurgent expansions. We were reminded that an earlier work \cite{Novaes:2014lha} also used the Painlev{\'e} transcendents and the GIL formula in the studies of scattering  of black holes and their quasinormal modes. 

This is a peculiar example of an intricate relation between gauge theory and gravity.

\subsection{Strong/weak coupling dualities}

The Painlev{\'e} equations are traditionally used as a testing ground for the resurgence ideas, attempting in reconstructing the non-perturbative description of a physical system by resumming its perturbation expansion in terms of trans-series. Our blowup formulas are, in many respects, a way of relating weak and strong coupling expansions of the same theory, and should probably be recast in the form of the resurgent expansion. See 
\cite{Dunne:2019aqp} for the recent discussion of closely related problems. 

\subsection{Whitham hierarchies}

It is tempting to conjecture, based on the possibility of subjecting ${\CalN}=2$ theory to a two dimensional $\Omega$-deformation, that the analogues of the isomonodromic deformations are universal companions of the algebraic integrable systems, even outside the domain of Hitchin systems. Perhaps the universal Whitham hierarchies \cite{Krichever:1992qe} are the answer (see \cite{Takasaki:1997fy} for some discussion to that effect). 
The first interesting examples would be the moduli spaces of ADE instantons on ${\BR}^{2} \times T^{2}$ and the moduli spaces of ADE monopoles on ${\BR}^{d+1} \times T^{2-d}$, which are the moduli spaces of vacua of quiver gauge theories, as was rigorously shown in \cite{NP1} using gauge theory analysis, and, in the $A$-type case in \cite{Cherkis:2000ft}, using string dualities.

\subsection{Higher genus, more points...}

In the appendix we discuss some generalizations to the case of higher rank gauge theories in four dimensions, higher rank current algebras in two dimensions, more punctures, higher genus. 

\subsection{Five dimensions, quantizations and topological strings}

The uplift of the blowup relations to the case of five dimensional theories \cite{NikFive} compactified on a circle, is very interesting from many perspectives. Mathematically the story is much more complicated (ref. 3 in \cite{NY}). 

In \cite{Aganagic:2003qj}, \cite{Grassi:2013qva}, \cite{Gu:2015pda}, and, 
more recently \cite{Grassi:2019coc} various features of topological strings on local Calabi-Yau manifolds has been observed to connect to quantum integrable systems, with the string coupling $g_{s}$ playing the role of the Planck constant. This is in contrast with the conjectures
in \cite{Nekrasov:2002qd} (mostly proven today) relating the topological string to the ${\ve}_{1} = - {\ve}_{2} = g_{s}$ case of the $\Omega$-deformed four dimensional theory, while quantum integrability
is found in the ${\ve}_{2} \to 0$, ${\ve}_{1} = {\hbar}$ NS limit \cite{Nekrasov:2009rc}. 

Now we know how to relate the $({\ve}_{1}, {\ve}_{2}) \to ( {\hbar}, - {\hbar})$ limit to the
$({\ve}_{1}, {\ve}_{2}) \to ( {\hbar}, 0)$ limit, using the blowup formulas. In the five dimensional version, we would be relating the theories at $(q_{1}, q_{2})$ to $(q_{1}, q_{2}/q_{1})$ and $(q_{1}/q_{2}, q_{2})$, so that the \emph{topological string} limit $(e^{g_{s}},e^{-g_{s}})$ would be related to the NS limit.  Hopefully the blowup formula will also shed some light on the mysteriously beautiful proposal of \cite{Hatsuda:2015qzx} to add the effective twisted superpotentials of theories with $S$-dual $q$-parameters.

Another possible extension of our story might be the search for the analogue of the $\yt$-coordinates in the topological $B$ model, as a way to parametrize the holomorphic anomaly equation \cite{Witten:1993ed}.

\appendix{}

\section{Blowup, conifold, and origami}

\subsection{Blowup of a plane}

From now on we are working with the local model of $M$, namely a copy of Euclidean space ${\BR}^4$ which we shall identify with ${\BC}^{2}$ by fixing one of its complex structures. We shall place the point $p$ at the origin $0 \in {\BC}^2$. The blowup $M'$ in this case is denoted by $\widehat{{\BC}^{2}}$, and can be defined as the space of pairs $(u, l)$, where $u \in {\BC}^2$, and $l \approx {\BC}^1$ is a complex line passing through $u$ and $0$. Forgetting $l$ defines the map:
\beq
p: \widehat{\BC}^{2} \longrightarrow {\BC}^{2} \, , \qquad p(u,l) = u
\eeq
For $u \neq 0$, $p^{-1}(u)$ consists of exactly one point, as there is only one line passing through two distinct points on a plane. For $u = 0$ the preimage $E$ is the space of all lines passing through a point on a plane, also known as the projective line ${\BC\BP}^1 = p^{-1}(0)$. This is the exceptional set $E$ we mentioned previously. 
We can cover ${\widehat{\BC}^{2}}$ with two coordinate charts $U_0$ and $U_{\infty}$, both isomorphic to ${\BC}^{2}$, with the transition function:
\beq
(z_{1}, l_{1}) \in U_0 \, \mapsto\, (l_2, z_2) \in U_{\infty} \, , \ z_{1}l_{1} = z_{2}, \ z_{2}l_{2} = z_{1}
\eeq
which is invertible when both $l_{1}, l_{2} \neq 0$ (and are mutual inverses, $l_{1} l_{2} = 1$), so that $U_{0} \cap U_{\infty} = {\BC} \times {\BC}^{\times}$. 

\subsubsection{K{\"a}hler quotient construction}

One description of $\widehat{{\BC}^{2}}$ is via the K\"ahler quotient of $\BC^3$ by the $U(1)$ action: 
\beq
(w_{1}, w_{2}, w_{3}) \approx (e^{\ii t} w_1 , e^{\ii t} w_2 , e^{-\ii t} w_{3} )
\label{eq:uoa}
\eeq
as ${\BC}^{3}//U(1)$:
\beq
|w_{1}|^{2} + |w_{2}|^{2} -|w_{3}|^{2} = {\zeta} > 0  \ .
\label{eq:momm}
\eeq
The coordinate patches $U_{0}$ and $U_{\infty}$ correspond to the loci where $w_{1}\neq 0$ and $w_{2} \neq 0$, respectively. Indeed, \eqref{eq:momm} implies that, for ${\zeta}>0$, $w_1$ and $w_2$ cannot vanish simultaneously. The holomorphic coordinates on $U_0$ are the holomorphic gauge invariants $z_1 = w_{3}w_1$ and $l_1 = w_2/w_1$, while the holomorphic coordinates on $U_{\infty}$ are the gauge invariants $l_2 = w_1/w_2$ and $z_2 = w_{3}w_2$. Restricting the flat metric on ${\BC}^3$ onto the surface $S$ defined by the Eq. \eqref{eq:momm} and projecting along the $U(1)$ action gives a metric on $M'={\widehat{\BC}^{2}}$, which is K{\"a}hler with the K{\"a}hler potential, which we express in terms of $z_1 = w_{3}w_1$ and $z_2 = w_{3}w_2$:
\begin{multline}
K_{\rm kq} (z, {\zb}) = {\rm Crit}_{V} \left(   e^{-2V} |z_{1}|^2 +e^{-2V} |z_{2}|^2 + e^{2V} + 2{\zeta} V \right)  = 
2{\ell} + {\zeta} \, {\rm log} \left(  {\ell} \right) \, , \\
{\ell} = \sqrt{|z_{1}|^2 + |z_{2}|^2+{\zeta}^2/4} - {\zeta}/2 \ .
\label{eq:kqpot}
\end{multline}
For small $|z_{1}|, |z_{2}|$ we can approximate $K_{\rm kq}$ by the sum of the Euclidean space K{\"a}hler potential $\propto |z_1|^2+|z_2|^2$ and that of a two-sphere $\propto {\rm log}( |z_{1}|^2+|z_{2}|^2 )$. 
The K{\"a}hler form ${\varpi}_{\rm kq} = dd^{c} K_{\rm kq}$ has a non-trivial period over the two cycle $E$ given by the equation $w_{3}=0$.  

{}Let $w_{a} = |w_{a}| e^{\ii {\varphi}_{a}}$, $a = 1, 2, 3$. 
Another structure we get from the restriction of the flat ${\BC}^{3}$ metric onto the surface $S$ is the Riemannian connection on the $U(1)$-bundle 
\beq
{\pi}: S \to M' 
\label{eq:pimap}
\eeq 
The corresponding connection form ${\bA}_{\rm kq}$ is also obtained by minimizing:
\begin{multline}
{\partial}_{\bA} \biggr\vert_{{\bA} = {\bA}_{\rm kq}} \, \left( \vert dw_{3}  - {\ii} {\bA} w_{3}  \vert^2 + 
\sum_{{\alpha}=1}^{2} \vert dw_{\alpha} + {\ii} {\bA} w_{\alpha} \vert^2 \right)  = 0  \, , \\
{\bA}_{\rm kq} = \frac{|w_{3}|^2 d{\varphi}_{3} - |w_{1}|^{2} d{\varphi}_{1} - |w_{2}|^{2} d{\varphi}_{2}}{w_{1}|^{2} + |w_{2}|^{2} + |w_{3}|^{2}} = d{\varphi}_{3} - \frac{r_{1}^{2} d{\theta}_{1} + r_{2}^{2} d{\theta}_{2}}{r_{1}^{2} + r_{2}^{2} + {\ell}^{2}}\, .
\end{multline}
Naturally, it obeys: $\iota_{v} {\bA}_{\rm kq} = 1$, where $v = -{\partial}_{\varphi_3} +{\partial}_{\varphi_1} + {\partial}_{\varphi_2} =  {\partial}/{\partial}t$ is the generator of the \eqref{eq:uoa} action, and $d{\bA}_{\rm kq} = {\pi}^{*}F_{\rm kq}$, with \eqref{eq:pimap}: 
\beq
F_{\rm kq} = - d\left( \frac{(r_{1}/r)^2}{1 + ({\ell}/r)^{2}} \right) \wedge d{\theta}_{1} 
- d\left( \frac{(r_{2}/r)^{2}}{1 + ({\ell}/r)^{2}} \right) \wedge d{\theta}_{2}
\label{eq:univconn}
\eeq 
where
$z_{1} = r_{1} e^{{\ii}{\theta}_{1}}\, , \ z_{2} = r_{2} e^{{\ii}{\theta}_{2}}$. The formula for the curvature $F_{\rm kq}$ gets a bit tricky when both $r_{1}$ and $r_{2}$ approach zero. On the patch $U_0$ 
the good coordinates are $z_1 = r_1 e^{{\ii}{\theta}_{1}}$, and $z_{2}/z_{1} = r e^{{\ii}{\theta}}$, 
so we need to rewrite \eqref{eq:univconn} in terms of $r = r_{2}/r_{1}$, ${\theta} = {\theta}_{2} - {\theta}_{1}$:
\beq
F_{\rm kq} =  d \left( \frac{{\lambda}_{1}^{2}}{1+ r^2 + {\lambda}_{1}^2}\right) \wedge d{\theta}_{1} - 
d \left( \frac{r^2}{1+r^2 + {\lambda}_{1}^{2}} \right) \wedge d{\theta}\ ,
\eeq
where 
\beq
{\lambda}_{1} = - \frac{\zeta}{2r_{1}} + \sqrt{ 1 + r^2 + \frac{{\zeta}^{2}}{4r_{1}^{2}}} \approx \frac{r_{1}}{\zeta} ( 1 + r^2 ) + \ldots \, , \ r_{1} \to 0
\eeq
Analogous expression holds on $U_{\infty}$ with $r_{1} \leftrightarrow r_{2}$, $r \leftrightarrow 1/r, {\theta} \leftrightarrow - {\theta}$, etc. 

\subsubsection{Torus action on $M'$}
 
The two torus $U(1) \times U(1)$ acts of ${\widehat{\BC}^{2}}$ by $(w_1, w_2, w_{3}) \mapsto (e^{\ii \vt_1} w_1, e^{\ii \vt_2} w_2, w_{3})$, which, in coordinate patches, looks like the standard $U(1) \times U(1)$ action on ${\BC}^2$ albeit in different bases of generators :
on $U_0$ it is given by $(e^{\ii \vt_1} , e^{\ii \vt_2}) \cdot (z_1 , l_1) = ( e^{\ii \vt_1} z_1, e^{\ii ({\vt}_2 - {\vt}_{1})} l_1 )$, while on $U_{\infty}$ it is given by:
$(e^{\ii \vt_1} , e^{\ii \vt_2}) \cdot (l_2 , z_2) = ( e^{\ii ({\vt}_{1}-{\vt}_{2})} l_2, e^{\ii {\vt}_2} z_2 )$. The fixed points of the $U(1)\times U(1)$ action on ${\widehat{\BC}^{2}}$, therefore, are: $p_0 = (0,0) \in U_0$, and $p_{\infty} = (0,0) \in U_{\infty}$. Recall, that since $l_{1,2} \neq 0$ on $U_{0} \cap U_{\infty}$, the points $p_{0}$ and $p_{\infty}$ are distinct. 
\vskip .1in
\centerline{\includegraphics[width=10cm]{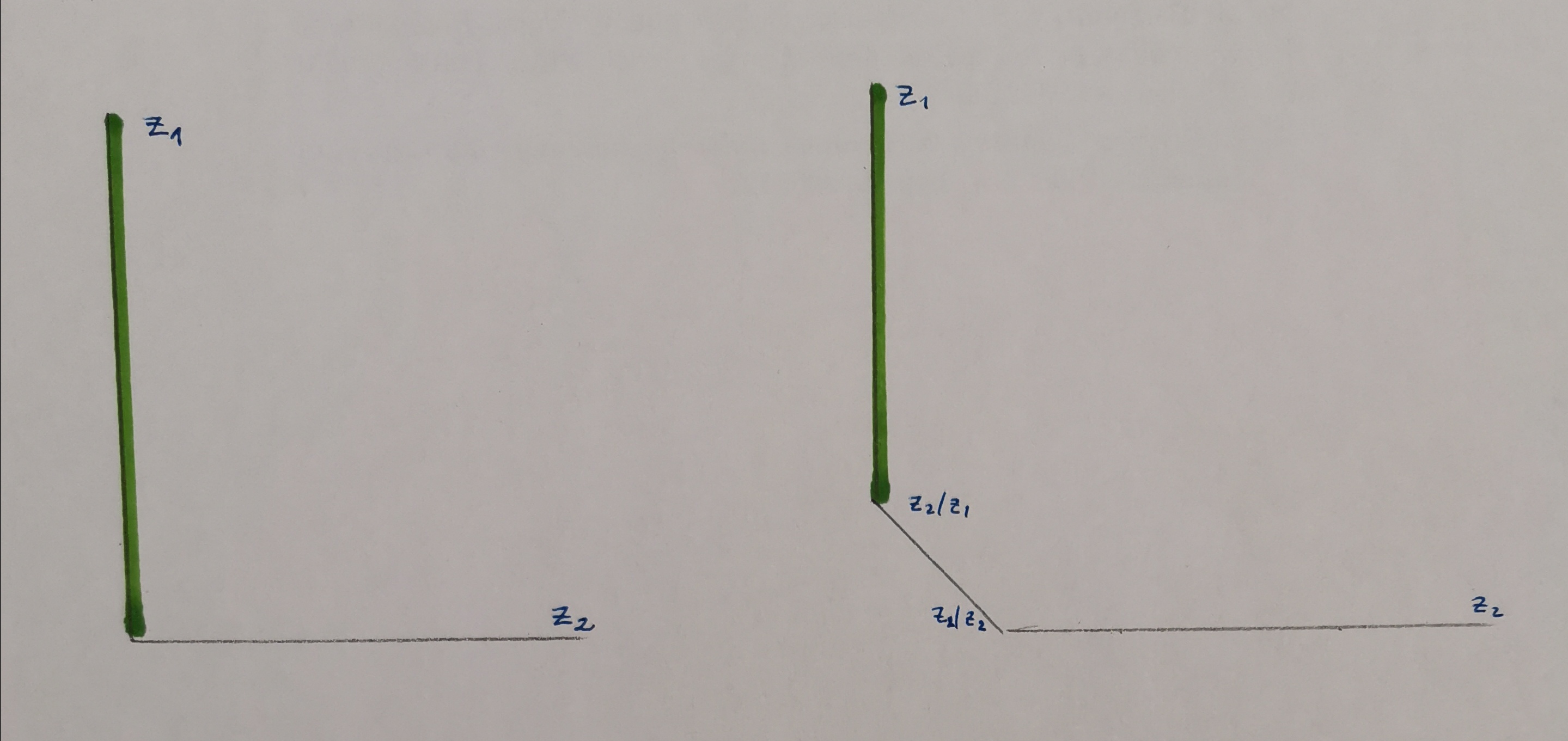}}
\vskip .1in
\centerline{Toric diagrams of ${\BC}^{2}$ and ${\widehat{\BC}^{2}}$}
\bigskip
It is not difficult to compute the generators of the $U(1) \times U(1)$ action. In the $z_{1}, z_{2}$
coordinates, the transformation $(z_{1}e^{{\ii}t {\ve}_{1}} , \ z_{2} e^{{\ii}t {\ve}_{2}})$ is generated
by the vector field
\beq
V = {\ve}_{1} {\partial}_{{\theta}_1} + {\ve}_{2} {\partial}_{{\theta}_{2}}\ , 
\eeq
which is a Hamiltonian vector field preserving the symplectic form ${\varpi}_{\rm kq} = {\partial}{\bar\partial} K_{\rm kq}$, with the Hamiltonian
function 
\beq
h = {\ve}_{1} h_{1} + {\ve}_{2} h_{2}  = \frac{{\ve}_{1} |z_{1}|^{2}+{\ve}_{2} |z_{2}|^2}{| z_{1} |^{2} + |z_{2}|^{2}}\,
\frac{{\zeta} + {\ell}}{2} 
\eeq
At the fixed point $p_0$ we need to use the $U_0$ parameterization:
\beq
h = \frac{{\ve}_{1} +{\ve}_{2} r^2}{1 + r^{2}}\,
\frac{{\zeta} + r_{1} {\lambda}_{1}}{2}, \
\eeq
so that
$h(p_{0}) =  {\ve}_{1}  {\zeta}/2$. 
Similarly, $h (p_{\infty}) = {\ve}_{2} {\zeta}/2$.

\subsubsection{Another metric and another gauge field}

Canonical as it is the K{\"a}hler quotient construction does not always produce nice results when the amount of underlying supersymmetry is too low (in plain English, things are not hyper-K{\"a}hler). In particular, the gauge field ${\bA}_{\rm kq}$ is not anti-self-dual in the K{\"a}hler metric defined by \eqref{eq:kqpot}. Fortunately, there exists a simple nudge to the formulas \eqref{eq:kqpot}, \eqref{eq:univconn}, making everything fall in place. We endow $M'$ with the K{\"a}hler metric ${\varpi}_{\zeta} =dd^{c} K_{\zeta}$, with
\beq
K_{\zeta} = \frac{|z_{1}|^{2} + |z_{2}|^{2}}{\zeta} + {\zeta} \, {\rm log} \left(  |z_{1}|^{2} + |z_{2}|^{2} \right)
\label{eq:burns}\eeq
so that the cohomology class of ${\varpi}_{\zeta}$ equals to that of ${\varpi}_{\rm kq}$. 
The metric corresponding to \eqref{eq:burns} is found in \cite{Burns} and is remarkable in that it is Weyl anti-self-dual, $W^{+}=0$ and scalar free. A larger class of metrics containing those on multi-blowups is in \cite{LeBrun}. In \cite{Braden:1999zp} a charge $1$
instanton was found:
\beq
{\bA} =  \frac{(r_{1}/r)^{2} d{\theta}_{1} + (r_{2}/r)^{2} d{\theta}_{2}}{1+(r/{\zeta})^2}
\eeq
This is the gauge field we use in the low energy description of gauge theory on $M'$. We have:
\beq
- \iota_{V} F_{\bA} = {\ve}_{1} d{\Sigma}_{1} + {\ve}_{2} d{\Sigma}_{2}
\eeq
with
\beq
{\Sigma}_{\alpha} = \frac{(r_{\alpha}/r)^{2}}{1+(r/{\zeta})^2}
\eeq

\subsection{Blowup and conifold}

It is often useful to view the four dimensional ${\CalN}=2$ gauge theory as the theory of low energy modes of open strings attached to $D3$ branes of $IIB$ string theory, in some supersymmetric ten dimensional background. The ten dimensional geometry ${\CalX}$ must contain a ${\BC} \approx {\BR}^{2}$ factor, transverse to the worldvolume of the branes, representing the complex line of eigenvalues of the scalars in the vector multiplets, i.e. ${\CalX} = {\BC} \times {\CalY}$. The simplest way to achieve supersymmetry is to demand ${\CalY}$ to be a Calabi-Yau fourfold. In \cite{NekBPSCFT} we took ${\CalY} = {\BC}^{4}$. The construction below corresponds to ${\CalY} = {\CalZ} \times {\BC}$, with
${\CalZ}$ the total space of the ${\CalO}(-1) \oplus {\CalO}(-1)$ bundle over ${\BC\BP}^1$, also known as a small resolution of the conifold singularity
\beq
x_1^2+x_2^2+x_3^2+x_4^2 = 0\, , \ x_i \in {\BC}
\label{eq:coni}
\eeq
where $x_{1} + {\ii} x_{2} = w_{1} w_{3} \, , \ x_{1} - {\ii} x_{2} = w_{2}  w_{4}$, $x_{3} + {\ii} x_{4} = w_{2}  w_{3}\, , \ x_{3} - {\ii} x_{4} = - w_{1}  w_{4}$ are the gauge invariant holomorphic coordinates on the K\"ahler quotient ${\BC}^{4}//U(1)$
of ${\BC}^{4}$ with the coordinates $w_{1}, w_{2}, w_{3}, w_{4}$
by the $U(1)$ action 
\beq
(w_{1}, w_{2}, w_{3}, w_{4}) \mapsto (e^{{\ii} t} w_{1}, e^{{\ii} t} w_{2}, e^{-{\ii} t}w_{3}, e^{-{\ii}t} w_{4})\ , 
\label{eq:u1action}
\eeq
where we choose the positive level ${\zeta}>0$ of the moment map
\beq
|w_{1}|^2+|w_{2}|^2 - |w_{3}|^2 - |w_{4}|^2 = {\zeta}
\label{eq:u1momcon}
\eeq
We observe that $M' = {\widehat{\BC}^2} \subset {\CalZ}$ is a hypersurface, defined by the equation homogeneous equation $w_{4}= 0$. This equation has a charge $-1$ under the gauge group of the K{\"a}hler quotient, so the normal bundle to $M'$ inside $\CalZ$ is topologically nontrivial, without global holomorphic sections, meaning a stack of $D$-branes wrapping $M'$ inside $\CalZ$ will not have infinitesimal deformations within $\CalZ$. 

However, if one uses a gauge-invariant function, e.g. $x_{1} - {\ii} x_{2}$, or $x_{3} + {\ii}x_{4}$, its zeroes define a two-component surface,  $M' \cup M''$, with $M''$ being a copy of ${\BC}^{2}$, a vanishing locus of $w_2$. 

Recall the crossed instanton construction of \cite{NekBPSCFT}. It  looks at the generalized gauge theory whose worldvolume is a union of two surfaces, each isomorphic to ${\BC}^{2}$ with some multiplicity, intersecting transversely inside ${\CalY} \approx {\BC}^{4}$. Such singular surface is invariant under the action of the Calabi-Yau torus $U(1)^3 \subset SU(4)$. In the context of the present paper we will be interested in the union of the surface defined by the equation $w=w_{4} = 0$ (with $w$ being the coordinate on the ${\BC}$-factor in 
${\CalY} = {\CalZ} \times {\BC}$, again with some multiplicity, corresponding to the rank of the gauge group, and the (multiple) surfaces located at $w_{1} = w_{3}= 0$, and at $w_{2} = w_{3} = 0$. In other words, there are two fundamental $qq$-characters in the $\Omega$-deformed theory on ${\widehat{\BC}^2}$, one inserted at the fixed point $p_0$ and another at the fixed point $p_{\infty}$.

\section{Instantons on ${\BR}^4$ and on ${\widehat{\BC}^{2}}$}

The instantons, i.e. minimal Euclidean action solutions to the Yang-Mills equations, were first found in \cite{Belavin:1975fg}. It was soon found that these solutions come in families, 
so that on a compact manifold $X$ the charge $k$ $G$-instantons  have 
\beq
4h_{G}^{\vee} k - \frac{{\chi}+{\sigma}}{2} {\rm dim}G
\eeq
moduli. The complete description of the moduli space of charge $k$ instantons on four-sphere $M^4 = S^4$ was given in \cite{ADHM}. We shall need a version of that construction, in which  $M^4 = {\BR}^4$, so that a priori one may get an infinite dimensional space of solutions. However, with appropriate boundary conditions imposed at infinity the corresponding moduli space is finite dimensional. 
The construction for gauge group $U(n)$ can be summarized as follows. Fix two complex vector spaces, $K$ and $N$, of dimensions $k$ and $n$, respectively, endowed with Hermitian metrics. Consider the space of quadruples $(B_1, B_2, I,J)$ where
$B_{1,2} \in {\rm End}(K,K)$, $I \in {\rm Hom}(N,K)$, $J\in {\rm Hom}(K,N)$, obeying
the equations ${\mu}_{\BC} = 0$ and ${\mu}_{\BR} = 0$:
\beq
{\mu}_{\BC} = [ B_{1}, B_{2} ] + IJ \, , \qquad {\mu}_{\BR} = [ B_{1}, B_{1}^{\dagger} ] + [ B_{2}, B_{2}^{\dagger} ] + II^{\dagger} - J^{\dagger} J 
\label{eq:mommps}
\eeq
The moduli space ${\CalM}(n,k)$ of \emph{framed} charge $k$ instantons with gauge group $SU(n)$ is isomorphic to 
\beq
{\CalM}^{\circ}(n,k) = \left( {\mu}_{\BC}^{-1}(0) \cap {\mu}_{\BR}^{-1}(0) \right)^{\circ} / U(k)
\label{eq:stbpnts}
\eeq 
where the term \emph{framed} means that ${\CalM}(n,k)$ parameterizes the solutions to the equation 
\beq
F_{A} + \star F_{A} = 0\, , \qquad F_{A} = dA + A \wedge A
\eeq
where $A$ is an $\mathfrak{su}(n)$-valued (i.e. $A = {\ii} a_{\mu} dx^{\mu}$, with $a_{\mu}$, ${\mu} =1,2,3,4$ being traceless $n \times n$ Hermitian matrices)  one-form on ${\BR}^{4}$, 
obeying
\beq
A(x) \longrightarrow h_{\infty}^{-1} dh_{\infty} \, , \qquad x \to \infty \, , \qquad h_{\infty}: S^{3}_{\infty} \to SU(n)
\eeq
modulo the equivalence relation
\beq
A \sim h^{-1} A h + h^{-1} dh \, , \qquad h \longrightarrow 1 + O(1/r^2)
\eeq
The charge $k$ is the degree of the map $g_{\infty}$, which can be computed as the integral:
\beq
k = \frac{1}{8\pi^2} \int_{S_{\infty}^{3}} {\rm Tr}( g_{\infty}^{-1} dg_{\infty} )^3
\eeq
The group $U(k)$ in \eqref{eq:stbpnts} acts via: 
$g \cdot (B_1, B_2, I , J) =  (g^{-1}B_{1} g, g^{-1} B_{2} g, g^{-1} I, J g)$, and the  symbol $\circ$ in \eqref{eq:stbpnts} means to keep only those $(B_1, B_2, I, J)$ solving the ${\mu}_{\BC} = {\mu}_{\BR} = 0$ equations,  whose the stabilizer is trivial, i.e. $g \cdot (B_1, B_2, I , J)  =  (B_1, B_2, I , J)$ iff $g =1$. 
Differential geometers often work with the partially compactified Uhlenbeck space 
\beq
{\CalM} (n, k) = \left( {\mu}_{\BC}^{-1}(0) \cap {\mu}_{\BR}^{-1}(0) \right)/ U(k)
\eeq
which parametrizes the so-called \emph{ideal  framed instantons}, i.e. the collections:
$([A'] ; {\bx})$ where $[A' ] \in {\CalM}(k-l, n)$, ${\bx} \in  Sym^{l}({\BR}^{4}) = \left( {\BR}^{4} \right)^{l} / S(l)$. However,  ${\CalM} (n, k)$  is not very useful for doing computations  since it is a singular contractible space, and we want to use topology to circumvent doing direct integration. The so-called Gieseker-Nakajima space \cite{NakHilb}
\beq
\widetilde{\CalM}(n,k) = \left( {\mu}_{\BC}^{-1}(0) \cap {\mu}_{\BR}^{-1}({\zeta} \cdot 1_{K}) \right)/ U(k)
\eeq
comes in handy. It parametrizes the torsion free sheaves $\CalN$ on ${\BC\BP}^{2}= {\BC}^{2} \cap {\BC\BP}^{1}_{\infty}$ with the trivialization at ${\BC\BP}^{1}_{\infty}$, 
${\CalN} \vert_{{\BC\BP}^{1}_{\infty}} \approx N \otimes {\CalO}_{{\BC\BP}^{1}_{\infty}}$. 
Physically it is the moduli space of $U(n)$ instantons (note the change of the gauge group)
on \emph{non-commutative} ${\BR}^{4}$ \cite{NekSch} which is (very) roughly the algebra generated by $z_1, z_2, {\zb}_1, {\zb}_2$ obeying $[z_1, z_2] = 0\, , [{\zb}_1, {\zb}_2] =0$, $[z_1, {\zb}_1 ] = {\theta}_1 \cdot 1, \  [z_2, {\zb}_2] = {\theta}_2 \cdot 1$ with $\theta_1 + \theta_2 = - \zeta$. 

\subsection{Instantons on blowup}
The ADHM construction extends to the blowup $\widehat{\BC}^2$ \cite{NY}. One fixes now three Hermitian spaces $N$, $K'$, $K''$ of dimensions $n$, $k'$, $k''$, respectively. 
Consider the space of $5$-tuples: $(b_1, b_2, b, i, j)$, where $b_{1,2}  \in  {\rm Hom}(K', K'' )$, $b \in {\rm Hom}(K'', K')$, $i \in {\rm Hom}(N, K'')$, $j \in {\rm Hom}(K', N)$, obeying  $m_{\BC} = 0$, $m_{\BR}' = {\zeta}' \cdot 1_{K'}$, $m_{\BR}'' = {\zeta}'' \cdot 1_{K''}$, where 
$m_{\BC} = b_{1} b b_{2} - b_{2} b b_{1} + ij : K' \to K''$, 
and
\beq
\begin{aligned}
& m_{\BR}' = - b_{1}^{\dagger} b_{1} - b_{2}^{\dagger} b_{2} - j^{\dagger} j + bb^{\dagger}\, , \\
& m_{\BR}'' = b_{1} b_{1}^{\dagger} + b_{2} b_{2}^{\dagger} + i i^{\dagger}  -b^{\dagger} b \, , \\
\end{aligned}
\eeq

The formula, which is (for $n=2$) equivalent to the main theorem in \cite{NY}, is
\beq
Z( {\vec a}, {\ve}_{1}, {\ve}_{2} , {\vec m} ; {\qe})  = \sum_{{\vec{\mathfrak{n}}} \in {\BZ}^{n-1} \subset {\BZ}^{n}}  Z( {\vec a} + {\vec{\mathfrak{n}}} {\ve}_{1}, {\ve}_{1}, {\ve}_{2}-{\ve}_{1} , {\vec m} + {\vec\rho} {\ve}_{1} ; {\qe}) Z( {\vec a} + {\vec {\mathfrak{n}}} {\ve}_{2} , {\ve}_{1}-{\ve}_{2}, {\ve}_{2} , {\vec m} + {\vec\rho} {\ve}_{2} ; {\qe})
\eeq
where ${\vec{\mathfrak{n}}} = (\mathfrak{n}_{1}, \ldots , \mathfrak{n}_{n}) \in {\BZ}^{n}$, 
$\mathfrak{n}_{1} + \ldots + \mathfrak{n}_{n} = 0$, and ${\vec\rho}$ is determined by 
the choice of the matter fields twists. 

\subsubsection{Perturbative check}

The classical and one-loop piece of $Z$ is given by \footnote{Note in passing, that we insist on using our normalizations, which differ from those in  \cite{Alday:2009aq}:
\beq
{\sE} \left[ - \sum_{{\alpha}< {\beta}} \frac{N_{\alpha}^{*}N_{\beta} q_{1}+
N_{\alpha}^{*}N_{\beta} q_{2}}{(1-q_{1}^{-1})(1-q_{2}^{-1})} \, \right]
\eeq}:
\beq
Z^{\rm pert} ( {\vec a}, {\ve}_{1}, {\ve}_{2} , {\vec m} ; {\qe}) = {\qe}^{-ch_{0}({\CalH}_{0})}
\, {\sE}\left[  \, \frac{-NN^{*}+N^{*}M}{(1-q_{1}^{-1})(1-q_{2}^{-1})}\, \right] \, , 
\eeq
where
\beq
{\CalH}_{0} = \frac{N}{(1-q_{1})(1-q_{2})}\, , \qquad ch_{0}({\CalH}_0) =  \sum_{\alpha=1}^{n} \frac{a_{\alpha} (a_{\alpha} - {\ve}_{1}- {\ve}_{2})}{2{\ve}_{1}{\ve}_{2}}   + \frac{n ( 1  + Q^2) }{12} \, , 
\eeq
with
\beq
Q^2 = \frac{({\ve}_{1}+{\ve}_{2})^2}{{\ve}_{1}{\ve}_{2}} 
\eeq 
The identity
\beq
\frac{1}{(1-q_{1}^{-1})(1-q_{1}q_{2}^{-1})} + \frac{1}{(1-q_{2}q_{1}^{-1})(1-q_{2}^{-1})} = \frac{1}{(1-q_{1}^{-1})(1-q_{2}^{-1})}
\eeq
implies the first check of \eqref{eq:blowbulk}: 
\beq
Z^{\rm pert}( {\vec a}, {\ve}_{1}, {\ve}_{2} , {\vec m} ; {\qe}) = Z^{\rm pert}( {\vec a}, {\ve}_{1}, {\ve}_{2}-{\ve}_{1} , {\vec m} ; {\qe}) Z^{\rm pert}( {\vec a}, {\ve}_{1}-{\ve}_{2}, {\ve}_{2} , {\vec m} ; {\qe}) 
\eeq

\subsection{ADHM construction for ${\CalN}=2^{*}$ theory}

In \cite{NekBPSCFT} we proposed a generalization of the ADHM construction for instantons
on ${\BR}^{4}$ which geometrizes the instanton counting in ${\CalN}=2$ gauge theory
with massive adjoint hypermultiplet. Namely, in addition to the $(B_1, B_2, I, J)$ matrices
one introduces the operators $Z_{1, 2} \in {\rm End}(K)$\footnote{In \cite{NekBPSCFT} we called them $B_{3}, B_{4}$}, so that the sextuple
$(B_{1}, B_{2}, Z_{1}, Z_{2}, I, J)$ obeys:
\beq
\begin{aligned}
& [ B_{1}, B_{2} ] + IJ + [Z_{1}, Z_{2}]^{\dagger} = 0 \, , \\
& [ B_{1}, Z_{1} ] -  [B_{2}, Z_{2}]^{\dagger} = 0 \, , \\
 & [ B_{1}, Z_{2} ] + [B_{2}, Z_{1}]^{\dagger} = 0 \, , \\
& II^{\dagger} - J^{\dagger} J + \sum_{a=1}^{2} \left( [B_{a}, B_{a}^{\dagger} ] + [Z_{a}, Z_{a}^{\dagger}] \right)  = {\zeta} \cdot 1_{K} \,  , \\
& Z_{1} I + (J Z_{2})^{\dagger} = 0 \, , \qquad Z_{2} I - (J Z_{1})^{\dagger} = 0 \, , \\
\end{aligned}
\label{eq:crossinst}
\eeq
modulo $U(K)$ action $(B_{a}, I, J) \mapsto ( g^{-1} B_{a} g, g^{-1} Z_{a} g, g^{-1} I, J g)$. 

\subsection{ADHM construction for ${\CalN}=2^{*}$ theory on the blowup}

We now propose the  analogous construction for the ${\CalN}=2^{*}$ theory on the blown
up plane ${\widehat{\BC}^{2}}$. The linear algebra data consists now of the maps:
\beq
i : N \to K''\, , \ j: K' \to N\, , \ b_{1,2}: K' \to K''\, , \ b_{3}, b_{4}: K'' \to K'\, , \ b^{'} \in {\rm End}(K^{'})\, , \ b^{''} \in {\rm End}(K^{''})\,  , 
\eeq
as in Fig. $1$ below
\vskip 0.5cm
\centerline{\includegraphics[width=8cm]{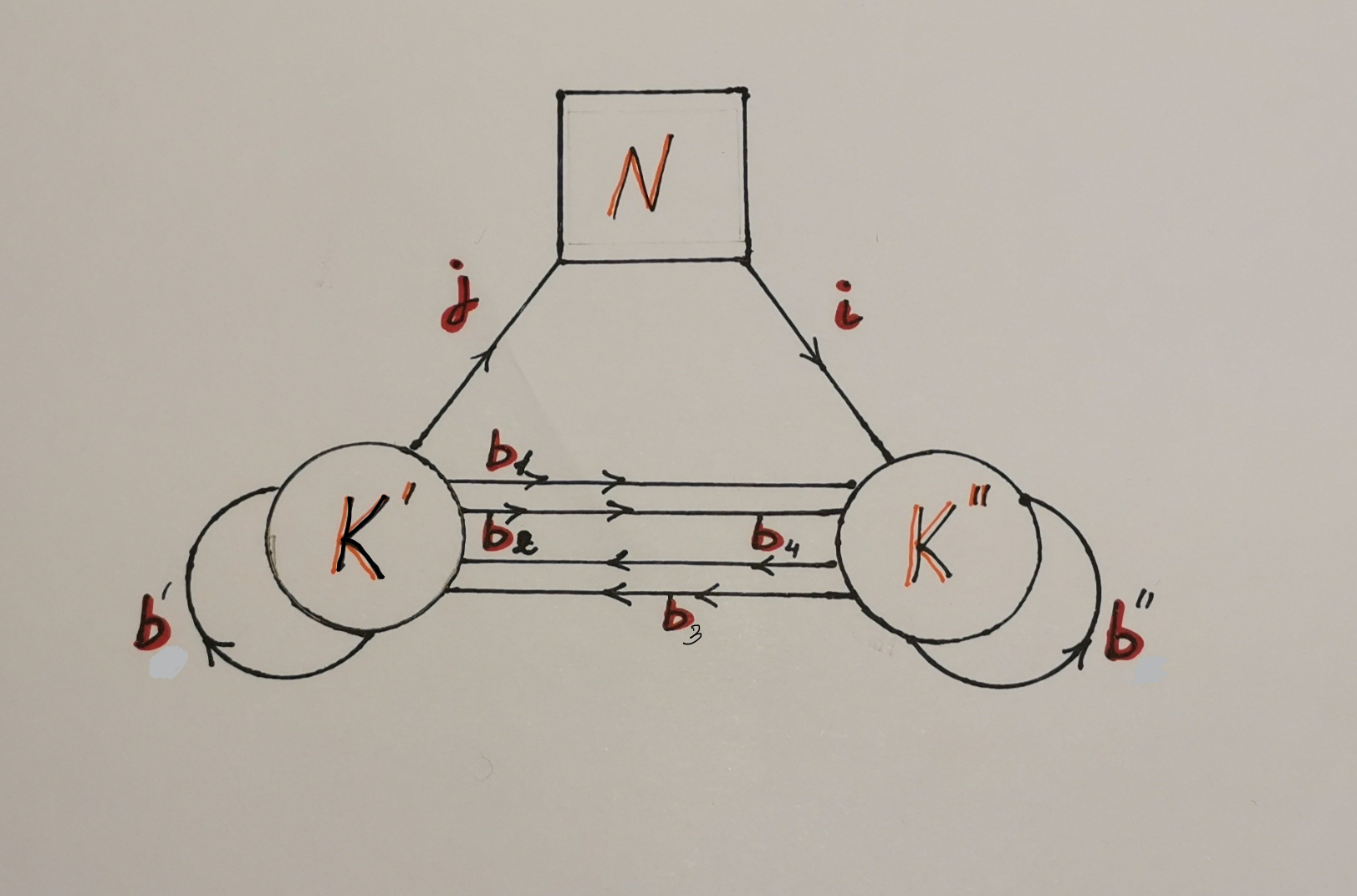}}
\centerline{Fig 1. Quiver description of instantons in ${\CalN}=2^{*}$ theory on ${\widehat{\BC}^{2}}$}
\vskip 0.5cm
while \eqref{eq:crossinst} generalize to 
\beq
\begin{aligned}
& b_{1} b_{3} b_{2} - b_{2}b_{3} b_{1} + \left( b' b_{4} - b_{4} b'' \right)^{\dagger} \, = \, - ij  \, , \\
& b_{1} b_{4} b_{2} - b_{2} b_{4} b_{1} - \left( b' b_{3} - b_{3} b''  \right)^{\dagger} \, = \, 0 \, , \\
& b_{1} b' - b'' b_{1}  -  \left(  b_{3} b_{2} b_{4} - b_{4} b_{2} b_{3} \right)^{\dagger} \, = \, 0 \, , \\
& b_{2} b' - b'' b_{2}  +  \left( b_{3} b_{1} b_{4} - b_{4} b_{1} b_{3} \right)^{\dagger}  \, =\, 0 \, , \\
& m_{\BR}''= ii^{\dagger} + b_{1} b_{1}^{\dagger} + b_{2} b_{2}^{\dagger}  - b_{3}^{\dagger} b_{3} - b_{4}^{\dagger} b_{4} + [b'', b^{''\dagger}] \,  = \, {\zeta}'' \cdot 1_{K''} \,  , \\
&  m_{\BR}' =  - j^{\dagger} j - b_{1}^{\dagger} b_{1} - b_{2}^{\dagger} b_{2}  + b_{3}b_{3}^{\dagger} + b_{4} b_{4}^{\dagger} +  [b', b^{'\dagger} ] \, = \, {\zeta}' \cdot 1_{K'} \,  , \\
& b'' i + (j b_{4})^{\dagger} = 0 \, , \qquad b_{4} i - (j b')^{\dagger}\, = \, 0 \, , \\
\end{aligned}
\label{eq:crossinstblown}
\eeq
The matrices $b_{1}, b_{2}, b_{3}, b_{4}$ and $b^{'}, b^{''}$ correspond, roughly, to the coordinates
$w_{1}, w_{2}, w_{3}, w_{4}$ and $w$ of the K\"ahler quotient construction of the
local Calabi-Yau fourfold ${\CalY} = {\CalZ} \times {\BC}$, with the resolved conifold ${\CalZ}$:
\beq
 (w_{1}, w_{2}, w_{3}, w_{4}, w) \mapsto ( e^{\ii\vartheta} w_{1}, e^{\ii\vartheta} w_{2}, e^{-\ii\vartheta} w_{3}, e^{-\ii\vartheta}w_{4}, w) \, , 
 \label{eq:cy4}
\eeq
where \eqref{eq:u1momcon} is imposed. 
The assignments of spaces $K'$, $K''$ and maps between them can be interpreted as
the cross-product construction of instantons on the quotient ${\BC}^{5}/{\BC}^{\times}$, with the specific choice of ${\BC}^{\times}$-representation in the framing space and the 
special choice of fractional charges, namely, $N$ corresponds to the trivial representation, while $K'$ and $K''$ correspond to the charge $-1$ and charge $0$ representations, respectively. The operators $b_1$, $b_2$ raise the charge by $+1$, as $w_1, w_2$ have charge $+1$ in \eqref{eq:cy4}, $b'$, $b''$ do not change the charge, as $w$ has the charge $0$, and $b_{3}, b_4$ lower the charge by $1$, as both $w_{3}$ and $w_4$ have the charge $-1$. 

The equations \eqref{eq:crossinstblown} have the $SU(2)$ symmetry
\beq
(b_{1}, b_{2}) \mapsto ({\alpha} b_{1} + {\beta}b_{2}, - {\bar\beta} b_{1} + {\bar\alpha} b_{2}), 
\label{eq:su2b1b2}
\eeq 
with $|{\alpha}|^2+ |{\beta}|^2 = 1$. The second $SU(2)$ symmetry
\beq
(b_{3}, b_{4}) \mapsto ({\alpha}' b_{3} + {\beta}'b_{4}, - {\bar\beta}' b_{3} + {\bar\alpha}' b_{4}), 
\label{eq:su2b3b4}
\eeq 
with $|{\alpha}'|^2+ |{\beta}'|^2 = 1$, is broken by the contribution of $ij$.  Geometrically, these equations describe the supersymmetric gauge field configurations on the brane located at the $w_4 = 0$, $w = 0$ surface inside ${\CalY}$, as can be seen from the holomorphic equations 
\beq
\begin{aligned}
& b_{1} b_{3} b_{2} - b_{2}b_{3} b_{1} + ij = 0\, , \qquad b_{1} b_{4} b_{2} - b_{2} b_{4} b_{1}  = 0 \, , \\
& b_{a} b' - b'' b_{a}  = 0 \, , a = 1, 2\, , \qquad b' b_{a} - b_{a} b''  = 0 \, , a = 3, 4\\
 &  b_{3} b_{a} b_{4} - b_{4} b_{a} b_{3}  = 0 \, , a = 1, 2\\
& b'' i  = 0 \, , \quad  j b_{4} = 0 \, , \quad b_{4} i = 0 \, , \quad j b' = 0 \, , \\
\end{aligned}
\label{eq:crossinstblownalg}
\eeq
which, as in \cite{NekBPSCFT} follow from the real (as opposed to ${\BC}$-linear) Eqs. \eqref{eq:crossinstblown}.

In addition, imposing the equations $m_{\BR}^{'} = {\zeta}^{'} \cdot 1_{K^{'}}$, $m_{\BR}^{''} = {\zeta}^{''} \cdot 1_{K^{''}}$ and dividing by $U(K') \times U(K'')$ can 
be replaced by imposing the stability conditions and dividing by the complex group
$GL(K') \times GL(K'')$. For ${\zeta}'> 0$, ${\zeta}''>0$ the stability condition reads:

\vskip 1cm

Any pair $(S', S'')$ of subspaces $S' \subset K'$, $S'' \subset K''$, such that $i(N) \subset S''$, 
$b'(S') \subset S'$, $b''(S'') \subset S''$, $b_{1,2}(S') \subset S''$, $b_{4}, b_{3} (S'') \subset S'$ coincides with $(K', K'')$. A simple consequence of this condition, and
the equations \eqref{eq:crossinstblownalg} is $K'' = {\BC}[b_{1}b_{3}, b_{2}b_{3}] i(N)$, $K' = 
{\BC}[b_{3} b_{1}, b_{3} b_{2} ] b_{3} i (N)$, $b_{4}= 0$, $b'=0$, $b''=0$, therefore
set-theoretically the moduli space of solutions to \eqref{eq:crossinstblown}
coincides with the moduli space of perverse coherent sheaves on 
$\widehat{{\BC}^{2}}$ as constructed in \cite{NY}. 

\subsection{Crossed instantons on the conifold and $qq$-characters on the blowup}

It is now easy to generalize \eqref{eq:crossinstblown} to the case of crossed instantons. It suffices to add another framing space $W$, the maps ${\tilde i}: W \to K'$, ${\tilde j}: K'' \to W$, and modify the Eqs. \eqref{eq:crossinstblown} to:
\beq
\begin{aligned}
& b_{1} b_{3} b_{2} - b_{2}b_{3} b_{1} + \left( b' b_{4} - b_{4} b'' \right)^{\dagger} \, = \, - ij - {\tilde j}^{\dagger} {\tilde i}^{\dagger} \, , \\
& b_{1} b_{4} b_{2} - b_{2} b_{4} b_{1}  - \left( b' b_{3} - b_{3} b''  \right)^{\dagger} \, = \, 0 \, , \\
& b_{1} b' - b'' b_{1}  -  \left(  b_{3} b_{2} b_{4} - b_{4} b_{2} b_{3} \right)^{\dagger} \, = \, 0 \, , \\
& b_{2} b' - b'' b_{2}  +  \left( b_{3} b_{1} b_{4} - b_{4} b_{1} b_{3} \right)^{\dagger}  \, =\, 0 \, , \\
& m_{\BR}''= ii^{\dagger}  + {\tilde i} {\tilde i}^{\dagger} + b_{1} b_{1}^{\dagger} + b_{2} b_{2}^{\dagger}  - b_{3}^{\dagger} b_{3} - b_{4}^{\dagger} b_{4} + [b'', b^{''\dagger}] \,  = \, {\zeta}'' \cdot 1_{K''} \,  , \\
&  m_{\BR}' =  - j^{\dagger} j -{\tilde j}^{\dagger}{\tilde j} - b_{1}^{\dagger} b_{1} - b_{2}^{\dagger} b_{2}  + b_{3}b_{3}^{\dagger} + b_{4} b_{4}^{\dagger} +  [b', b^{'\dagger} ] \, = \, {\zeta}' \cdot 1_{K'} \,  , \\
& b'' i + (j b_{4})^{\dagger} = 0 \, , \qquad b_{4} i - (j b')^{\dagger}\, = \, 0 \, , \\
& b_{3} b_{2} {\tilde i} + ({\tilde j} b_{1})^{\dagger} = 0 \, , \qquad b_{3} b_{1} {\tilde i} - ({\tilde j} b_{2})^{\dagger} = 0 \, , \\
\end{aligned}
\label{eq:crcrossinstblown}
\eeq
which again have the symmetry \eqref{eq:su2b1b2}, and again imply the ${\BC}$-linear equations. 

{}We leave the analysis of the moduli space of solutions to \eqref{eq:crcrossinstblown} to further study. We conjecture an analogue of the compactness theorem of part II in \cite{NekBPSCFT}
holds, opening another possibility for the proof of blowup formulas with and without surface defects.

\section{A little bit of geometry and representation theory}

\subsection{Another face of $G$ and $H$}

The Lie algebra of $G$ (and $H$, of course), can be realized in the space of \emph{densities}
on a (real) line. These are formal expressions $f({\gamma}) d{\gamma}^{-s}$, for some complex
number $s \in {\BC}$. Let us apply the fractional-linear transformation 
\beq
{\gamma} \mapsto \frac{a {\gamma} + b}{c {\gamma} +d}\ , 
\eeq
which acts on $f$ by
\beq
f({\gamma}) d{\gamma}^{-s} \mapsto f \left( \frac{a{\gamma}+b}{c{\gamma}+d} \right) \left( c{\gamma} +d \right)^{2s} d{\gamma}^{-s} \ . 
\label{eq:formalsl2}
\eeq
Now, for complex $s$ the expression $(c {\gamma}+ d)^{2s}$ is not well-defined. However,
for infinitesimal $a \approx 1$, $b \approx 0$, $c \approx 0$, $d \approx 1$ there
is a well-defined branch, which means that the generators $J^{+}, J^{0}, J^{-}$ of $Lie(G)$ 
are well-defined. They act on the coefficient function $f({\gamma})$ via:
\beq
J^{+} = - {\gamma}^{2} {\partial}_{{\gamma}} + 2 s {\gamma}\, , \ J^{0} = {\gamma}{\partial}_{{\gamma}} - s\, , \ J^{-} = {\partial}_{{\gamma}}
\label{eq:sl2}
\eeq
The quadratic Casimir
\beq
{\hat C} = 
J^{-} J^{+} + J^{0} (J^{0} +1) 
\eeq
is a central element, acting by multiplication by $s(s+1)$. 
 
{}The well-known Verma  $Lie(G)$-modules $V_{s}^{\pm}$
 consist of $f$ of the form ${\gamma}^{n} d{\gamma}^{-s}$ and 
${\gamma}^{2s-n} d{\gamma}^{-s}$, with $n \geq 0$, respectively. Another module $M_{s}$ consists of $f$ of the form ${\gamma}^{s+n} d{\gamma}^{-s}$, with $n \in {\BZ}$. 
These three types of $Lie(G)$-representations are characterized by the existence of the vector, annihilated by $J^{\pm}$, or $J^{0}$, respectively. 

Finally, a twisted version of the module $M_{s}$, which we denote by $M_{s}^{a}$, with $s , a \in {\BC}$, consists of the densities of the form ${\gamma}^{s+a+n} d{\gamma}^{-s}$, with the same generators \eqref{eq:sl2}. For special values of $s$ and $a$ these representations are not irreducible. For example, $V_{s}^{+} \subset M_{s}^{-s}$, $V_{s}^{-} \subset M_{s}^{s}$, 

In \cite{NT}, in the higher rank generalizations of our story, 
we encounter all these representations  (see also the Appendix D). 

\subsubsection{Integrable representations}

By restricting $s$ and the type of functions
$f({\gamma})$ one can build the unitary representations of some real forms of $G$. For example, 
for $2s \in {\BZ}_{\geq 0}$,  and $f({\gamma})$ polynomials of degree $2s$, 
we get a spin $s$ representation of $SU(2) \subset G$. The invariant Hermitian product
is given by:
\beq
\langle f, g \rangle = \int {\bar f}({\bar {\gamma}}) d{\bar {\gamma}}^{-s} g({\gamma}) d{\gamma}^{-s} \left( \frac{d{\gamma}d{\bar {\gamma}}}{(1+{\gamma}{\bar{\gamma}})^{2}} \right)^{s+1}
\label{eq:su2r}
\eeq
where the integral is over ${\BC\BP}^{1} = {\BC} \cap {\infty}$. As another example, taking 
$2s + 1 \in {\ii}{\BR}$, and requiring  $f \in L^2 ({\BR})$ makes up a unitary 
$SL(2, {\BR})$ representation with the Hermitian inner product
\beq
\langle f, g \rangle = \int_{\BR} {\bar f}( {\gamma}) d{\gamma}^{-{\bar s}} g({\gamma}) d{\gamma}^{-s} 
\label{eq:sl2r}
\eeq
where we use $s+{\bar s} = - 1$. 

\subsubsection{Invariants and intertwiners}

Much of the theory below is built using the $G$-invariance of the cross-ratio
\beq
\frac{({\gamma}_{2} - {\gamma}_{1})({\gamma}_{3}-{\gamma}_{4})}{({\gamma}_{3}-{\gamma}_{1})({\gamma}_{2}-{\gamma}_{4})}\, , 
\label{eq:crossr}
\eeq
and its infinitesimal version: 
\beq
\frac{d{\gamma}_{1} \otimes d{\gamma}_{2}}{({\gamma}_{1}-{\gamma}_{2})^{2}} \ . 
\label{eq:prime}
\eeq
For example, \eqref{eq:su2r} is built using \eqref{eq:prime} with ${\gamma}_{1} = {\gamma}$, ${\gamma}_{2} = - {\bar {\gamma}}^{-1}$. 

Given three spins $s_{1}, s_{2}, s_{3}$, we build the $Lie(G)$-invariant
\beq
I_{123} = \left( {\gamma}_{1} - {\gamma}_{2} \right)^{s_{1} +s_{2} -s_{3}} \left( {\gamma}_{1} - {\gamma}_{3} \right)^{s_{1} +s_{3} -s_{2}} \left( {\gamma}_{2} - {\gamma}_{3} \right)^{s_{2} +s_{3} -s_{1}} d{\gamma}_{1}^{-s_{1}} d{\gamma}_{2}^{-s_{2}} d{\gamma}_{3}^{-s_{3}}
\label{eq:3inv}
\eeq 
Depending on the relative positions of the three points ${\gamma}_{1}, {\gamma}_{2}, {\gamma}_{3}$, the expression
in \eqref{eq:3inv} can be viewed as an element of the tensor product of three representations of $Lie(G)$. 
For example, in the domain $|{\gamma}_{1}| \ll |{\gamma}_{2}| \ll |{\gamma}_{3}|$ (this condition is $Lie(G)$-invariant, not $G$-invariant, obviously), 
\beq
I_{123} \in V_{s_{1}}^{+} \otimes M_{s_{2}}^{s_{3}-s_{1}} \otimes V_{s_{3}}^{-}
\eeq

 \subsection{Antipodal maps and flipping spins}
 
 The map ${\gamma}_{+} \mapsto {\gamma}_{-}$ we discussed in \eqref{eq:pmtr} is in fact a semiclassical version of the transformation $V_{s}^{\pm} \to V_{-s-1}^{\pm}$
 \beq
 f({\gamma}) d{\gamma}^{-s} \mapsto {\tilde f}({\tilde\gamma}) d{\tilde\gamma}^{s+1} = 
 \oint f({\gamma}) d{\gamma}^{-s} \left( \frac{d{\gamma}d{\tilde\gamma}}{({\gamma}-{\tilde\gamma})^{2}} \right)^{s+1} 
 \label{eq:flipspin}
 \eeq
 In the limit ${\kappa} \to \infty$, $s = {\vt}{\kappa}$ with ${\vt}$ finite, the integral 
\eqref{eq:flipspin} is dominated by a saddle point, giving rise to the canonical transformation
generated by \eqref{eq:flipgf}.   

 \subsubsection{Flipping spins in the $\yt$-coordinates}
 
 For completeness we list the generating functions ${\sigma}_{\xi} ({\yt}, {\yt}_{\xi})$ of the canonical transformations $({\yt}, p_{\yt}) \mapsto ({\yt}_{\xi}, p_{{\yt}_{\xi}})$, for ${\xi} = 0, {\qe}, 1, {\infty}$, corresponding to the changes of one sign ${\vt}_{\xi} \mapsto - {\vt}_{\xi}$ and ${\gamma}_{+, {\xi}} \mapsto {\gamma}_{-,{\xi}}$\footnote{Recall that this is a complexification of the antipodal map of a two dimensional sphere}:
 \beq
 \begin{aligned}
&  {\sigma}_{0}({\yt}, {\yt}_{0}) = {\vt}_{0} \, {\rm log} \left( \frac{{\yt}\, {\yt}_{0}}{  ({\yt} - {\yt}_{0})^2} \right) + \left( {\vt}_{\qe} - {\vt}_{1} + {\vt}_{\infty} \right) \, {\rm log} \left( \frac{\yt}{{\yt}_{0}} \right) \ , \\
&   {\sigma}_{\qe}({\yt}, {\yt}_{\qe}) = {\vt}_{\qe} \, {\rm log} \left( \frac{{\yt} \, {\yt}_{\qe}\,  ({\yt}-1)\, ({\yt}_{\qe}-1)}{({\yt} - {\yt}_{\qe})^{2}} \right) + \left( {\vt}_{0} - {\vt}_{1} + {\vt}_{\infty} \right) \, {\rm log} \left( \frac{({\yt}_{\qe}-1) {\yt}}{({\yt}-1) {\yt}_{\qe}} \right) \, , 
\\
&   {\sigma}_{1}({\yt}, {\yt}_{1}) = {\vt}_{1} \, {\rm log} \left( \frac{(1-{\yt})\, (1-{\yt}_{1})}{({\yt} - {\yt}_{1})^{2}}  \right)  + 
\left( {\vt}_{0} - {\vt}_{\qe} + {\vt}_{\infty} \right) {\rm log} \frac{1-{\yt}_{1}}{1-{\yt}} 
+ 2{\vt}_{\infty} \, {\rm log} \left( \frac{\yt}{{\yt}_{1}} \right) \, , \\
&   {\sigma}_{\infty}({\yt}, {\yt}_{\infty}) = 2 {\vt}_{\infty}\, {\rm log} \left( \frac{{\yt}\, {\yt}_{\infty}}{{\yt}-{\yt}_{\infty}} \right) \ ,\\
\end{aligned}
\eeq
so that
\beq
p_{{\yt}_{\xi}} = - \frac{{\partial} {\sigma}_{\xi}}{{\partial} {\yt}_{\xi}} \, , \qquad p_{\yt} = \frac{{\partial} {\sigma}_{\xi}}{{\partial} {\yt}}\ . 
\label{eq:yyxi}
\eeq

\section{Higher rank generalizations I: orbits and modules}
 
 The orbit ${\CalO}_{\vt}$ of $G$ has two obvious generalizations to the higher rank case $G_{r} = SL(r+1, {\BC})$ with $r>1$.

 \subsection{Minimal orbit ${\CalO}_{\tt min}^{r, {\nu}}$} 
 
 Let ${\CalV} = {\BC}^{r+1}$ be the defining representation of $G_{r}$. 
 The first generalization of ${\CalO}_{\vt}$, 
 which we shall call ${\CalO}_{\tt min}^{r, {\nu}}$  (the subscript {\tt min} will be explained later), consists of the pairs
 $(u,v)$, with $v \in {\CalV}$, $u \in {\CalV}^{*}$ such that 
 \beq
 u (v) = {\nu} \in {\BC} \ , 
 \label{eq:moma1}
 \eeq
 modulo the ${\BC}^{\times}$-action
 \beq
 (u, \ v ) \to ( t\, v , \ t^{-1}\, u ) \ , \qquad t \in {\BC}^{\times}
 \label{eq:tora1}
 \eeq
 The $u-v$ description allows for a very simple description of the holomorphic symplectic forms
 on ${\CalO}_{\tt min}^{r, {\nu}}$. It is obtained from the
 $2$-form 
 \beq
 du\, (^{\wedge} dv) := \sum_{a=1}^{r+1} du_{a} \wedge dv^a
 \eeq
 by the symplectic reduction with respect to the  ${\BC}^{\times}$-action \eqref{eq:tora1},  for which \eqref{eq:moma1} is the moment map equation. The space ${\CalO}_{\tt min}^{r, {\nu}}$ is an orbit of $G_r$, which acts in the obvious way: $g: (u , \ v) \mapsto ( u g^{-1} , \ g v )$ for $g \in G_r$. This action is generated by 
the moment map ${\mu}_{\tt min}: \, {\CalO}_{\tt min}^{r, {\nu}} \longrightarrow Lie(G_{r})^{*}$ which reads as follows:
 \beq
 {\mu}_{\tt min} (u, v) = {\scriptstyle{\frac{\nu}{r+1}}}\, {\bf 1}_{\CalV} - v \otimes u 
 \label{eq:mumin}
 \eeq 
 The map \eqref{eq:mumin} makes ${\CalO}_{\tt min}^{r, {\nu}}$ a coadjoint orbit.  
 For ${\nu} \neq 0$ we have two maps $p_{\pm}: {\CalO}_{\tt min}^{r, {\nu}} \to {\BC\BP}^{r}$, generalizing the $p_{\pm}$ maps \eqref{eq:ppm}  of the $r=1$ case, $p_{-}(u,v) = [u] \in {\BP}({\CalV}^{*})$, $p_{+}(u,v) = [v] \in {\BP}({\CalV})$. In fact, we can identify:
 \beq
 {\CalO}_{\tt min}^{r, {\nu}} \approx {\BP} ({\CalV} ) \times {\BP}({\CalV}^{*}) \backslash {\CalC}
 \eeq
 where ${\CalC}$ is the incidence subvariety, consisting of the pairs $(h, l)$, $l \subset {\CalV}$, 
 ${\rm dim}(l) = 1$, $h \subset {\CalV}$, ${\rm codim}(h) = 1$, such that $l \subset h$.
  
 {}For $\nu = 0$ one supplements the equation $u(v) = 0$ with one of the two stability conditions: either $u \neq 0$, or $v \neq 0$. For either stability condition
 the orbit ${\CalO}_{0}^{r}$ is isomorphic to $T^{*} {\BC\BP}^{r}$.
 
 \subsubsection{$Lie(G_r)$ module $M_{\nu}$}
 
 The following $Lie(G_r)$-module (representation) is associated to ${\CalO}_{\tt min}^{{\nu},r}$ \footnote{The precise meaning of the word ``associated'' is worth another paper. Prospective graduate students seeking a project on topological strings are welcome to apply their effort}. Consider the space of formal expressions
 \beq
 f( u_{1}, u_{2} , \ldots , u_{r+1} ) = ( u_{1}\ldots u_{r+1} )^{\frac{\nu}{r+1}} {\psi} \left( u_{1}/u_{r+1} , \ldots , u_{r}/u_{r+1} \right)
 \label{eq:fnu}
 \eeq
As in the discussion around \eqref{eq:formalsl2} we apply to \eqref{eq:fnu} an $G_{r}$-transformation
\beq
a: ( u_{i} ) \mapsto \left( \sum_{j=1}^{r+1} a_{ij} u_{j} \right) \, , \ {\rm Det}(a) = 1
\eeq
acting by: ${\psi} \mapsto {\psi}^{a}$
\beq
{\psi}^{a} ({\gamma}) = {\psi} \left( \frac{a_{i1} {\gamma}_{1} + \ldots + a_{i,r} {\gamma}_{r} + a_{i,r+1}}{a_{r+1, 1} {\gamma}_{1} + \ldots + a_{r+1, r} {\gamma}_{r} + a_{r+1, r+1}} \right) \prod\limits_{i=1}^{r+1}  \left( a_{i1} {\gamma}_{1}/{\gamma}_{i} + \ldots + a_{i,r} {\gamma}_{r}/{\gamma}_{i} + a_{i,r+1} \right)^{\frac{\nu}{r+1}}
\label{eq:fagamma}
\eeq
where ${\gamma}_{i} = u_{i}/u_{r+1}$, ${\gamma}_{r+1} = 1$. Again, \eqref{eq:fagamma} is not well-defined for general $\nu$, for general $a \in G_r$. Again, for infinitesimal $a_{ij} = {\delta}_{ij} + {\ve} {\xi}_{ij}$, ${\ve} \to 0$ we have a preferred branch, leading to the action of $Lie(G_r)$ via first order differential operators 
\beq
J_{\xi} {\psi} = \sum_{i,j = 1}^{r+1}  {\xi}_{ij} u_{i}  \left( \frac{\partial}{\partial u_{j}} {\psi} + \frac{\nu {\psi}}{(r+1) u_{j}}   \right)  
\label{eq:Jacmin}
\eeq
where $\xi$ is a traceless $r+1 \times r+1$ matrix. 
The representation \eqref{eq:Jacmin}
is realized using the minimal number $r$ of degrees of freedom, hence the term ``minimal''. The generators
\eqref{eq:Jacmin} correspond to \eqref{eq:mumin} under the naive quantization rule:
\beq
v_{i} \mapsto \frac{\partial}{\partial u_{i}}\, , \qquad ( - {\mu}_{\tt min} )_{ij} \mapsto J_{ij} 
\eeq
with the ordering prescription which pushes $v$'s to the right of $u$.  In comparing with the formulae \eqref{eq:sl2}, with $s = {\nu}/2$,  one should remember the normalization $f ({\gamma} ) = {\gamma}^s {\psi}({\gamma})$.  
The representation $M_{\nu}$ is important in applications, as its zero-weight subspace $M_{\nu}[0]$, i.e. the subspace of vectors, annihilated by the generators, corresponding to the diagonal matrices $J_{E_{i,i} - E_{i+1,i+1}}$, is one-dimensional. 

\subsection{Regular orbit}

 The second generalization, which we shall call ${\CalO}_{\tt reg}^{{\vec\nu},r}$ (the subscript {\tt reg} will be explained later), with ${\vec\nu} = ({\nu}_{1}, \ldots , {\nu}_{r}) \in {\BC}^{r}$, 
 is the quotient of the space of $r$-tuples $({\bu}, {\bv}) = (u^{i}, v_{i})_{i=1}^{r}$, $v_{i} \in {\CalV}$, $u^{i} \in {\CalV}^{*}$, obeing 
 \beq
 u^{i}(v_{j}) - {\nu}_{i} \, {\delta}_{j}^{i} = 0 \ , 
 \label{eq:cons2}
 \eeq
 modulo the $({\BC}^{\times})^{r}$ symmetry, acting by 
 \beq
  (u^{i}, v_{i})_{i=1}^{r} \mapsto (t_{i} u^{i}, t_{i}^{-1} v_{i})_{i=1}^{r} ,  \qquad
  t_{i} \in {\BC}^{\times} \ . 
 \label{eq:tora2} 
 \eeq
 Note that $({\bu}, {\bv})$ defines the $r+1$'st pair $(u^{0} \in {\CalV}^{*}, v_{0} \in {\CalV})$,
 by $u^{i}(v_{0}) = 0$, $u^{0}(v_{i}) = 0$ for all $i = 1, \ldots, r$.
 
 {}We would like 
 to have a symplectic description of ${\CalO}_{\tt reg}^{{\vec\nu},r}$. 
 Let us start with the symplectic form
 \beq
d {\bu} \left(^{\wedge} d{\bv} \right) \, = \,  \sum_{i=1}^{r} du^{i}\, (^{\wedge} dv_{i})
 : = \sum_{i=1}^{r} \sum_{a=1}^{r+1} du_{a}^{i} \wedge dv^a_{i}
 \eeq  
 on the space of all $r$-tuples $({\bu}, {\bv})$. We would like to interpret 
 \eqref{eq:cons2}
 and taking the quotient \eqref{eq:tora2} as the symplectic quotient. The fact that the Poisson brackets of the equations \eqref{eq:cons2} are non-zero, for ${\vec\nu} \neq 0$,
 on the surface of solutions to \eqref{eq:cons2}, is an obstacle. However, there is a simple well-known fix for that. 
 Consider the group $B_{r}$ of invertible upper-triangular $r \times r$ matrices 
 $\Vert b_{i}^{j} \Vert_{i,j=1}^{r}$, i.e. $b_{i}^{j} = 0$ for $i < j$\footnote{$
 \sum\limits_{j \leq k \leq i}
 b_{k}^{j} \left( b^{-1} \right)^{k}_{i} = {\delta}^{i}_{j}$}, acting on the space
 of $r$-tuples: 
 \beq
 b: ( {\bu}, {\bv}) \mapsto \left( {\bu}^{b}, {\bv}^{b} \right) = \left( \sum_{k\geq i} 
 b_{k}^{i}
 u^{k}, \sum_{k \leq i} \left( b^{-1} \right)_{i}^{k} v_{k} \right)_{i=1}^{r}
 \eeq
 The $B_r$-action is generated by the upper-triangular matrices-valued moment map 
 \beq
 {\mu}_{\tt B} = \Vert u^{i}(v_{j}) \Vert_{i \geq j}\ . 
 \eeq
 Now, since the space of diagonal matrices
 is preserved by the conjugation by the upper-triangular matrices (in other words, the 
 eigenvalues of an upper-triangular matrix are read off its diagonal), we can perform
 the Hamiltonian reduction  at the non-zero level of ${\mu}_{\tt B}$ set to be a diagonal 
 matrix ${\bnu} = {\rm diag} ({\nu}_{1} , \ldots , {\nu}_{r})$, meaning we impose the equations
 \beq
 u^{i}(v_{j} ) - {\nu}_{i}\ {\delta}_{j}^{i}\, = 0 \, , \qquad r \geq  i \geq j \geq 1
 \eeq 
 We can define two more complex lines: ${\BC} u^{r+1} \subset {\CalV}^{*}$ and
 ${\BC} v_{r+1} \subset  {\CalV}$, by:
 \beq
 u^{i} (v_{r+1}) = 0\, , \ u^{r+1} (v_{i}) = 0 \, , \ i = 1, \ldots , r
 \label{eq:extra}
 \eeq
{}The group $B_r$ preserves two flags:
 \beq
 \begin{aligned}
& 0 = {\CalV}_{0} \subset {\CalV}_{1} \subset \ldots \subset {\CalV}_{r} \subset {\CalV}
\ , \\
& \qquad\qquad {\CalV}_{i} \ =\ \sum_{k \leq i} \, {\BC}\,v_{k} \, , \ {\rm and} \\
& 0 = {\CalU}^{0} \subset {\CalU}^{1} \subset \ldots \subset {\CalU}^{r} 
\subset {\CalV}^{*} 
\ , \\
& \qquad\qquad {\CalU}^{i} \ =\ \sum_{k \geq i} \, {\BC}\, u^{k} \, . 
\end{aligned}
\label{eq:twoflags}
\eeq
 These flags are compatible in the sense that ${\CalU}^{i+1} ({\CalV}_{i} ) = 0$, ${\tt L}_{i} = {\rm ker}\ {\CalU}^{i} / {\CalV}_{i-1}$
 is one-dimensional,  spanned by $v_i$, for $i = 1, \ldots , r+1$, as well as
 ${\underline{\tt L}}^{i} = {\CalU}^{i}/\, {\rm ker}\ {\CalV}_{i-1}$ is one-dimensional, spanned by $u^i$, for $i = 1, \ldots , r+1$. Thus,\footnote{If the collection
 of $r$ pairs $(u^{i}, v_{i})_{i=1}^{r}$ of vectors and covectors obeys
 $u^{i}(v_{j}) = $ for $i > j$ and $u^{i}(v_{i}) = {\nu}_{i}$, then by the strictly 
 upper-triangular transformation:
 \beq
 \begin{aligned}
& u^{i} \mapsto u^{i} + \sum\limits_{k > i} {\xi}^{i}_{k}\, u^{k} \, , \\
& v_{i} \mapsto v_{i} - \sum\limits_{i > k} {\tilde\xi}^{k}_{i}\, v_{k}
 \end{aligned}
 \eeq 
 with ${\xi} - {\tilde \xi} - {\xi}{\tilde \xi} = 0$, 
 we can achieve $u^{i}(v_{j}) = 0$ for $i < j$. Indeed, 
 \beq
 u^{i}(v_{j}) \mapsto  u^{i}(v_{j}) + \sum\limits_{j \geq k > i} {\xi}^{i}_{k}\, u^{k}(v_{j})
 -  \sum\limits_{j > k \geq i} {\tilde\xi}^{k}_{j}\, u^{i} (v_{k} ) 
 - \sum\limits_{j > l \geq k > i } {\xi}^{i}_{k}{\tilde\xi}^{l}_{j} \, u^{k}(v_{l}) = 0\, , 
 \eeq
  is solved recursively:
  \beq
  \begin{aligned}
&   {\xi}^{i}_{i+1} = {\tilde\xi}^{i}_{i+1} = \frac{u^{i}(v_{i+1})}{{\nu}_{i} -  {\nu}_{i+1}} \, , \ i = 1, \ldots r -1 \\
& {\tilde\xi}^{i}_{i+2} = {\xi}^{i}_{i+2} - {\xi}^{i}_{i+1} {\xi}^{i+1}_{i+2} = \frac{u^{i}(v_{i+2}) - \frac{1}{{\nu}_{i+1}-{\nu}_{i+2}}
u^{i}(v_{i+1}) u^{i+1}(v_{i+2}) }{{\nu}_{i} -{\nu}_{i+2}}\, , \ i=1, \ldots , r-2 
\\
& \qquad \ldots \qquad \ , 
\end{aligned}
\eeq
assuming ${\nu}_{i} \neq {\nu}_{j}$, $i \neq j$.} 
 \beq
 {\CalO}_{\tt reg}^{{\vec\nu},r} = {\mu}_{\tt B}^{-1}({\bnu})/B_{r}
 \label{eq:borel}
 \eeq
 The moment map 
 ${\mu}: \, {\CalO}_{\tt reg}^{{\vec\nu},r}\longrightarrow Lie(G_{r})^{*}$
  is given by:
 \begin{multline}
 {\mu}_{\tt reg} ({\bu}, {\bv}) ={\vt}_{r+1} \, {\bf 1}_{\CalV} - \sum_{i=1}^{r}
 v_{i} \otimes u^{i} = 
 \sum_{i=1}^{r+1} \ {\vt}_{i} \, v_{i} \otimes v^{i\vee}\, , \\
 {\vt}_{i} =  {\vt}_{r+1} - {\nu}_{i} \, , \ i = 1, \ldots , r \, , {\vt}_{r+1} = \frac{{\nu}_{1}+\ldots + {\nu}_{r}}{r+1} \\
 v^{i\vee} (v_{j} ) = {\delta}_{j}^{i} \, , \quad i,j = 1, \ldots , r+1 
 \end{multline}  
 Another well-known presentation of ${\CalO}_{\tt reg}^{{\vec\nu},r}$ is the $A_r$-type quiver variety. 
 Namely, consider the sequence of complex vector spaces ${\bV}_{1}, {\bV}_{2}, \ldots , {\bV}_{r}$, such that ${\rm dim}_{\BC} {\bV}_{i} = i$, and linear maps:
 \beq
 B_{i} : {\bV}_{i} \to {\bV}_{i+1}\, , \ {\tilde B}_{i}: {\bV}_{i+1} \to {\bV}_{i}
 \eeq
 obeying:
 \beq
 {\tilde B}_{i} B_{i} - B_{i-1} {\tilde B}_{i-1} = {\zeta}_{i} \cdot {\bf 1}_{{\bV}_{i}}\, , \ i = 1, \ldots , r\, , 
 \label{eq:mombibti}
 \eeq
 where ${\zeta}_{i} \in {\BC}$, ${\bV}_{r+1} = {\CalV}$, and we identify:
 \beq
 \left( B_{i}, {\tilde B}_{i} \right)_{i=1}^{r}\, \sim \, \left( g_{i+1}^{-1} B_{i} g_{i}, g_{i}^{-1} {\tilde B}_{i} g_{i+1} \right)_{i=1}^{r}\, , \qquad \left( g_{i} \right)_{i=1}^{r} \in \times_{i=1}^{r}  GL({\bV}_{i})
 \label{eq:gibibti}
\eeq 
The previous description is recovered quite simply:
 \beq
 {\mu}_{\tt reg} ({\bu}, {\bv}) ={\vt}_{r+1} \, {\bf 1}_{\CalV} - B_{r} {\tilde B}_{r}
\eeq 
so that $v_{r+1} = {\rm ker}{\tilde B}_{r} \in {\CalV}$, $u^{r+1} = {\rm im}B_{r}^{\perp} \in {\CalV}^{*}$, and assuming ${\zeta}_{i} \neq 0$ for all $i = 1, \ldots , r$, 
\beq
\begin{aligned}
& {\BC} v_{1} = B_{r}B_{r-1} \ldots B_{1} {\psi}_{0} \, , \\
& \qquad\qquad\qquad 0 \neq {\psi}_{0} \in {\bV}_{1}\\
& {\BC} v_{i} = B_{r} B_{r-1} \ldots B_{i} {\psi}_{i-1}\, , \, \\
& \qquad\qquad\qquad 0 \neq {\psi}_{i-1} \in {\bV}_{i}, \ {\tilde B}_{i-1} {\psi}_{i-1} = 0\, , i = 2, \ldots , r \\
\end{aligned}
\eeq
Analogously, 
\beq
\begin{aligned}
& {\BC} u^{1} = {\chi}_{0} {\tilde B}_{1}{\tilde B}_{2} \ldots {\tilde B}_{r} \, , \\
& \qquad\qquad\qquad  0 \neq {\chi}_{0} \in {\bV}_{1}^{*}\\
& {\BC} u^{i} = {\chi}_{i-1} {\tilde B}_{i} {\tilde B}_{i+1} \ldots {\tilde B}_{r}\, , \\
& \qquad\qquad\qquad 0 \neq {\chi}_{i-1} \in {\bV}_{i}^{*}, \ {\chi}_{i-1}B_{i-1} = 0\, , i = 2, \ldots , r \\
\end{aligned}
\eeq
so that
\beq
 {\mu}_{\tt reg} ({\bu}, {\bv}) v_{i} = 
 {\vt}_{i} v_{i} \, , \ u^{i} {\mu}_{\tt reg} ({\bu}, {\bv}) =  {\vt}_{i} u^{i}
\eeq
with
\beq
{\vt}_{i} = {\vt}_{r+1} - {\zeta}_{r} - \ldots - {\zeta}_{i} \ .
\eeq
The maps $p_{+}: {\CalO}_{\tt reg}^{{\vec\nu},r} \to Fl( 1, 2, 3, \ldots , r , {\CalV})$, $p_{-}: {\CalO}_{\tt reg}^{{\vec\nu},r} \to Fl( 1, 2, 3, \ldots , r , {\CalV}^{*})$, generalizing \eqref{eq:ppm}, send the set
$( B_i, {\tilde B}_{i} )_{i=1}^{r}$ obeying \eqref{eq:mombibti} modulo \eqref{eq:gibibti}, to
the flags
\beq
V_{1}  = B_{r}B_{r-1} \ldots B_{1} \, {\bV}_{1}\, \subset 
V_{2} = B_{r}B_{r-1}\ldots B_{2} \, {\bV}_{2} \, \subset  \ldots \, \subset V_{r} = B_{r}\, {\bV}_{r}   \subset {\CalV}
\eeq
and
\beq
U_{1} =  {\bV}_{1}^{*} \, {\tilde B}_{1}{\tilde B}_{2} \ldots {\tilde B}_{r}  \subset U_{2} = {\bV}_{2}^{*} \, {\tilde B}_{2} \ldots {\tilde B}_{r}  \subset \ldots \subset U_{r} =  {\bV}_{r}^{*}\, {\tilde B}_{r} \subset {\CalV}^{*}
\eeq

\subsection{Verma module}

To the regular orbit ${\CalO}_{\tt reg}^{{\vec \nu}, r}$ one associates the highest weight Verma module $V_{\vec\nu}^{+}$ (and, dually, the lowest weight Verma module $V_{\vec\nu}^{-}$), which is generated by the vector $v_{\vec\nu}^{\pm}$, annihilated
by the $Lie(G_r)$ generators $J_{\xi}$ corresponding to the strictly upper (lower)-triangular matrices, which is the eigenvector of the diagonal traceless matrices $J_{E_{i,i} - E_{i+1,i+1}} v_{\vec\nu} = {\nu}_{i} v_{\vec\nu}$, $i = 1, \ldots ,r$. 

\section{KZ, BPZ and Okamoto}

For completeness, we remind some basic details about the $\mathfrak{sl}_{2}$ Knizhnik-Zamolodchikov equation. We also explain that Okamoto transformations are the classical limit of the KZ-BPZ
correspondence \cite{FZ}.

\subsection{Conformal blocks}

The genus $0$, $p+3$-point conformal block in the theory with level $k$ 
$\widehat{\mathfrak{sl}}_{2}$ current algebra is a vector 
\beq
{\bf\Psi} \in \left( \bigotimes\limits_{i=-1}^{p+1} V_{s_{i}}^{\gamma}\otimes V_{{\Delta}_{i}}^{z} \right)^{Lie(G) \times Lie(H)}
\label{eq:conbl}
\eeq
where
$V^{\gamma}_{s}$ stands for $Lie(G)$-representation by differential operators \eqref{eq:sl2} in the $\gamma$-variable, and $V^{z}_{\Delta}$ stands for the $Lie(H)$-representation by differential
operators \eqref{eq:sl2} in the $z$-variable with $s = \Delta$, the conformal dimension. 
The conformal dimension $\Delta$ is related to $s$ via the celebrated \cite{KZ} relation
\beq
{\Delta}_{i} = \frac{s_{i}(s_{i}+1)}{\kappa}\, , \ i = -1, \ldots, p+1
\label{eq:confdim}
\eeq
with $\kappa = k +2$. 
In the analysis below we don't specify what kind of representations we mean by $V_{s}$, what kind of completion is used in the tensor product \eqref{eq:conbl}. We want ${\bf\Psi}$ to be both $Lie(H)$ and $Lie(G)$-invariant. 

Using \eqref{eq:prime} and \eqref{eq:crossr}
we build the formal invariants
\beq
{\bf\Psi} = \prod_{-1 \leq i<j \leq p+1} \left\{ \left( \frac{dz_{i} dz_{j}}{z_{ij}^{2}} \right)^{D_{ij}} 
 \left( \frac{d{\gamma}_{i} d{\gamma}_{j}}{{\gamma}_{ij}^{2}} \right)^{S_{ij}} \right\} \, 
\times \, {\Psi} \left( {\bf y}; \, {\bf \qe} \right) 
\label{eq:wzwinv}
\eeq
with ${\bf y} = ( y_{a} )_{a=1}^{p}$, ${\bf \qe} = ( {\qe}_{a} )_{a=1}^{p}$, 
\beq
{\gamma}_{ij} = {\gamma}_{i} - {\gamma}_{j} \, , \qquad  
y_{a} = \frac{{\gamma}_{a,-1} {\gamma}_{0,p+1}}{{\gamma}_{0,-1} {\gamma}_{a,p+1}} 
\eeq
and similarly 
\beq
z_{ij} \equiv z_{i,j} = z_{i} - z_{j} \, , \qquad {\qe}_{a} = \frac{z_{a,-1} z_{0,p+1}}{z_{0,-1} z_{a,p+1}}
\eeq 
provided the symmetric matrices $S = S^{t}$, $D = D^{t}$, 
obey
\beq
\sum_{j \neq i}\, S_{ij} = - s_{i} \, , \ \sum_{j \neq i}\, D_{ij} = - {\Delta}_{i}
\label{eq:sdel}
\eeq
We have much freedom in the choice of these symmetric matrices\footnote{For example,  kinematic invariants $S_{ij} = {\vec p}_{i} \cdot {\vec p}_{j}$ of a set of $p+3$ momenta ${\vec p}_{i} \in {\BC}^{D}$ in a $D\geq p+3$ dimensional scattering problem, obeying
\beq
{\vec p}_{i} \cdot {\vec p}_{i}\ = \ s_{i} \, , \qquad \sum_{i} {\vec p}_{i} = 0 \ .
\eeq
solve \eqref{eq:sdel}. Similarly, for the $\Vert D_{ij} \Vert$ matrix.} 

Single out three points, e.g. $z_{-1}, z_{0}$ and $z_{p+1}$, then a possible choice of $D_{ij}$ is
\begin{multline}
D = - {\half} {\Delta} \left( e_{0} \otimes e_{-1}^{t} + e_{-1} \otimes e_{0}^{t} \right) + \\
+ \left\{  - {\bf\Delta} + {\half} {\Delta} ( e_{0} + e_{-1} )  \right\} \otimes e_{p+1}^{t} - e_{p+1} \otimes \left\{  - {\bf\Delta} + {\half} {\Delta} ( e_{0} + e_{-1} )  \right\}^{t}  \, , \\
{\rm where} \qquad {\bf\Delta} = \sum_{i=-1}^{p} {\Delta}_{i} \, e_{i} \, , \qquad {\Delta} = - {\Delta}_{p+1} + \sum_{i=1}^{p} {\Delta}_{i} \ . 
\label{eq:choicedij}
\end{multline}
\subsection{Knizhnik-Zamolodchikov connection}

The KZ equation (in genus zero) is 
\beq
{\nabla} {\bf\Psi} = 0
\label{eq:kze1}
\eeq where ${\nabla}$ is a (projective) flat connection
over ${\CalM}_{0,p+3}$ on the infinite rank vector bundle with the fibers ${\CalH} = (V_{s_{-1}}^{\gamma} \otimes \ldots \otimes V_{s_{p+1}}^{\gamma} )^{Lie(G)}$, explicitly:
\beq
{\nabla}_{i} = {\kappa} \frac{\partial}{{\partial} z_{i}} + \sum_{j \neq i} \frac{J^{+}_{i}J^{-}_{j} + J^{+}_{j} J^{-}_{i} + 2 J^{0}_{i} J^{0}_{j} }{z_{i}- z_{j}}  \, , \ i = -1, \ldots , p+1
\label{eq:kze}\eeq
The equations \eqref{eq:kze} are compatible, $[{\nabla}_{i}, {\nabla}_{j}] = 0$, for any value of $\kappa$, which means that the two sets of equations, cf. \eqref{eq:comm}, \eqref{eq:poissh} hold:
\beq
{\partial}_{z_{i}} {\hat H}_{j} - {\partial}_{z_{j}} {\hat H}_{i} = 0 \, , \qquad [ {\hat H}_{i}, {\hat H}_{j} ] = 0 
\label{eq:kzflat}\eeq
with
\beq
{\hat H}_{i} \, = \, \sum_{j\neq i} \frac{1}{z_{ij}} \left( {\gamma}_{ij}^{2} 
{\partial}^{2}_{{\gamma}_{i}{\gamma}_{j}} - 2 {\gamma}_{ij}(s_{i} {\partial}_{{\gamma}_{j}} - s_{j} {\partial}_{{\gamma}_{i}}) - 2 s_{i}s_{j} \right) \ .
\eeq
The Eqs. \eqref{eq:kzflat} follow from the celebrated relations \cite{ArnoldBr} 
\beq
d {\rm log}(z_{i}-z_{j}) \wedge d {\rm log} (z_{i}-z_{k} ) + d{\rm log} (z_{j}-z_{i}) \wedge d{\rm log} (z_{j}-z_{k}) + d{\rm log}(z_{k}-z_{i}) \wedge d {\rm log}(z_{k}-z_{j} ) = 0 
\eeq

The general higher genus generalization of the KZ connection  \cite{Bernard, Axelrod, Losev:1991gn} and its quasiclassical limits $\kappa \to \infty$ \cite{Ivanov:1994ab, Ivanov:1996bg} are an involved story.

\subsubsection{Degenerate cases}

Let us take $p=2$, with $s_{3} = \frac 12$. The algebra $\mathfrak{sl}_{2}$ has a two-dimensional representation $W$, which is a subrepresentation of $V_{\frac 12}^{\gamma}$:
\beq
W = {\BC} d{\gamma}^{-\frac 12}  \oplus {\BC}  {\gamma} d{\gamma}^{-\frac 12} 
\eeq
So we restrict the conformal block ${\Psi}$ onto this subrepresentation. In other words we assume the linear dependence of the conformal block ${\bf\Psi}d{\gamma}_{3}^{-\frac 12}\, dz_{3}^{-\frac{3}{4{\kappa}}}\, \prod_{i=-1}^{2} d{\gamma}_{i}^{-s_{i}} dz_{i}^{-{\Delta}_{i}}$ on  ${\gamma}_3$:
\begin{multline}
{\bf\Psi} = \left( \frac{{\gamma}_{3,2}}{{\gamma}_{1,2}}  {\Psi}_{0} (y, z ; {\qe}) 
+ \frac{{\gamma}_{3,-1}}{{\gamma}_{1,-1}} {\Psi}_{1} (y, z ; {\qe})  \right) \,  \times \, z_{2,1}^{2{\Delta}_{2}}{\gamma}_{2,1}^{2s_{2}} \, z_{0,-1}^{2{\Delta}_{-1}}{\gamma}_{0,-1}^{2s_{-1}}\\
\left( \frac{z_{3,2}}{z_{2,1}} \right)^{\frac{3}{2{\kappa}}} \,
 \left( \frac{z_{1,0} z_{1,-1}}{z_{0,-1}} \right)^{\frac{3}{4{\kappa}}+{\Delta}_{1}-{\Delta}_{2}} \, 
\times \, \left(
\frac{{\gamma}_{1,0} {\gamma}_{1,-1}}{{\gamma}_{0,-1}} \right)^{s_{1}-s_{2} + \half}\left( \frac{z_{1,0}z_{0,-1}}{z_{1,-1}} \right)^{{\Delta}_{0}-{\Delta}_{-1}} \left( \frac{{\gamma}_{1,0}{\gamma}_{0,-1}}{{\gamma}_{1,-1}} \right)^{s_{0}-s_{-1}}\ ,
\label{eq:5wzw}
\end{multline}
where we recall \eqref{eq:confdim}, and 
\beq
y = \frac{{\gamma}_{0,-1} {\gamma}_{1,2}}{{\gamma}_{1,-1} {\gamma}_{0,2}} \, , \quad  {\qe} = \frac{z_{0,-1} z_{1,2}}{z_{1,-1} z_{0,2}} \, , \quad z = \frac{z_{3,-1} z_{1,2}}{z_{1,-1} z_{3,2}} 
 \eeq
 are the $G \times H$ invariants (the remaining invariant, 
 ${\zeta} = ({\gamma}_{3,-1}/{\gamma}_{3,2})({\gamma}_{1,2}/{\gamma}_{1,-1})$, enters only linearly). 
 The $G \times H$ invariance allows to set ${\gamma}_{-1}=0, {\gamma}_{1}=1, {\gamma}_{2}= \infty$, $z_{-1}= 0, z_{1}=1, z_{2} = \infty$, giving us the equation
 \beq
 {\kappa} \frac{\partial}{\partial z} {\bf\Psi} = \left( \frac{{\hat A}_{0}}{z} + \frac{{\hat A}_{\qe}}{z-{\qe}} + \frac{{\hat A}_{1}}{z - 1} \right) {\bf\Psi}
 \eeq
 with ${\bf\Psi} = \left( \begin{matrix} {\Psi}_{0} \\ {\Psi}_{1}  \end{matrix} \right)$, 
 and ${\hat A}_{z_{i}}$ are two by two matrices of differential operators in the $y$-variable. The generalization to arbitrary $p$ is straightforward. 
  
 \subsubsection{From KZ to BPZ in $\yt$ variables}
 
 Let us consider the accompanying case $p=1$, with the same spins $s_{-1}, s_{0}, s_{1}, s_{2}$, as above. The conformal block ${\bf\Psi}\prod_{i=-1}^{2} d{\gamma}_{i}^{-s_{i}} dz_{i}^{-{\Delta}_{i}}$ can be now written as:
\begin{multline}
{\bf\Psi} ({\gamma}_{-1}, {\gamma}_{0}, {\gamma}_{1}, {\gamma}_{2} ; z_{-1}, z_{0}, z_{1}, z_{2} )  = {\yt}^{a}({\yt}-1)^{b} ({\yt}-{\qe})^{c} \, {\psi} ({\yt} ; {\qe}) \, \times \, z_{1,2}^{2{\Delta}_{2}} z_{0,-1}^{2{\Delta}_{0}}\,
{\gamma}_{1,2}^{2s_{2}} {\gamma}_{0,-1}^{2s_{0}} \\ \,
 \, \left( \frac{z_{1,0} z_{1,-1}}{z_{0,-1}} \right)^{{\Delta}_{1}-{\Delta}_{2}} \, 
\left( \frac{z_{1,0}}{z_{0,-1} z_{1,-1}} \right)^{{\Delta}_{0}-{\Delta}_{-1}} 
 \, \left( \frac{{\gamma}_{1,0} {\gamma}_{1,-1}}{{\gamma}_{0,-1}} \right)^{s_{1}-s_{2}} \, 
  \left( \frac{{\gamma}_{1,0}}{{\gamma}_{0,-1} {\gamma}_{1,-1}} \right)^{s_{0}-s_{-1}} \label{eq:4wzw}
\end{multline}
with 
\beq
{\yt} = \frac{{\gamma}_{0,-1}{\gamma}_{2,1}}{{\gamma}_{1,-1}{\gamma}_{2,0}} \, , \ {\qe} = \frac{z_{0,-1}z_{2,1}}{z_{1,-1} z_{2,0}} \, , 
\eeq 
\beq
2a = -1 - s_{-1} -s_{0} +s_{1} - 3s_{2}, 2b = - 1 +s_{-1} -s_{0} -s_{1} + s_{2}, 2c = 1+ s_{-1} +s_{0} + s_{1} + s_{2} \ .
\label{eq:abc}
\eeq 
 The KZ equation assumes a form of the simple partial differential equation:
 \begin{multline}
 \frac{1}{\kappa} \frac{\partial {\psi}}{\partial {\qe}} = - \frac{1}{{\kappa}^2} \frac{y (y-1)(y-{\qe})}{{\qe}({\qe}-1)} {\partial}^{2}_{yy} {\psi} \ - \\ -  \left( \frac{{\hat\vt}_{0}^2 -\frac{1}{4 {\kappa}^{2}}}{y (1-{\qe})} + \frac{ \frac 14 (1 + \frac{1}{\kappa})^2 - ({\hat\vt}_{\qe} + \frac{1}{\kappa})^2}{y-{\qe}} +  
 \frac{\left( {\hat\vt}_{1}^{2} - \frac{1}{4{\kappa}^{2}} \right) y}{(y-1) {\qe}} + \frac{y \left( ({\hat\vt}_{\infty} - {\half})^2- \frac{1}{4{\kappa}^{2}} \right)}{{\qe}(1-{\qe})} 
  \right) {\psi}
 \label{eq:bpz6}
 \end{multline}
 where ${\hat\vt}$'s are given by the Eqs. \eqref{eq:thehat}, 
 with  $s_{-1} = {\kappa}{\vt}_{0}, s_{0} = {\kappa} {\vt}_{\qe}, s_{1} = {\kappa} {\vt}_{1}, s_{2} = {\kappa}{\vt}_{\infty}$. 
 
 The Eq. \eqref{eq:bpz6} is none other than the BPZ equation \cite{BPZ} for the Liouville
 conformal block with one degenerate field and four generic primary fields.  
 
  \subsubsection{From KZ to BPZ in the separated variables}
 
 Now let us now consider the general $p$ case. What follows goes, in a way, back to \cite{FZ, Fateev, sklyanin} and has been rediscovered and discussed in various publications, e.g. in \cite{Sasha, Feigin:1994in}, was connected to free field representations \cite{Dotsenko:1984nm, SV, FreeFields, Babujian:1993tm, Ribault:2005wp}, to 
 $T$-duality and Fourier-Mukai transform \cite{Gorsky:1999rb}, RG flows \cite{Frenkel:2015rda} etc.\footnote{We should warn the reader, that in the absence of the microscopic derivation, the conjectures based on plausible arguments of holomorphy, classical limits, or analogies with the two dimensional CFT, may prove wrong. For example, the two realizations of surface defects, using quiver gauge theory with fine tuned masses, i.e. the vortex strings, and the regular defects, which could be realized using orbifolds, are not different points on the renormalization group trajectory, but rather are different points in the K{\"a}hler moduli space of the surface theory. In other words they are related by the generalized flop transitions \cite{Jeong:2018qpc, SJNN}, and both are needed to fully package the geometry of the infrared theory, including the effective twisted superpotential.} 
 
 The reason we present it here is to elucidate a few subtleties usually not presented, and to resolve an apparent contradiction with the classical case. 
 
 Let us present the solution ${\Psi}({\gamma}_{-1}, {\gamma}_{0} , \ldots , {\gamma}_{p+1}; z_{-1}, z_{0} , \ldots , z_{p+1})$ as an integral:
 \beq
 {\Psi}({\vec\gamma}, {\vec z}) = {\CalF} \left[ {\hat\Psi} \right]({\vec\gamma}, {\vec z}) : =  \int_{\Gamma} d^{p+2}{\beta} \ {\hat\Psi} ( {\vec\beta} ; {\vec z} ) \, {\exp} \left( \sum_{i=-1}^{p} {\beta}_{i} ( {\gamma}_{i} - {\gamma}_{p+1} ) \right)
 \label{eq:fourier}  \eeq
 over an appropriate $p+2$-dimensional contour $\Gamma$. In order to manipulate the generators $J^{a}_{i}$ acting on $\Psi$ we form the generating functions $J^{a}(z)$, operator-valued meromorphic functions of an auxiliary variable $z$:
 \beq
 J^{a}(z) = \sum_{i=-1}^{p+1} \frac{1}{z-z_{i}} J_{i}^{a}
 \eeq
 We also form the generating function of KZ connections \eqref{eq:kze}:
 \beq
 {\bf D}(z) =  \sum_{i=-1}^{p+1} \frac{1}{z-z_{i}} {\nabla}_{i}
 \label{eq:genkze}
 \eeq
 The following relations are easy to verify:
 \beq
{\bf D}(z) {\Psi} = {\CalF} \left[  \left( {\CalD}(z) -  \sum_{i=-1}^{p+1} \frac{s_{i}(s_{i}+1)}{(z-z_{i})^{2}}  +
 {\bf j}^{0}(z)^{2} - {\partial}_{z}\, {\bf j}^{0}(z) + {\bf j}^{-}(z) {\bf j}^{+}(z) \right) {\hat\Psi} \right] 
 \label{eq:kztrans}
 \eeq 
 with 
 \beq
 \begin{aligned}
&  {\CalD}(z) = \sum_{i=-1}^{p+1} \frac{\kappa}{z-z_{i}} \frac{\partial}{{\partial}z_{i}}\ , \\
& {\bf j}^{0}(z) = {\CalV}(z) - \sum_{i=-1}^{p+1} \frac{s_{i}+1}{z-z_{i}} \, \\
& {\CalV}(z) = - \sum_{i=-1}^{p} \frac{1}{z-z_{i}} {\beta}_{i} \frac{\partial}{{\partial}{\beta}_{i}} \ .
\end{aligned} 
 \label{eq:j0}
 \eeq
 The first order differential operators ${\CalD}(z), {\CalV}(z)$ are
 vector fields on  the space ${\BC}^{2(p+2)}$ with the coordinates 
 \[ 
 ({\beta}_{-1}, \ldots , {\beta}_{p}; z_{-1}, \ldots , z_{p}) \ . \]
 The operator
 ${\bf j}^{-}(z)$ acts by multiplication by
 \beq
 {\beta}(z) = \sum_{i=-1}^{p} \left( \frac{{\beta}_{i}}{z-z_{i}} - \frac{{\beta}_i}{z-z_{p+1}} \right)  \ .
 \eeq 
Finally,  ${\bf j}^{+}(z)$ is a second order differential operator in $\beta$'s whose explicit form we don't need\footnote{\[ {\bf j}^{+}(z) = - \sum_{i=-1}^{p} \frac{1}{z-z_{i}} \left( {\beta}_{i} \frac{\partial^2}{\partial \beta_i^2} + 2 ( s_{i}+1) \frac{\partial}{\partial \beta_i} \right) \]}, since we are going to use a clever trick \cite{sklyanin}. Namely, we pass to the new variables: 
 $({\beta}_{-1} , \ldots , {\beta}_{p}) \mapsto ( b ; w_{0}, \ldots , w_{p})$, 
 where
 ${\beta}(w_{a}) = 0$, i.e.
 \beq
 {\beta}(z) = b\ \frac{W(z)}{Z(z)}\, , \ {\rm with} \ W (z) = \prod\limits_{a=0}^{p} ( z-w_{a})\, , \ Z(z) = \prod\limits_{i=-1}^{p+1}(z-z_{i})\, , 
 \eeq 
and $b = \sum_{i=-1}^{p} (z_i - z_{p+1}) {\beta}_{i}$.  
It is a simple matter to express the vector fields ${\CalD}(z), {\CalV}(z)$ in the new coordinate system.  We observe:
 \begin{multline}
 {\CalL}_{{\CalD}(u)} {\beta}(v) = {\kappa} \frac{\partial}{\partial v} \frac{{\beta}(u) - {\beta}(v)}{u-v} \, = \, \\
 \frac{{\beta}(v)}{b}  ({\CalL}_{{\CalD}(u)}b)  + \sum_{a=0}^{p} \frac{{\beta}(v)}{w_{a}-v}  ({\CalL}_{{\CalD}(u)} w_{a} )  + \sum_{i=-1}^{p+1} \frac{{\beta}(v)}{v-z_{i}}  ({\CalL}_{{\CalD}(u)} z_{i} ) \, , \\ {}{\CalL}_{{\CalV}(u)} {\beta}(v) = \frac{{\beta}(u) - {\beta}(v)}{u-v} + \frac{{\beta}(u)}{v-z_{p+1}}\, = \, \\
 \frac{{\beta}(v)}{b}  ({\CalL}_{{\CalV}(u)}b)  + \sum_{a=0}^{p} \frac{{\beta}(v)}{w_{a}-v}  ({\CalL}_{{\CalV}(u)} w_{a} )
 \end{multline}
 with 
 $u, v$ auxiliary parameters, 
 hence
 \beq
 {\CalV}(z) = - {\beta}(z)(z-z_{p+1}) \left[ \frac{\partial}{\partial b} + \sum_{a=0}^{p} \frac{1}{(z - w_{a}) (w_{a}-z_{p+1}){\beta}^{\prime}(w_{a})} \frac{\partial}{\partial w_{a}} \right]
\eeq 
 and
 \beq
 \frac{1}{\kappa} {\CalD}(z) =  {\beta}(z) \frac{\partial}{\partial b} + \sum_{i=-1}^{p+1} \frac{1}{z-z_{i}} \frac{\partial}{\partial z_i} + \sum_{a=0}^{p} 
\frac{1}{z-w_{a}} \left(  1  - \frac{{\beta}(z)}{{\beta}^{\prime}(w_{a}) (z-w_{a})} \right) \frac{\partial}{\partial w_{a}} \ . 
 \eeq
 Note that the presence of the $w_a$-derivatives in the ${\CalD}(z)$ vector field is the reflection of the \emph{background dependence} of the $w$-coordinates we alluded to in the section $6$.

 {}Let us now solve the KZ equations
 \beq
 {\bf D}(z) {\Psi} = 0
 \label{eq:kzbeta}
 \eeq
together with the global $Lie(G)$ constraints. The $\sum_{i=-1}^{p+1} J_{i}^{-} {\Psi} = 0$
is obeyed trivially in the $\beta$-representation, 
the $\sum_{i=-1}^{p+1} J_{i}^{0} {\Psi} = 0$ equation translates to $[z^{-1}] {\bf j}^{0}(z) {\hat\Psi} = - {\hat\Psi}$.  

Now comes the clever trick \cite{sklyanin}. The left hand side of \eqref{eq:kzbeta}, as a function of $z$, has only the first order poles at $z = z_i$, being equal to \eqref{eq:genkze}. Moreover, with \eqref{eq:j0solve} in place, at large $z$ it goes to zero faster than $z^{-2}$, 
 therefore  the right hand side of \eqref{eq:kzbeta} has the form:
 \beq
 {\bf D}(z){\bf\Psi} = {\CalF} \left[  \frac{W(z)}{Z(z)} \sum_{a=0}^{p} \frac{Z(w_{a})}{W^{\prime}(w_{a}) (z-w_{a})} {\bf D}(w_{a}){\Upsilon} \right]
 \eeq
It remains to compute ${\bf D}(w_{a})$. First note that, as $z \to w_{a}$, 
 the vector field ${\CalV}(z)$ approaches  $-{\partial}_{w_{a}}$. Next, in computing ${\bf j}^{0}(z){\bf j}^{0}(z)$ for $z \to w_{a}$ one should be careful in that the variable $z$ is not acted upon by the derivatives in $w_{a}$'s and $z_i$'s. Fortunately the term ${\partial}_{z} {\bf j}^{0}(z)$ in \eqref{eq:kztrans} precisely cancels this discrepancy leading to  
\begin{multline}
 {\mathfrak D}_{a} := 
- {\bf D}(z\to w_{a}) = 
 {\nabla}_{a}^{2} +  \sum_{b\neq a} 
 \frac{\kappa}{w_{a}-w_{b}}  \left( \frac{\partial}{\partial w_{a}}  - \frac{\partial}{\partial w_{b}} \right) - \\
 - \sum_{i=-1}^{p+1} \left( \frac{s_{i}(s_{i}+1)}{(w_{a}-z_{i})^{2}} + \frac{\kappa}{w_{a} - z_{i}} \left( \frac{\partial}{\partial z_i} + \frac{\partial}{\partial w_{a}} \right)  \right)
 \label{eq:kzbpz2}
 \end{multline}
 with
 \beq
 {\nabla}_{a} = \frac{\partial}{\partial w_{a}} + \sum_{i=-1}^{p+1} \frac{s_{i} +1}{w_{a} - z_{i}}
 \eeq
 The equation $\sum_{i=-1}^{p+1} J^{+}_{i} {\bf\Psi}=0$ is, in fact, contained in the equations ${\mathfrak D}_{a} {\Upsilon} = 0$, since 
\beq
[z^{-3}]{\bf D}(z) {\bf\Psi} =  {\CalF} \left[ b^{2-S} \left( [z^{-1}]{\bf j}^{+}(z) \right) {\Upsilon} \right]
\eeq
assuming  
we already imposed the global $Lie(H)$-constraints:
\beq
L_{-1} {\bf\Psi} =  [z^{-1}] {\CalD}(z){\bf\Psi}= 0 \, , \  L_{0}{\bf\Psi} = \frac{1}{\kappa}[z^{-2}]{\CalD}(z){\bf\Psi} - {\Delta} {\bf\Psi} = 0 
\label{eq:lm0}
\eeq 
with 
\beq
 {\Delta} = \sum_{i=-1}^{p+1} {\Delta}_{i} \ , 
 \eeq
 implying
 \begin{multline}
 {\hat\Psi}( {\vec\beta}; {\vec z} ) = b^{1-S} {\Upsilon} ({\vec w}; {\vec z})\, ,  \\
 {\Upsilon} ({\vec w}; {\vec z}) = 
 (z_{p}-z_{-1})^{S-1+{\Delta}} \ {\Xi}( \frac{w_{0}-z_{-1}}{z_{p} - z_{-1}}, \ldots , \frac{w_{p} -z_{-1}}{z_{p}-z_{-1}} ; \frac{z_{0}-z_{-1}}{z_{p}-z_{-1}}, \ldots , \frac{z_{p+1}-z_{-1}}{z_{p}- z_{-1}})
 \label{eq:j0solve}
 \end{multline}
 with
 \beq
 S = \sum_{i=-1}^{p+1} (s_{i}+1)  \, , \
 \eeq
and
\begin{multline}
L_{+} {\Upsilon} = \\ \sum_{i=-1}^{p+1} \left(  z_{i}^{2} \frac{\partial \Upsilon}{\partial z_{i}} - (2 {\Delta}_{i}+ S - 1) z_{i} {\Upsilon} \right) + \left(  \sum_{a=0}^{p} w_{a}^{2} \frac{\partial \Upsilon}{\partial w_{a}}  + (S-1) w_{a}{\Upsilon} \right) = 0
\label{eq:lplus}
\end{multline}
is the remaining global $Lie(H)$-invariance constraint. Together, \eqref{eq:lm0} and \eqref{eq:lplus} signify that the $w$-variables transform under the $H$-action together with the $z$-parameters, with some effective conformal dimensions assigned both to $z_i$'s and $w_a$'s. 

{}One can further define
\begin{multline}
{\Upsilon}(w_{0}, \ldots ,w_{p}; z_{-1} , \ldots , z_{p+1}) = \\
\prod_{-1 \leq i < j \leq p+1} (z_{i}  - z_{j})^{s_{i} +s_{j} + 2 - \frac{\kappa}{2}} \,  \prod_{a, i} ( w_{a} - z_{i} )^{\frac{\kappa}{2} - s_{i} - 1}
\prod_{a< b} (w_{a} - w_{b} )^{-\frac{\kappa}{2}}  \ {\tilde\Upsilon}( {\vec w}; {\vec z}) 
\end{multline}
to map \eqref{eq:kzbeta} precisely to the BPZ equations \cite{BPZ} 
\begin{multline}
\frac{\partial^2 {\tilde\Upsilon}}{{\partial} w_{a}^{2}}  =  \sum_{b \neq a} \left( \frac{\kappa}{w_{a}- w_{b}}\frac{\partial}{{\partial} w_{b}} + \frac{{\kappa}( {\scriptstyle{\frac 34}} {\kappa}- {\half})}{(w_{a} - w_{b})^{2}} \right) {\tilde\Upsilon}
 + \\
 \sum_{i=-1}^{p+1} \left( \frac{{\kappa}}{w_{a} - z_{i}} \frac{\partial}{\partial z_{i}} + 
 \frac{s_{i}(s_{i}+1) + \frac{\kappa}{2} \left( 1 - \frac{\kappa}{2} \right)}{(w_{a}-z_{i})^{2}} \right) {\tilde\Upsilon} 
 \label{eq:kzbpz3}
 \end{multline}
 This relation is usually interpreted in some kind of coset realization of minimal models
from the WZNW theory, using free fields \cite{FreeFields}. Since we established this relation 
for complex values of  all parameters, spins, levels, it cannot be explained in this way. Instead, it means that the analytic continuation of the WZNW theory, being actually a four dimensional theory, is a Liouville/Toda theory, coupled to a sigma model on the flag variety. The Liouville/Toda theory is the BPS/CFT dual of the supersymmetric gauge theory we used in our miscroscopic computations of the partition functions and expectation values of the surface defect, so that, in particular, ${\kappa} = n {\ve}_{2}/{\ve}_{1}$ for $SU(n)$ theory. 
We plan to return to this elsewhere. 

\subsubsection{NS limits}

The two limits, ${\ve}_{2} \to 0$ and ${\ve}_{1} \to 0$,  are of special interest. In the ${\ve}_{2} \to 0$ limit, ${\kappa} \to 0$ and \eqref{eq:kzbpz3} becomes truly separated, so that $w_a$'s don't talk to each other. In this way one obtains the $SL_2$-opers \cite{BD} (in this case simply second order differential operators with regular singularities \cite{Fuchs})  as a way to package the low-energy 
information of the gauge theory, as in \cite{NRS, Jeong:2018qpc}. The KZ/BPZ relation becomes, in this limit, a relation between the Gaudin quantum integrable system (a genus zero version of the Hitchin system), and the monodromy data of the $SL_2$-opers. For integral $s_i$
this relation was explored in \cite{Feigin:1994in}. In \cite{NRS} a proposal was made for general $s_i$'s.

The opposite,  ${\ve}_{1} \to 0$, limit corresponds to the quasiclassical limit ${\kappa}\to \infty$ \cite{Kolya}. 

In this limit the Eq. \eqref{eq:bpz6}, with the ansatz ${\psi} = e^{\kappa S}$ becomes the Hamilton-Jacobi equation of the Painlev{\'e} VI, as 
 in \cite{FZ, Litvinov:2013sxa}.  More generally, let $s_{i} = {\kappa}{\vt}_{i}$, $i = -1, \ldots, p$, $s_{p+1} = \frac 12$, and let us look for the solutions of the KZ equation in the WKB form:
\beq
{\bf\Psi} = e^{{\kappa}S ({\gamma}_{-1}, \ldots, {\gamma}_{p}; z_{-1}, \ldots , z_{p})} \, {\chi}({\gamma}_{-1}, \ldots, {\gamma}_{p}, {\gamma}_{p+1}; z_{-1}, \ldots , z_{p+1} ) \eeq
where 
\beq
{\chi}({\gamma}_{-1}, \ldots, {\gamma}_{p}, {\gamma}_{p+1}; {\bz} )  = \, {\chi}_{-}({\gamma}_{-1}, \ldots, {\gamma}_{p}; {\bz} )  + {\gamma}_{p+1} {\chi}_{+}({\gamma}_{-1}, \ldots, {\gamma}_{p}; {\bz} )  + O(1/{\kappa})
\eeq
Then the leading term in $\kappa$, the equations \eqref{eq:kze1} become:
\beq
\frac{\partial S}{\partial z_i} = \sum_{j\neq i} \frac{{\rm Tr}A_{i}A_{j}}{z_{i}-z_{j}} \, , \ i = -1, \ldots , p
\label{eq:schleshj}
\eeq
with $A_{i} = A( {\beta}_{i} \frac{{\gamma}_{i}^{2}-1}{2} - {\vt}_{i}{\gamma}_{i} ,  {\beta}_{i} \frac{{\gamma}_{i}^{2}+1}{2\ii} + {\ii}{\vt}_{i}{\gamma}_{i}, {\vt}_{i} - {\beta}_{i}{\gamma}_{i} )$, cf. \eqref{eq:orb1}, ${\beta}_{i} = {\partial}S/{\partial\gamma}_{i}$, 
 and
 \beq
 \frac{\partial}{\partial z_{p+1}} \left( \begin{matrix} {\chi}_{+} \\ {\chi}_{-} \end{matrix} \right)
 = \sum_{i} \frac{A_{i}}{z_{p+1}-z_{i}} \left( \begin{matrix} {\chi}_{+} \\ {\chi}_{-} \end{matrix} \right) \ .
\label{eq:horsec} \eeq
 The Eq. \eqref{eq:schleshj} is none other than our friend
 the Hamilton-Jacobi form of the Schlesinger system, while 
 \eqref{eq:horsec} shows $\Upsilon$ is the horizontal section of the meromorphic
 connection 
 \beq
 {\partial}_{z} + \sum_{i} \frac{A_{i}}{z-z_{i}}
 \eeq 
A much more thorough account of these matters can be found in \cite{Kolya, Harnad:1994fk}. 
   
 {}On the other hand, setting ${\tilde\Upsilon} = e^{{\kappa}{\tilde S}({\vec w}; {\vec z})}$ in \eqref{eq:kzbpz3}  and sending $\kappa \to \infty$, we arrive at:
 \begin{multline}
\left( \frac{\partial {\tilde S}}{{\partial} w_{a}} \right)^{2}  = \sum_{b \neq a} \left( \frac{1}{w_{a}- w_{b}}\frac{\partial {\tilde S}}{{\partial} w_{b}} + \frac{3}{4 (w_{a} - w_{b})^{2}} \right)
 + \\
 \sum_{i=-1}^{p+1} \left( \frac{1}{w_{a} - z_{i}} \frac{\partial \tilde S}{\partial z_{i}} + 
 \frac{{\vt}_{i}^2  - \frac{1}{4}}{(w_{a}-z_{i})^{2}} \right)\, , \\ 
 a = 0, \ldots , p \ .
 \label{eq:kzbpz34}
 \end{multline}
Clearly, \eqref{eq:kzbpz34} do not look like the Painlev{\'e}-Schlesinger equations, as we have $p+1$ variables $w_{a}$, as opposed to $p$ variables in the $\yt$-representation. Indeed, unlike the classical case where we could fix a gauge where ${\beta}_{\infty} = 0$, i.e. $w_{p} = z_{p+1}$, in the quantum case this is not possible, the $J^{+}$ generator of the $Lie(G)$ symmetry being represented, in the $\beta$-representation, by a second order differential operator. However, the solution to the equation ${\mathfrak D}_{0} {\Upsilon} = 0$ at fixed $w_{1}, \ldots, w_{p}$, as a function of $w_{0}$, can be fixed once the initial conditions are chosen. Physically it means 
bringing  one of the degenerate fields ${\CalV}_{(2,1)}$ to one of the primaries or another degenerate field, and extracting the asymptotics following from the operator product expansion. 
Since the equation is of the second order, one needs two initial conditions, which corresponds to the fact that the fusion of the generic primary field with ${\CalV}_{(2,1)}$ contains exactly two primaries\footnote{It is useful to perform this exercise even in the case $p=0$, where this two-fold
degeneracy is exhibited in the monodromy of the hypergeometric function $_{2}F_{1}$}. 

Thus, the PVI equation is recovered in the limit e.g. $w_{0} \to \infty$ with the shift of ${\vt}_{\infty}$, as in \cite{Litvinov:2013sxa}. 
In this way the Okamoto transformation \eqref{eq:wmap1} is recovered asymptotically.

 \section{Higher rank generalizations II: flat connections}
 
 We shall now describe the moduli space ${\CalM}^{\rm alg}_{r,p}$ of meromorphic $G_r$-connections on the $p+3$-punctured sphere with the residues $A_{0}$ and $A_{\infty}$ belonging to the orbits
 ${\CalO}_{\tt reg}^{{\vec\nu}^{(0)}, r}$ and ${\CalO}_{\tt reg}^{{\vec\nu}^{({\infty})}, r}$ with some generic
 vectors ${\vec\nu}^{(0)}, {\vec\nu}^{({\infty})} \in {\BC}^{r}$, and 
 $A_{z_{0}}, A_{z_{1}}, \ldots , A_{z_{p}}$ belonging to ${\CalO}_{\tt min}^{{\nu}_{0}, r}, {\CalO}_{\tt min}^{{\nu}_{1}, r}, \ldots , {\CalO}_{\tt min}^{{\nu}_{p}, r}$, respectively, with ${\nu}_{0}, {\nu}_{1}, \ldots, {\nu}_{p} \in {\BC}$:
 \beq
 {\CalA}(z) = \frac{{\mu}_{\tt reg}({\bu}^{(0)}, {\bv}^{(0)})}{z} + \sum_{b=0}^{p} \frac{{\mu}_{\tt min} (u^{(z_{b})}, v^{(z_{b})})}{z-z_{b}}
 \eeq
 so that the sum of the residues vanishes:
 \beq
 {\mu}_{\tt reg}({\bu}^{(0)}, {\bv}^{(0)}) + {\mu}_{\tt reg}({\bu}^{({\infty})}, {\bv}^{({\infty})}) +  \sum_{b=0}^{p} {\mu}_{\tt min} (u^{(z_{b})}, v^{(z_{b})}) = 0 
 \eeq
As before, there are several coordinate systems we can put on the moduli space
\beq
{\CalM}^{\rm alg}_{G_{r}, p} = \left( {\CalO}_{\tt reg}^{{\vec\nu}^{(0)}, r} \times    {\CalO}_{\tt min}^{{\nu}_{0}, r} \times {\CalO}_{\tt min}^{{\nu}_{1}, r} \times  \ldots \times {\CalO}_{\tt min}^{{\nu}_{p}, r} \times {\CalO}_{\tt reg}^{{\vec\nu}^{({\infty})}, r} \right) // G_{r}
\label{eq:mpmdsp}
\eeq
The analogues of the $\gamma$, $\yt$-coordinates are the invariants built
out of $v$'s only. In the basis $(v_{i})$, where $A_{0}$ is lower-triangular, while $A_{\infty}$ is upper-triangular, and $u^{(1)} = \sum_{i=1}^{r+1} v_{i}$ (this is the higher rank analogue of the $G$-gauge where ${\gamma}_{0} = 0$, ${\gamma}_{1} = 1$, and ${\gamma}_{\infty} = {\infty}$) we define simply
\beq
{\yt}_{a}^{m} = v_{(z_{a})}^{m}
\eeq
where $1 \leq m \leq r$, $1 \leq a  \leq p$. More invariantly, define the `tau-functions'
\beq
\begin{aligned}
& {\tau}_{0, {\xi}} =  {\psi}^{({\infty})}_{r} \wedge \ldots \wedge {\psi}^{({\infty})}_{1} \wedge v_{({\xi})}\, , \\
& {\tau}_{i, {\xi}} = {\psi}^{(0)}_{i} \wedge {\psi}^{(0)}_{i-1} \wedge \ldots \wedge {\psi}^{(0)}_{1} \wedge {\psi}^{({\infty})}_{r-i} \wedge \ldots \wedge {\psi}^{({\infty})}_{1} \wedge v_{({\xi})}\, , \\
& \qquad\qquad\qquad\qquad\qquad\qquad i = 1, \ldots , r-1 \\
& {\tau}_{r, {\xi}} = {\psi}^{(0)}_{r} \wedge {\psi}^{(0)}_{r-1} \wedge \ldots \wedge {\psi}^{(0)}_{1} \wedge  v_{({\xi})}\, 
\end{aligned}
\label{eq:hirota}
\eeq
where, in terms of the $(B^{(0)}, {\tilde B}^{(0)}), (B^{({\infty})}, {\tilde B}^{({\infty})})$ data attached to the residues $A_{0}$ and $A_{\infty}$ respectively, 
\beq
{\rm Span} ( {\psi}^{({\xi})}_{1}, \ldots , {\psi}^{({\xi})}_{i} ) = {\bV}_{i}^{({\xi})}\, , \qquad {\xi} = 0, {\infty}
\eeq 
The tau-functions \eqref{eq:hirota} are not gauge-invariant, they are defined up to a scaling. However, the ratios
\beq
{\yt}_{a}^{m} = \frac{{\tau}_{m-1, z_{a}} {\tau}_{m, z_{0}}}{{\tau}_{m, z_{a}} {\tau}_{m-1, z_{0}}}\, , \ m = 1, \ldots, r\, , \ a = 1, \ldots, p
\eeq
are well-defined meromorphic functions on ${\CalM}_{r,p}^{\rm alg}$. 

Let us  now specify to the $p=1$ case. The analogue of the polygon length coordinates $\pm {\ell}$ are the eigenvalues
${\ell}_{i}$, $i = 1, \ldots , r$, and ${\ell}_{r+1} = - {\ell}_{1} - \ldots - {\ell}_{r+1}$ of the sum $A_{0} + A_{\qe}$, which are equal to the eigenvalues of $A_{1} + A_{\infty}$. 
Let us package the eigenvalues of $A_{0}$, $A_{0} + A_{\qe} = - A_{1} - A_{\infty}$, $A_{\infty}$
with the help of the characteristic polynomials:
\beq
\begin{aligned}
& {\CalL} (z)  = {\rm Det} ( A_{0} + A_{\qe} - z ) = \prod\limits_{l=1}^{r+1} ({\ell}_{l} - z )\, , \\
& {\CalA}_{0}(z) =  {\rm Det} (A_{0} - z ) = \prod\limits_{l=1}^{r+1} ( {\vt}^{(0)}_{l} - z ) \, , \\
& {\CalA}_{\infty} (z) = {\rm Det} (A_{\infty} - z ) = \prod\limits_{l=1}^{r+1} ( {\vt}^{({\infty})}_{l} - z ) \ .\\
\end{aligned}
\label{eq:charpol}
\eeq
The simple identities follow from the special form of $A_{\qe}$, $A_{1}$:
\beq
\begin{aligned}
& \frac{{\CalL} (z)}{{\CalA}_{0} (z - {\nu}_{\qe})} =  
 1 - u_{({\qe})} \frac{1}{A_{0} + {\nu}_{\qe} - z} v^{({\qe})}   \, , \quad
\frac{(-)^{N} {\CalL} (-z)}{{\CalA}_{\infty} (z - {\vt}_{1})} = 
 1 - u_{(1)} \frac{1}{A_{\infty} + {\nu}_{1} - z} v^{(1)}    \, , \\
 & \frac{{\CalA}_{0} (z)}{{\CalL} ( z + {\nu}_{\qe} )} =  1 + u_{({\qe})} \frac{1}{A_{0} + A_{\qe} - {\nu}_{\qe} - z} v^{({\qe})}\, , \quad \frac{{\CalA}_{\infty} (z)}{{\CalL}  ( z+ {\nu}_{1} )} = 1 + u_{(1)} \frac{1}{A_{1} + A_{\infty} - {\nu}_{1} - z} v^{(1)} \, , \\\\
 \end{aligned}
 \eeq
 implying the residues:
\beq
\begin{aligned}
& \sum_{i} u_{({\qe})} (v_{i}) u^{i} = u_{({\qe})} \\
& u^{i}(v^{(1)})u_{(1)}(v_{i}) = - \frac{{\CalA}_{\infty} ({\ell}_{i} - {\nu}_{1})}{{\CalL}^{\prime}({\ell}_{i})} \, , \\
& u^{i}(v^{({\qe})})u_{({\qe})}(v_{i}) = - \frac{{\CalA}_{0} ({\ell}_{i} - {\nu}_{\qe})}{{\CalL}^{\prime}({\ell}_{i})} \, , \\
\end{aligned}
\eeq
where $v_{i} \in {\CalV}$, $u^{i} \in {\CalV}^{*}$ are the eigenvectors of $A_{0} + A_{\qe} = - A_{1} - A_{\infty}$, with the eigenvalue ${\ell}_{i}$. 

The analogue of the angles $\pm \theta$ are the complex angular variables ${\Theta}_i$
defined by:
\beq
e^{{\Theta}_i} = \frac{u^{i}(v^{(1)})}{u^{i}(v^{({\qe})})} \frac{u^{i+1}(v^{({\qe})})}{u^{i+1}(v^{(1)})}
\, , \qquad i = 1, \ldots, r
\label{eq:mulpol}
\eeq
from which we compute the Hamiltonian
\beq
H_{\qe} = \frac{h_{0\qe}}{\qe} + \frac{h_{1\qe}}{{\qe}-1}
\label{eq:gaudin4}
\eeq
with
\beq
h_{0{\qe}} = \frac{1}{2} \left( - {\vec \nu}_{0}^{2} - {\scriptstyle{\frac{1}{r+1}}} {\nu}_{\qe}^{2} +  \sum_{i=1}^{r+1} {\ell}_{i}^{2} \right) 
\eeq
and
\beq
h_{1{\qe}} = - (r+1) {\nu}_{\qe}{\nu}_{1} + \sum_{m,n=1}^{r+1} \frac{{\CalA}_{\infty} ({\ell}_{n} - {\nu}_{1})}{{\CalL}^{\prime}({\ell}_{n})} \frac{{\CalA}_{0} ({\ell}_{m} - {\nu}_{\qe})}{{\CalL}^{\prime}({\ell}_{m})} e^{{\theta}_{m} - {\theta}_{n}}
\eeq
where ${\theta}_{i} - {\theta}_{i+1} = {\Theta}_{i}$. 

This construction can be inductively generalized to the case of general $p$. At the $a$'th step, take $r$ of 
the eigenvalues ${\ell}_{i}^{(a)}$, $i = 1, \ldots , r$, ${\ell}_{r+1}^{(a)} = - {\ell}_{1}^{(a)} - \ldots - {\ell}_{r}^{(a)}$ of 
\beq
{\CalL}_{a} = A_{0} + A_{z_{0}} + \ldots + A_{z_{a-1}} = 
- \left( A_{z_{a}} + A_{z_{a+1}} + \ldots + A_{z_{p}} + A_{\infty} \right)\ , 
\eeq
and the angles
\beq
e^{{\Theta}_i^{(a)}} = \frac{{\bu}^{i}(v^{(z_{a})})}{{\bu}^{i}(v^{(z_{a-1})})} \frac{{\bu}^{i+1}(v^{(z_{a-1})})}{{\bu}^{i+1}(v^{(z_{a})})}
\, , \qquad i = 1, \ldots, r
\label{eq:mulpolp}
\eeq
where ${\bu}^{i} {\CalL}_{a} = {\ell}_{i}^{(a)} {\CalL}_{a}$ are the eigenvectors corresponding to the chosen eigenvalues. 

Finally, the analogues of ${\alpha}, {\beta}$-coordinates on the moduli space ${\CalM}^{\rm loc}_{r,p}$ of local systems, i.e. the monodromy data, i.e representations of the fundamental group 
of the $p+3$-punctured sphere into $G_{r}$, such that the conjugacy class of the monodromy around $z_{a}$ is that of ${\exp} \, 2{\pi}{\ii}A_{z_{a}}$, while those around $0$, $\infty$ are equal to the conjugacy class of ${\exp} \, 2{\pi}{\ii} A_{0, \infty}$, respectively, are described in \cite{Jeong:2018qpc}. Another coordinate system, based on the spectral networks, was proposed earlier in \cite{Hollands:2017ahy}. 

We have therefore all the necessary ingredients to formulate the higher rank, general $p$ 
analogue of the GIL formula.  It relates the tau-functions of the higher rank Schlesinger deformations to the $c=1$ conformal blocks of Toda conformal theories, in agreement with the conjectures \cite{Wyllard:2009hg}:
\begin{multline}
{\tau}_{0, p}^{r} ( {\qe}_{1}, \ldots , {\qe}_{p} ; {\vec\alpha}, {\vec\beta}; {\vec\nu}^{(0)} , {\nu}_{0}, \ldots , {\nu}_{p}, {\vec\nu}^{({\infty})} ) \ = \\
\sum_{{\mathfrak{n}}_{1}, \ldots , {\mathfrak{n}}_{p} \in {\BZ}^{r} \subset {\BZ}^{r+1}}  e^{\sum_{a=1}^{p} {\mathfrak{n}}_{a} \cdot {\vec\beta}_{a}} \, {\CalZ}_{A_{p}} \left( ( {\vec\alpha}_{a} + {\mathfrak{n}}_{a} ) {\hbar} ; {\vec m}_{1}, {\mu}_{1}, \ldots , {\mu}_{p-1} , {\vec m}_{p} ; {\qe}_{1}, \ldots , {\qe}_{p} ; {\hbar}, - {\hbar} \right) 
\label{eq:rpconj}
\end{multline}
where on the right-hand side we have the partition functions of the linear quiver gauge theories with the $A_{p}$-type quiver \cite{NP1} with the gauge groups $U(n_{i})$, $n_{i} = r+1$ at each node, with $r+1$ fundamental hypermultiplets of masses ${\vec m}_{1}$, and ${\vec m}_{p}$ at the nodes $U(n_{1})$ and $U(n_{p})$, respectively, and the bi-fundamental hypermultiplets charged under $U(n_{i}) \times U(n_{i+1})$, of mass ${\mu}_{i}$. The relations between the masses and the monodromy data ${\vec\nu}^{(0,{\infty})}$, ${\nu}_{a}$ can be found in \cite{Wyllard:2009hg, NP1, NekBPSCFT}.  

To support the conjecture \eqref{eq:rpconj}  we mention that in Ref. \cite{NT} it is shown that the quantum version of \eqref{eq:gaudin4} is a) equivalent to the Knizhnik-Zamolodchikov equation for ${\widehat{sl(r+1)}}$ four-point conformal block; b) equivalent to the non-perturbative Dyson-Schwinger equation obeyed by the surface defect in the $SU(r+1)$ four dimensional ${\CalN}=2$ super-Yang-Mills theory with $2(r+1)$ fundamental hypermultiplets. 

\section{Genus one and ${\CalN}=2^{*}$ theory}

It was demonstrated in \cite{NekBPSCFT} that the partition function of the regular
surface defect in the $SU(r+1)$ gauge theory with adjoint hypermultiplet, also known as the 
${\CalN}=2^{*}$ theory, obeys the Knizhnik-Zamolodchikov-Bernard \cite{Bernard} equation for the torus 
one-point conformal block with the minimal representation ${\CalV}_{\nu}$:
\beq
(r+1) {\ve}_{1}{\ve}_{2} \frac{d}{d{\tau}} {\bf\Psi} = \left( \sum_{i=1}^{r+1} \frac{{\ve}_{1}^{2}}{2} \frac{{\partial}^2}{{\partial q}_{i}^{2}} + m (m+{\ve}_{1}) \sum_{1 \leq i < j \leq r+1} 
\, {\wp} ( q_{i} - q_{j} ; {\tau} ) \right) {\bf\Psi}
\label{eq:kzb}
\eeq
In the limit ${\ve}_{1} \to 0$, with the rest of the parameters kept fixed the equation \eqref{eq:kzb}
degenerates, via the familiar ansatz ${\bf\Psi} \sim e^{{\ve}_{2} S(q, a/{\ve}_{2}; m/{\ve}_{2}, {\tau})/{\ve}_{1} + \ldots}$, to 
the Hamilton-Jacobi equation 
\beq
(r+1) {\ve}_{2} \frac{dS}{d{\tau}} = H_{\rm eCM} \left( \frac{\partial S}{\partial q_i}, q_i ; m, {\tau} \right)
\label{eq:g1iso}
\eeq 
with $H_{\rm eCM}$ the Hamiltonian of $r+1$-particle elliptic Calogero-Moser model \cite{Calogero}. 
Let us explain the meaning of \eqref{eq:g1iso} in the context of the isomonodromic deformation (cf. \cite{Ivanov:1994ab, Ivanov:1996bg, Levin:1997tb, Takasaki:1997gg, 
Krichever:2001cx}). 

We work on the elliptic curve $E_{\tau} = {\BC} / {\BZ} \oplus {\tau}{\BZ}$, with the coordinate $z \sim z +a + b {\tau}$, for $a,b \in {\BZ}$, which is isomorphic to the standard two-torus $T^{2} = {\BR}^{2}/{\BZ} \oplus {\BZ}$, with the coordinates $(x,y)$, $x \sim x + 1$, $y \sim y+1$. The relation between the complex and the real descriptions is simply:
\beq
z = x  + y {\tau}\, , \ \label{eq:holc}
\eeq 
We are going to study a family of flat connections on $T^{2} \backslash p$,  where the point $p$ is at $z = 0$. Moreover we assume the conjugacy class of the holonomy of the flat connection around $p$ is fixed, with the eigenvalues of multiplicity $(N-1, 1)$, respectively. In other words, 
we allow for a $\delta$-function singularity in the curvature $F_{xy}$,
\beq
F_{xy} = {\partial}_{x} A_{y} - {\partial}_{y} A_{x} + [A_{x}, A_{y}] = {\mu}_{\rm min}(u,v) {\delta}(x,y)
\label{eq:curvdelta}
\eeq
for some $u, v, {\nu}$, as in \eqref{eq:mumin}, and $r = N-1$. There are two ways of solving \eqref{eq:curvdelta}. We can fix a gauge \cite{Gorsky:1993dq} in which $A_{x}$ (equivalently, $A_{y}$) is diagonal $x$-($y$-)independent matrix. But we would like to fix a gauge in a $\tau$-dependent way, and explore the $\tau$-dependence of what follows. So, we fix the gauge, as in \cite{GN}: 
\beq
A_{\bar z} = \frac{2\pi\ii}{{\tau} - {\bar\tau}} {\rm diag}\left( q_{1}, \ldots , q_{r+1} \right) \ , 
\label{eq:azb}
\eeq
meaning that ${\bar\partial} - A_{\bar z} d{\bar z}$ defines a holomorphic vector bundle on $E_{\tau}$, isomorphic to a direct sum of $r+1$ line bundles. 
Then, for $A_{z}$ we get the equation
\beq
{\bar\partial}_{\bar z} A_{z} - [ A_{\bar z}, A_{z} ] \propto  {\mu}_{\rm min}(u,v)  {\delta}^{(2)}(z, {\bar z} ) 
\eeq
which is identical to the equation for a Higgs field in the punctured analogue of Hitchin's equation \cite{GN, Nekrasov:1995nq}. The solution is unique up to a shift of $A_{z}$ by the constant diagonal matrix (cf. \eqref{eq:holc}):
\beq
A_{z} = 2{\pi}{\ii} \sum_{i=1}^{r+1}  \left( p_{i} - \frac{q_{i}}{{\tau}-{\bar\tau}} \right)\, E_{ii} + 
{\nu} \sum_{i \neq j} \frac{{\vt}_{11} (z + q_{ij}; {\tau}) {\vt}_{11}^{\prime}(0; {\tau})}{{\vt}_{11}(z; {\tau}) 
{\vt}_{11}(q_{ij}; {\tau})} \, 
e^{2\pi\ii y q_{ij}} \, E_{ij}
\label{eq:az}
\eeq
where $q_{ij} = q_{i} - q_{j}$, and we used the odd theta function
\beq
{\vt}_{11}(z; {\tau}) = \sum_{r \in {\BZ}+\frac 12} e^{2\pi\ii (z+\frac 12) r + {\pi}{\ii}{\tau}r^2}
\eeq
obeying the heat equation
\beq
{\partial}_{\tau} {\vt}_{11}  = \frac{1}{4\pi\ii} {\partial}_{z}^{2} {\vt}_{11}
\eeq
We use\footnote{Note that our definition of $\wp$ differs by a $\tau$-dependent constant from the standard one}:
\beq
{\xi}(z; {\tau}) = {\partial}_{z} {\rm log}{\vt}_{11}(z; {\tau}) \, , \qquad {\wp}(z) = - {\partial}_{z}{\xi}(z)
\eeq
The family \eqref{eq:azb}, \eqref{eq:az} obeys:
\beq
{\delta}A_{x} = D_{x}{\ep}\, , \ {\delta}A_{y} = D_{y}{\ep}
\label{eq:isom}
\eeq
where the ${\delta}$-variation means varying $\tau$ and $\bar\tau$ while
keeping $x,y$ fixed. Then \eqref{eq:isom} holds, with  
with
\beq
{\ep} = {\nu}{\delta}{\tau}  \sum_{i\neq j} \left[ {\wp}(q_{ij}) E_{ii} + 
\left( {\xi} (z+q_{ij}) - {\xi}(q_{ij}) + 2\pi \ii y \right)
\frac{{\vt}_{11} (z + q_{ij}) {\vt}_{11}^{\prime}(0)}{{\vt}_{11}(z) 
{\vt}_{11}(q_{ij})} \, 
e^{2\pi\ii y q_{ij}} \, E_{ij} \right]
\label{eq:compg1}
\eeq
and the variations of the $q_i$ and $p_i$ parameters according to
\beq
{\delta}q_{i} = p_{i}{\delta}{\tau}\, , \ {\delta}p_{i} =  {\nu}^{2} \sum_{j\neq i}
{\wp}^{\prime}(q_{ij}) {\delta}{\tau}
\eeq
This is nothing but the Hamiltonian flow with the $\tau$-dependent Hamiltonian
\beq
H_{\rm eCM} = \frac 12 \sum_{i=1}^{r+1} p_{i}^{2}  - {\nu}^{2} \sum_{1\leq i< j \leq r+1} {\wp} ( q_{ij}; {\tau}) 
\eeq
Note that in verifying \eqref{eq:isom}, \eqref{eq:compg1} the relations 
\beq
{\wp}(z_1) + {\wp}(z_{2}) + {\wp}(z_{3})  - \left( {\xi}(z_1) + {\xi}(z_{2}) + {\xi}(z_{3}) \right)^{2} = 
- \frac{{\vt}_{11}^{\prime\prime\prime}(0)}{{\vt}_{11}^{\prime}(0)}\ , 
\eeq
for $z_{1} + z_{2} + z_{3} = 0 \in E_{\tau}$, 
and
\beq
- \frac{{\vt}_{11}(z_{1}+z_{2}){\vt}_{11}(z_{1}+z_{3}){\vt}_{11}(z_{2}+z_{3}){\vt}_{11}^{\prime}(0)}{{\vt}_{11}(z_{1}){\vt}_{11}(z_{2}){\vt}_{11}(z_{3}){\vt}_{11}(z_{4})} = {\xi} (z_{1}) + {\xi}(z_{2})+{\xi}(z_{3})+{\xi}(z_{4})
\eeq
for $z_{1} + z_{2} + z_{3} +z_{4} = 0 \in E_{\tau}$, are used. For the general theory of the related
functional equations and their r{\^o}le in the theory of Calogero-Moser systems see
\cite{Calogero, BB}.  The genus one isomonodromy problem is analyzed, in a different gauge, in 
\cite{Krichever:2001cx, Levin:1997tb}. For the relation to the KZB equations, see
\cite{Losev:1991gn, Ivanov:1994ab, Ivanov:1996bg}. It is straightforward to generalize this analysis
to the case of $p$ punctures, specifically for $p$ minimal ones. The formulas for $A_{z}$
are very similar to the formulas for the genus one Higgs field \cite{Nekrasov:1995nq, Krichever:2001cx, Levin:1997tb}.

{}Naturally we expect the isomonodromic $\tau$-function for the genus one with $p$ punctures
$z_1, \ldots , z_{p}$, at least
in the case of the residues $A_i = {\rm res}_{z_{i}} \ A_{z}dz$ of the poles of the meromorphic connection belonging to the minimal orbits ${\CalO}^{\nu_{i}, r}_{\rm min}$,   to be related to the partition function of the ${\ve}_{1} = - {\ve}_{2} = {\hbar}$ affine ${\hat A}_{p-1}$ quiver theory (i.e. circular quiver with $p$ nodes, with the gauge group $SU(n_{0}) \times SU(n_{1}) \times \ldots \times SU(n_{p-1})$, with $n_{i} = r+1$ for all $i$),  
\begin{multline}
{\tau}_{1, p}^{r} \left( {\vec\alpha}_{1}, {\vec\beta}_{1}, \ldots , {\vec\alpha}_{p}, {\vec\beta}_{p} ;  {\tau}, z_{1}, \ldots, z_{p} \right) \sim \\
\sum_{\mathfrak{n}_{1}, \ldots ,  \mathfrak{n}_{p} \in {\BZ}^{r}}\, e^{{\vec\beta}_{i} \cdot  {\vec{\mathfrak{n}}}_{i}} \, {\CalZ}_{{\hat A}_{p-1}} ( ( {\vec{\alpha}}_{i} + {\vec{\mathfrak{n}}}_{i}) {\hbar}; m_{1}, \ldots , m_{p} ; {\hbar}, - {\hbar} ; {\qe}_{1}, \ldots , {\qe}_{p} ) 
\end{multline}
 where $m_i$'s are the masses of the bi-fundamental hypermultiplets charged under the $SU(n_{i}) \times SU(n_{i+1})$, related to the eigenvalues $(r{\nu}_{i}, - {\nu}_{i}, \ldots , - {\nu}_{i} )$ of the residues $A_i$ and the monodromy parameters ${\vec\alpha}_{i}, {\vec\beta}_{i}$ are defined analogously to the genus zero case studied in \cite{Jeong:2018qpc}, ${\qe}_{i}$
 are the instanton factor associated with the $SU(n_i)$ gauge group, 
 \beq
 {\qe}_{1}\ldots {\qe}_{p} = e^{2\pi\ii \tau}\, , \ {\qe}_{i} = e^{2\pi \ii ( z_{i} - z_{i+1})}\, , \qquad i = 1, \ldots, p-1
 \eeq
 and $\sim$ stands for the $U(1)$-factors ($\vec\alpha, \vec\beta$-independent functions of ${\qe}_i$'s).

 \vskip 1cm
 \centerline{$\bullet\sim\bullet\sim\bullet\sim\bullet\sim\bullet\sim\bullet$}

\end{document}